\documentclass{article}
\usepackage{authblk}
\usepackage[utf8]{inputenc}
\usepackage[T1]{fontenc}
\usepackage[a4paper]{geometry}
\usepackage{graphicx}
\usepackage[textwidth=8em,textsize=small]{todonotes}
\usepackage{amsmath,amsthm,amssymb}
\usepackage{natbib}
\usepackage{hyperref}
\usepackage[nameinlink]{cleveref}
\usepackage{subcaption}
\usepackage{bbm}
\usepackage{amsfonts}
\usepackage{mathtools}
\usepackage[titletoc,toc]{appendix}
\usepackage{url}
\usepackage[ruled, vlined, linesnumbered]{algorithm2e}
\usepackage{algpseudocode}
\usepackage{multirow}
\usepackage{booktabs}
\usepackage[normalem]{ulem}
\usepackage{threeparttable}
\usepackage{enumitem}
\usepackage{array}
\usepackage{setspace}
\usepackage{orcidlink}
\newcommand\blank[1]{\multicolumn{#1}{l|}{}}

\definecolor{darkblue}{HTML}{0c7dbb}
\definecolor{darkgreen}{rgb}{0,0.48,0.65}
\hypersetup{
	colorlinks = true,
	citecolor=darkgreen,
	filecolor=black,
	linkcolor=darkblue,
	urlcolor=blue
}
\providecommand{\keywords}[1]
{
	\small
	\textbf{\textit{Keywords---}} #1
}

\DeclareMathOperator*{\argmin}{arg\,min}

\newtheorem{theorem}{Theorem}[section]

\newtheorem{lemma}[theorem]{Lemma}   
\Crefname{algocf}{Algorithm}{Algorithms}
\Crefname{lemma}{Lemma}{Lemmas}
\title{Bayesian Inference General Procedures for A Single-subject Test Study}
\author[1,2]{Jie Li\orcidlink{0000-0001-8353-1322}\footnote{Addresses for correspondence: Jie Li, School of Mathematics, Statistics and Actuarial Science, University of Kent, Canterbury CT2 7NF.\ \textbf{Email}:~\href{jl725@kent.ac.uk}{jl725@kent.ac.uk}}}
\author[2,3]{Gary Green}
\author[2]{Sarah J. A. Carr}
\author[4]{Peng Liu}
\author[1]{Jian Zhang}
\affil[1]{School of Mathematics, Statistics and Actuarial Science, University of Kent, Canterbury, CT2 7NF, UK}
\affil[2]{Innovision IP Ltd., 50 Seymour Street, London, England, W1H 7JG}
\affil[3]{York Neuroimaging Centre, University of York, Innovation Way, York, YO10 5NY, UK}
\affil[4]{Department of Mathematical Sciences, Loughborough University, Loughborough, LE11 3TU, UK}

\begin{document}
\maketitle
\begin{abstract}
	Abnormality detection in identifying a single-subject which deviates from the majority of a control group dataset is a fundamental problem. Typically, the control group is characterised using standard Normal statistics, and the detection of a single abnormal subject is in that context. However, in many situations, the control group cannot be described by Normal statistics, making standard statistical methods inappropriate. This paper presents a Bayesian Inference General Procedures for A Single-subject Test (BIGPAST) designed to mitigate the effects of skewness under the assumption that the dataset of the control group comes from the skewed Student \( t \) distribution. BIGPAST operates under the null hypothesis that the single-subject follows the same distribution as the control group. We assess BIGPAST's performance against other methods through simulation studies. The results demonstrate that BIGPAST is robust against deviations from normality and outperforms the existing approaches in accuracy nearest to the nominal accuracy 0.95.
	BIGPAST can reduce model misspecification errors under the skewed Student
	$t$ assumption by up to 12 times, as demonstrated in Section~\ref{subsec:Model_misspecification_error}. We
	apply BIGPAST to a Magnetoencephalography (MEG) dataset consisting of an
	individual with mild traumatic brain injury and an age and gender-matched
	control group. For example, the previous method failed to detect abnormalities
	in 8 brain areas, whereas BIGPAST successfully identified them, demonstrating
	its effectiveness in detecting abnormalities in a single-subject.

	\keywords{Bayesian inference, Skewed Student's \( t \) distribution, Single-subject test, Magnetoencephalography (MEG), Jeffreys prior}

\end{abstract}
\section{Introduction}\label{sec:Introduction}
Abnormality detection is a fundamental task in many scientific fields, including Medicine, Psychology, Econometrics, Telehealthcare, Neuroscience
as well as other fields for Industry 5.0~\citep{crawfordComparisonSingleCase2007,dasEldocareEEGKinect2023,higbeeComparisonGB2Skewed2024,kim2023cyber,ktenaGenerativeModelsImprove2024,yinFLSNMVOEdgeComputing2024}. One example of such abnormality detection lies in the detection of traumatic brain injury (TBI), which is a fundamental problem, with over 2.8 million individuals a year in America presenting at an Emergency room~\citep{Taylor2017}. We exemplify the use of skewed Student \( t \) statistics in the study of brain activity as a potential biomarker for the identification of a brain injury. We use magnetoencephalography (MEG) to measure brain activity and compare single individual to normal controls from age and gender-matched cohorts. Existing literature typically assumes that the dataset from the control group follows a Normal distribution. However, extensive practical evidence shows that the Normal distribution assumption is often violated.
This violation is frequently observed in brain Magnetoencephalography (MEG) datasets. Therefore, the current approaches using Normal statistics could not be directly applied. Unlike the existing literature, we assume that the dataset from the control group belongs to a more general distributional family - the skewed Student \(t\) distribution, which includes the Normal distribution and Student \(t\) as special cases. The benefit of this skewed Student \(t\) distributional assumption is that it could cover a variety of empirical behaviours and is thus more suitable for practical needs.
This generalisation is essential since many brain Magnetoencephalography (MEG) datasets show a significant level of asymmetry in the observation distribution and hence are highly skewed. It is also the case that the high skewness not only appears in a MEG dataset but also in other fields, such as Cybersecurity~\citep{kim2023cyber}, Medicine~\citep{ktenaGenerativeModelsImprove2024},
Econometric~\citep{higbeeComparisonGB2Skewed2024}, Neuroscience~\citep{zimmermannArbitraryMethodologicalDecisions2024},
Telehealthcare~\citep{dasEldocareEEGKinect2023} and Internet of Thing (IoT)~\citep{yinFLSNMVOEdgeComputing2024}
to name only a few.

The main focus of this research is the adaptability of statistical models
under the skewed student $t$ distribution. Specifically, the statistical
challenges arise from the interplay between the mechanism of data observation
from a single-subject, the type of hypothesis (two-sided, less, greater),
and the sign of the skewness parameter, which we will discuss one by one
in the following paragraphs.

\textbf{The mechanism of data observation in Type I error and Power test}. Under the assumption that the dataset observed from the control group follows a Normal distribution, several methods exist to compare a single-subject against a control group.~\citet{crawfordComparingIndividualTest1998} proposed a \( t \)-score test defined as follows:
\begin{equation*}
	t=\frac{x^{*}-\bar{x}}{s/\sqrt{(n+1)/n}},
\end{equation*}
where \( x^{*} \) represents the observation of the single-subject, \( \bar{x} \) denotes the average of \( n \) observations from the control group, and \( s \) stands for the standard deviation of the dataset from the control group. The test score \( t \) follows a Student \( t \) distribution with \( n-1 \) degrees of freedom. Let \( \beta \) be the prespecified significance level, define the \( p \) value as \(\mathbb{P}(t>|t_{1-\beta/2}(n-1)|)\) where \(t_{1-\beta/2}(n-1)\) is the \((1-\beta/2)\)-th critical value of Student \( t \) distribution with \( n-1 \) degrees of freedom. If the \( p \) value is less than \(\beta\), then one can claim that \(x^{*}\) has an abnormality. Moreover, under the assumption of a skewed Student \( t \) distribution,~\citet{crawfordTestingDeficitSinglecase2006} examined the effects of departures from normality on testing for a deficit in a single-subject via using a Leptokurtic distribution with different skewness parameters. They asserted that the \( t \) score~\citep{crawfordComparingIndividualTest1998} outperforms the \( z \) score in terms of Type I error, and the \( t \) score shows robustness as good as the result under the Normal assumption. Additionally,~\citet{crawfordMethodsTestingDeficit2006} investigated the statistical power of testing for an abnormality in a single-subject using Monte Carlo simulations. They found that the power of \(t\) score performs slightly worse than the Normal assumption results.

The simulation results for Type I error in \citet{crawfordTestingDeficitSinglecase2006} and power in~\citet{crawfordMethodsTestingDeficit2006} led to the assertion that the \( t \) score~\cite{crawfordComparingIndividualTest1998} is robust even when the underlying distribution is skewed Student \(t\) distribution. However, the simulation settings for generating the samples for a single-subject in \citet{crawfordTestingDeficitSinglecase2006},~\citet{crawfordMethodsTestingDeficit2006} are not comprehensive to evaluate the performance of the \(t\) score under skewed Student \(t\) distribution.  The samples of the single-subject may come from either the distribution of the control group or a different distribution. Let \(c:d\) be the ratio of the samples from the control group to those from a different distribution. In \citet{crawfordTestingDeficitSinglecase2006}, the ratio is \(100:0\); and in~\citet{crawfordMethodsTestingDeficit2006}, the ratio is \(0:100\). Both are special cases for evaluating the Type I error and Power, respectively.

\begin{table}[h]
	\small\sf\centering
	\caption{The data generation mechanism of single-subject observations in \citet{crawfordTestingDeficitSinglecase2006},~\citet{crawfordMethodsTestingDeficit2006} respectively. `NA' means that the observations in the cell are not available. `Positive' means the single-subject is claimed as abnormal, while `Negative' means the single-subject is claimed as normal.\label{T1}}
	\begin{minipage}[t]{0.45\textwidth}
		\centering
		\begin{tabular}{|l|l|c|c|}
			\cline{3-4}
			\blank{2} & \multicolumn{2}{c|}{Predicated} \\
			\cline{3-4}
			\blank{2} & Positive & Negative \\
			\hline
			\multirow{2}{*}{Actual} & Positive & NA & NA \\
			\cline{2-4}
			& Negative & \( n_{01} \) & \( n_{00} \) \\
			\hline
		\end{tabular}
		\subcaption{Type I error evaluation in \citet{crawfordTestingDeficitSinglecase2006}}\label{f1}
	\end{minipage}
	\hfill
	\begin{minipage}[t]{0.45\textwidth}
		\centering
		\begin{tabular}{|l|l|c|c|}
			\cline{3-4}
			\blank{2} & \multicolumn{2}{c|}{Predicated} \\
			\cline{3-4}
			\blank{2} & Positive & Negative \\
			\hline
			\multirow{2}{*}{Actual} & Positive & \( n_{11} \) & \( n_{10} \) \\
			\cline{2-4}
			& Negative & NA & NA \\
			\hline
		\end{tabular}
		\subcaption{Power evaluation in~\citet{crawfordMethodsTestingDeficit2006}}\label{f2}
	\end{minipage}
\end{table}

\textbf{For \(\mathbf{c:d=100:0}\)}, \citet{crawfordTestingDeficitSinglecase2006} generates all single-subject observations from the distribution of the control group to evaluate the Type I error only, see~\Cref{f1}. In~\Cref{f1}, \( n_{01} \) is the number of false positives (FP) while \( n_{00} \) is the number of true negatives (TN). The false positive rate (FPR, i.e., Type I error) is estimated as \( n_{01}/(n_{01}+n_{00}) \). As there are no actual positive observations in~\Cref{f1}, the true positive rate (TPR, i.e., power) can not be estimated given the simulation settings in \citet{crawfordTestingDeficitSinglecase2006}. Therefore, the comparison of Type I error rates in \citet{crawfordTestingDeficitSinglecase2006} is not thoroughly discussed, as the study did not control for statistical power.
Conversely, \textbf{for \(\mathbf{c:d=0:100}\)},~\citet{crawfordMethodsTestingDeficit2006} generate all single-subject observations from a distribution that is different from the distribution of the control group to evaluate the power only, see~\Cref{f2}. In~\Cref{f2}, \( n_{11} \) is the number of true positives (TP) while \( n_{01} \) is the number of false negatives (FN). The true positive rate (TPR, i.e., power) is estimated as \( n_{11}/(n_{11}+n_{10}) \). However, the Type I error is unavailable based on the observations. Again,~\citet{crawfordMethodsTestingDeficit2006} did not fully discuss the comparison of the power evaluation due to the absence of Type I error.  Neither~\citet{crawfordTestingDeficitSinglecase2006} nor~\citet{crawfordMethodsTestingDeficit2006} discussed that the ground truth observations of a single-subject are well-mixed. For instance,  if 50\% of the single-subject observations are generated from the underlying distribution of the control group data and the other 50\% of observations come from a different distribution, the Type I error and power of the \( t \) score test would differ significantly. Direct evidence illustrated in~\Cref{fig:figs/Poweralpha3df5.pdf} shows that the median of powers of the Crawford-Garthwaite (CG) method and BIGPAST under two-sided test are around 0.62 and 0.96, respectively. To address this issue, all the simulation settings (unless specified) of the single-subject in this paper are well-mixed, i.e., the single-subject observations are generated from a mixture of the null and alternative distribution. The well-mixed setting is more realistic and can provide a comprehensive evaluation of the performance of BIGPAST test under skewed Student \( t \) assumption.

\begin{figure}[ht]
	\centering
	\begin{subfigure}[bt]{0.48\textwidth}
		\centering
		\includegraphics[width=\textwidth]{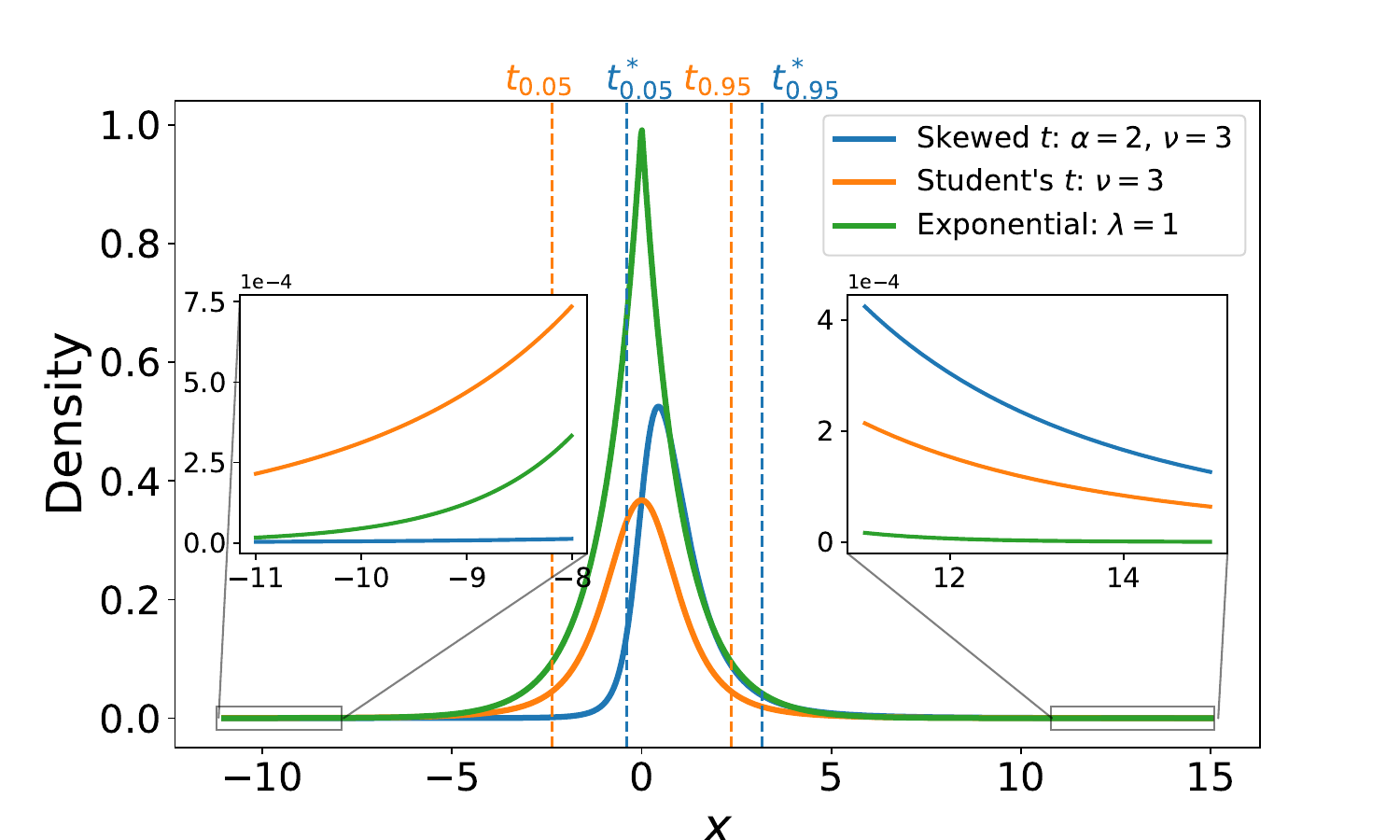}
		\caption{Positive skewness: \( \alpha=2 \) with \( \nu=3 \)}\label{fig:figs/skew_positive_alpha2_nv3.pdf}
	\end{subfigure}
	\begin{subfigure}[bt]{0.48\textwidth}
		\centering
		\includegraphics[width=\textwidth]{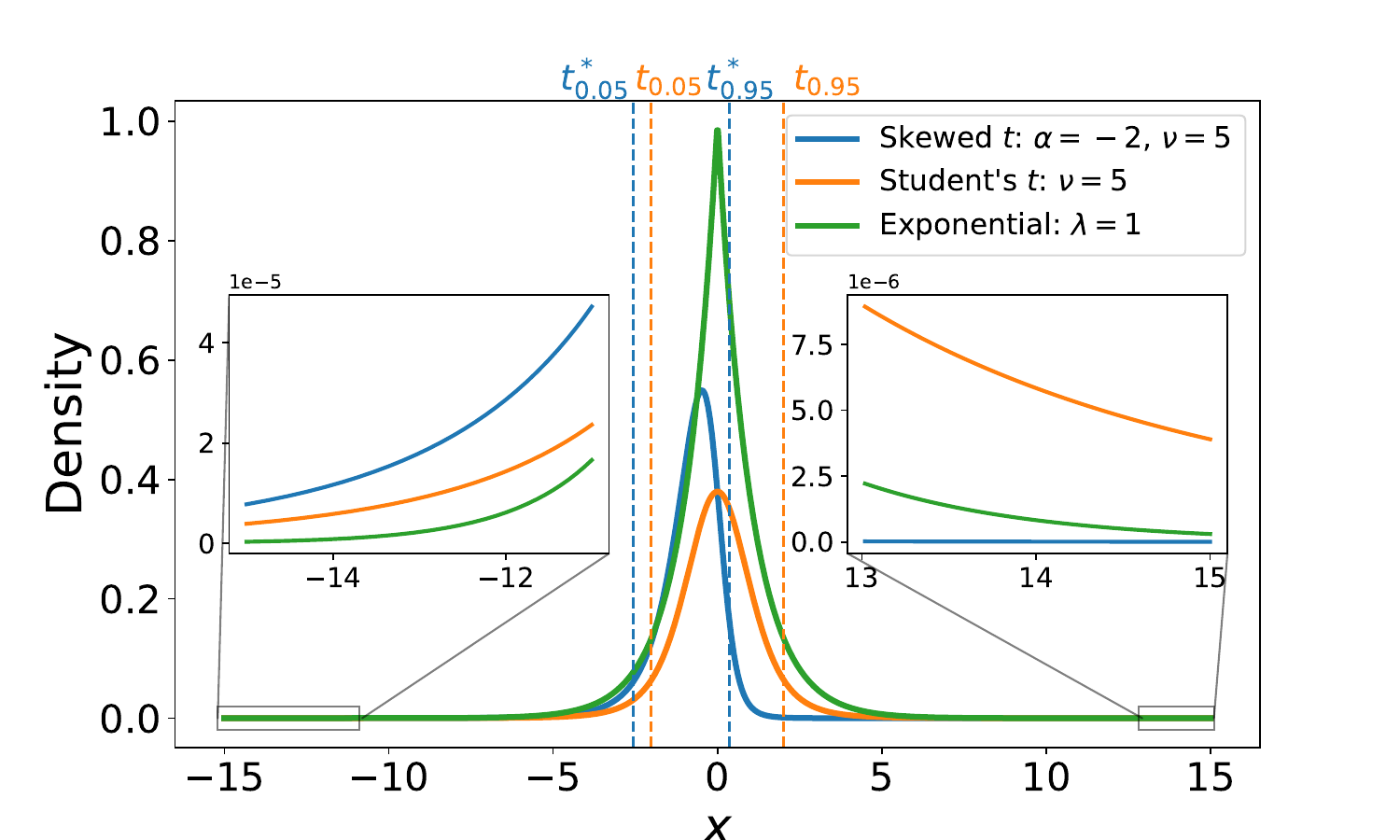}
		\caption{Negative skewness: \( \alpha=-2 \) with \( \nu=5 \)}\label{fig:figs/skew_negative_alpha-2_nv5.pdf}
	\end{subfigure}
	\caption{The heavy and light tails of the standard skewed Student \( t \) distribution for both positive and negative skewness parameters. The orange lines represent the standard Student \( t \) density function, the green lines represent the curve of \( \exp({-{\left\lvert x \right\rvert}}) \). The vertical lines are quantiles: \( t_{0.05}, t_{0.95} \) (resp.~\( t_{0.05}^{*}, t_{0.95}^{*} \)) are 5\% and 95\% quantiles of the standard Student \( t \) distribution (resp.~the standard skewed Student \( t \) distribution). In (a), \( t_{0.05}=-2.353, t_{0.95}=2.353 \), \( t_{0.05}^{*}=-0.398, t_{0.95}^{*}=3.169 \). In (b), \( t_{0.05}=-2.015, t_{0.95}=2.015 \), \( t_{0.05}^{*}=-2.568, t_{0.95}^{*}=0.370 \).}\label{fig:heavy_light_tails}
\end{figure}

Before delving into the rationale behind the assumption of the skewed Student \( t \) distribution, it is essential to review the definitions of heavy tails and light tails within the context of the skewed Student \( t \) distribution. Let \( F(x) \) be the cumulative distribution function of a random variable \( X \). The right tail distribution is denoted as \( \overline{F}(x)\coloneqq\Pr(X>x) \) while the left tail distribution is denoted as \( \underline{F} (x)\coloneqq\Pr(X\leq x) \). \( F \) is said to have a right heavy tail if \( \lim_{x \to +\infty}\exp({tx})\overline{F}(x) =\infty\) for all \( t>0 \), and have a right light tail if \( \exp({tx})\overline{F}(x) =O(1)\) for large positive \( x \) and some constant \( t>0 \). Similarly, \( F \) is said to have a left heavy tail if \( \lim_{x \to -\infty}\exp({tx})\underline{F}(x) =\infty\) for all \( t>0 \), and have a left light tail if \(\exp({tx})\underline{F}(x) =O(1)\) for small negative \(  x  \) and some constant \( t>0 \). In other words, light tails decay to zero much faster than the exponential distribution, whereas heavy tails decay much slower, see~\Cref{fig:heavy_light_tails}.

It is well-known that the Student \( t \) distribution is symmetric and has two heavy tails. The skewed Student \( t \) distribution is a generalisation of the Student \( t \) distribution~\citep{azzaliniSkewnormalRelatedFamilies2014}. Let \( f(z|\alpha,\nu) \) be the skewed Student \( t \) density function~\citep{azzaliniDistributionsGeneratedPerturbation2003}, defined as follows:
\begin{equation}\label{eq:skewed_t_def}
	f(z|\alpha,\nu) \coloneqq 2t(z|\nu)T(w|\nu+1), \quad z\in\mathbb{R}, \alpha\in\mathbb{R}, \nu>0,
\end{equation}
where \( r(z,\nu)=\sqrt{(\nu+1)/(\nu+z^2)} \) and \( w= \alpha z r(z,\nu)\), \( t(\cdot|\nu) \) represents the density function of the \( t \) distribution with degrees of freedom \( \nu \), and \( T(\cdot|\nu+1) \) represents the cumulative density function of the \( t \) distribution with degrees of freedom \( \nu+1 \). Let \( \xi \) and \( \omega  \) be the location and scale parameters, with the substitution \( z=(x-\xi)/\omega \) in~\Cref{eq:skewed_t_def}, then the skewed Student \( t \) density function is given by
\begin{equation}\label{eq:skewed_Student_t}
	f(x|\alpha,\nu,\xi,\omega) \coloneqq \frac{1}{\omega}f\left( \frac{x-\xi}{\omega}|\alpha,\nu \right).
\end{equation}

The skewness parameter \( \alpha \) changes the symmetry of Student \( t \) distribution given the degrees of freedom \( \nu \). When \( \alpha=0 \), \( \nu \) is finite, the skewed Student \( t \) distribution reduces to Student \( t \) distribution which has heavy tails. When \( \alpha=0 \) and \( \nu \to +\infty \), the skewed Student \( t \) distribution reduces to the Normal distribution which has lighter tails; When \( |\alpha|\to\infty\) and \( \nu < +\infty \), the skewed Student \( t \) distribution reduces to the half Student \( t \) distribution which can be tuned to allow for a heavy tail (\( \nu\to 1 \)) or a light tail (large \( \nu \)). Therefore, for other settings of (\( \alpha, \nu \)), the skewed Student \( t \) distribution may have one light tail and one heavy tail.  The sign of \( \alpha \) dictates whether the heavy tail is on the left or right side of the distribution, see~\Cref{fig:heavy_light_tails}. In~\Cref{fig:figs/skew_positive_alpha2_nv3.pdf}, the skewness is positive, and the standard skewed Student \( t \) distribution has a right heavy tail and a left light tail. Conversely, in~\Cref{fig:figs/skew_negative_alpha-2_nv5.pdf}, the skewness is negative, the standard skewed Student \( t \) distribution has a right light tail and a left heavy tail, see the zoomed insets in~\Cref{fig:figs/skew_positive_alpha2_nv3.pdf,fig:figs/skew_negative_alpha-2_nv5.pdf} respectively.

\textbf{The effect of hypothesis (two-sided, less, greater)}. In this paper, we consider three kinds of alternative hypotheses: ``less than'' (\( < \)), ``greater than'' (\( > \)) and ``two-sided''. For clarity, we refer to the alternative hypothesis with symbol (\( < \)) as the \textit{one-sided test (less)}, and the alternative hypothesis with symbol (\( > \)) as the \textit{one-sided test (greater)}. Under the assumption of symmetric distributions (e.g.,\ Normal or Student \( t \)), one-sided tests, either ``less'' or ``greater'', share the same absolute critical value given the same significance level. However, under the assumption of the skewed Student \( t \) distribution, the absolute critical value of the one-sided test (less) is different from the absolute critical value of the one-sided test (greater) due to the relationship between skewness and heavy tails. Hence, under the assumption of a skewed Student \( t \) distribution, there are six combinations of hypothesis tests: the one-sided test (less), the one-sided test (greater), and the two-sided test, each with either a positive or negative skewness parameter.

\textbf{The interaction effect of hypothesis type and skewness sign}. Let us look into the parameter settings in \citet{crawfordTestingDeficitSinglecase2006}. They set \( \alpha<0 \) and only considered the one-sided test (less). The observations of the control group and single-subject are generated from the same skewed Student \( t \) distribution with the same skewness parameter. Furthermore, the observations of the single-subject are all negative, as described in~\Cref{f1}. They used the \( t \) statistic to test the abnormalities of the single-subject.~\Cref{fig:figs/skew_negative_alpha-2_nv5.pdf} describes a similar skewness parameter setting as in \citet{crawfordTestingDeficitSinglecase2006}. Given the significance level 0.05, the critical value of the one-sided test (less) for the \( t \) test is \( t_{0.05}=-2.015 \) while the critical value of the one-sided test (less) for the skewed \( t \) test is \( t_{0.05}^{*}=-2.568\). Apparently, \( t_{0.05}>t^{*}_{0.05} \), which implies that if some single-subject observation \( x^{*} \) is claimed as abnormal by \( t_{0.05}^{*} \), then \( x^{*} \) must be claimed as abnormal by \( t_{0.05} \).

\textbf{Motivation to reduce the model misspecification error}. Let's define the \textit{model misspecification error} for the one-sided test (less) as the probability \( \Pr_t(t_{0.05}^{*}\leq x\leq t_{0.05}) \) where \( \Pr_{t}(\cdot) \) is the probability under the Student \( t \) distribution. In~\Cref{fig:figs/skew_negative_alpha-2_nv5.pdf}, the model misspecification error for the one-sided test (less) is \( \Pr_{t}(-2.568\leq x\leq-2.015)\approx0.0249 \). However, if we consider the one-sided test (greater) under the same settings, the model misspecification error for the one-sided test (greater) is \( \Pr_{t}(0.0370\leq x\leq2.015)\approx0.3131 \). The model misspecification error for the one-sided test (greater) is almost 12 times higher than that for the one-sided test (less). This means that the one-sided test (less) is more robust than the one-sided test (greater) when \( \alpha<0 \). Under the settings of \( \alpha<0 \), one-sided test (less) and not well-mixed observations of single-subject, \citet{crawfordTestingDeficitSinglecase2006} claimed that the \( t \) score is robust in terms of Type I error even under skewed Student \( t \) distribution. However, this assertion does not hold if the one-sided test (less) is replaced with the one-sided test (greater). For the power comparison,~\citet{crawfordMethodsTestingDeficit2006} make a similar statement. However, by using a similar discussion to the above, we can find that the robustness statement for the power only holds for the specific settings in their paper.

Moreover, we also investigate the model misspecification error between Normal and skewed Student \( t \) distributions in~\Cref{fig:model_misspecification}.
The \( z \) score will always have model misspecification error under the skewed Student \( t \) distribution assumption. Neither the \( z \) score test, the \( t \) score test, nor even the Crawford-Garthwaite Bayesian method~\citep{crawfordComparisonSingleCase2007} can eliminate the model misspecification error.

The rationale for this paper's assumption of the skewed Student \( t \) distribution for the underlying proposal can be attributed to the fact that it can help eliminate model misspecification errors, provided that the underlying distribution is a special case of the skewed Student \( t \) distribution. These special cases encompass the regular Student \( t \) distribution (\( \alpha=0 \)), skewed Normal distribution (\( \nu\to\infty \)), half Normal distribution (\( \alpha\to\pm\infty,\nu\to\infty\)), and Normal distribution (\( \alpha=0,\nu\to\infty\)) as special cases. To show the elimination of model misspecification error, simulation results with a Normal assumption but using BIGPAST method can be found in~\Cref{tab:fpr_acc_less_comparison_mixed_two-sided_normal}.

We can conclude that the interplay between skewness (whether positive or negative), the type of hypothesis (`two-sided', `less', or `greater'), and the single-subject observations (whether well-mixed or not) significantly influences both the Type I error rate and the power of the \( t \) score test under the assumption of the skewed Student \( t \) distribution. The key contributions are:
\begin{itemize}
	\item This paper introduces a Bayesian Inference General Procedures for A Single-subject Test (BIGPAST) framework, which employs the nested sampling technique under the skewed Student \( t \) assumption to fully discuss the test performance in different settings (i.e.,\ the combinations of the sign of \( \alpha \), type of hypothesis and well-mixed single-subject observations).
	\item We propose a Jeffreys prior, which is defined by explicit mathematical formulas. We evaluate BIGPAST's performance against the existing approaches through simulation studies. The results demonstrate that BIGPAST is robust against departures from normality and surpasses the existing methods based on the assumption of a Normal distribution in terms of Type I error, power and accuracy.
\end{itemize}

The remainder of the paper is organised as follows. In~\Cref{sec: Bayesian Inference on Comparison a single-subject with control group}, we introduce the Bayesian inference procedures for comparing a single-subject with a control group under the skewed Student \( t \) distribution. In~\Cref{sec:Simulation}, we conduct a comprehensive set of simulation studies to evaluate the performance of the proposed BIGPAST method. In~\Cref{sec:Real_data_analysis}, we apply the BIGPAST method to a MEG dataset comprising an individual with mild traumatic brain injury and a control group. Finally, we summarise our findings and discuss future research directions in~\Cref{sec:discussion}.

\section{Bayesian Inference Procedures for Single-subject's Abnormality Detection}\label{sec: Bayesian Inference on Comparison a single-subject with control group}

In this section, we introduce Bayesian inference procedures to compare a single-subject with a control group, assuming that the control group observations follow a skewed Student \( t \) distribution. First, we should verify whether the assumption holds. For practical applications, we recommend employing a modified version of the general goodness-of-fit test~\citep{chenGeneralPurposeApproximate1995} tailored explicitly for the skewed Student \( t \) distribution in the~\Cref{subsec:Goodness of fit for skew distribution}. From now on, we assume that the control group data follows a skewed Student \( t \) distribution in the subsequent sections.

\subsection{Bayesian Inference General Procedures for a Single-subject Test (BIGPAST)}\label{subsec:Bayesian_inference_procedures}
Let \( x^{*} \) denote the observation of a single-subject. We aim to test whether \( x^{*} \) is a random sample from the same distribution as the control group. Let \( f_{c}(\alpha,\nu,\xi,\omega) \) represent the skewed Student \( t \) density function of \( x_{1},x_{2},\ldots,x_{n} \). The hypothesis testing can be formulated as follows:
\begin{equation*}
	\mathcal{H}_{0}: x^{*} \sim f_{c}(\alpha,\nu,\xi,\omega)\quad v.s.\quad \mathcal{H}_{1}: x^{*} \not\sim  f_{c}(\alpha,\nu,\xi,\omega),
\end{equation*}
where \( \not\sim \) represents that \( x^{*} \) is not an observation from \( f_{c}(\alpha,\nu,\xi,\omega) \). We propose the Bayesian inference general procedures for a single-subject test under the null hypothesis \( \mathcal{H}_{0} \). The procedures are as follows:
\begin{enumerate}
	\item Choose the proposed Jeffery prior \( \boldsymbol{\pi}^{J}(\alpha,\nu,\xi,\omega) \) (defined below) as the joint prior of \( \alpha,\nu,\xi,\omega \).
	\item Compute the posterior distribution of \( \alpha,\nu,\xi,\omega \) based on the control group sample \(\mathbf{x}=(x_{1},x_{2},\ldots,x_{n}) \) and the prior \( \boldsymbol{\pi}^{J}(\alpha,\nu,\xi,\omega) \). By Bayes's Theorem, we have
	\begin{equation*}
		\boldsymbol{\pi}(\alpha,\nu,\xi,\omega|\mathbf{x})\propto l(\mathbf{x}|\alpha,\nu,\xi,\omega)\boldsymbol{\pi}^{J}(\alpha,\nu,\xi,\omega),
	\end{equation*}
	where \( l(\mathbf{x}|\alpha,\nu,\xi,\omega)\coloneqq \prod_{i=1}^{n}f_{c}(x_{i}|\alpha,\nu,\xi,\omega) \) is the likelihood function.
	\item Draw \( m \) samples of \( \alpha,\nu,\xi,\omega \)  from the posterior distribution \( \boldsymbol{\pi}(\alpha,\nu,\xi,\omega|\mathbf{x}) \) using the Metropolis-Hastings algorithm~\citep{hastingsMonteCarloSampling1970}.~\Cref{alg:mh_adjust} lists the details of sampling, the samples are denoted as \( \alpha_{j},\nu_{j},\xi_{j},\omega_{j} \), \( j=1,2,\ldots,m \).
	\item For each \( \alpha_{j},\nu_{j},\xi_{j},\omega_{j} \), draw \( s \) random samples from the skewed Student \( t \) distribution, denoted as \( x_{1j},x_{2j},\ldots,x_{sj} \), \( j=1,2,\ldots,m \). Sort the observations \( x_{1j},x_{2j},\ldots,x_{sj} \), \( j=1,2,\ldots,m \), in ascending order and label them as \( x_{(1)},x_{(2)},\ldots, x_{(B)} \) where \( B=ms \).
	\item Given the significance level \( \beta \) and \( \mathbf{x} \), the credible interval of \( x^{*} \)  is \([x_{(\lfloor B\beta/2\rfloor)}, x_{(\lceil B(1-\beta/2)\rceil)}] \), where \( \lfloor \cdot \rfloor \) and \( \lceil \cdot \rceil \) are the floor and ceiling functions, respectively. If \( x^{*} \) is not in the credible interval, then reject the null hypothesis \( \mathcal{H}_{0} \) at the significance level \( \beta \).
\end{enumerate} %

\begin{algorithm}
	\caption{Metropolis-Hastings Algorithm}\label{alg:mh_adjust}
	\KwData{Initial parameters: $(\alpha_{0},\nu_{0},\xi_{0},\omega_{0})$, data: $\mathbf{x}$, size $m_{0}=10000$, burn-in: $b=0.4$, step size: $\delta=0.5$.}
	\KwResult{Samples from the posterior distribution}
	$n_{0} \gets 4$\;
	$\Theta \gets \text{zeros matrix with dimension:}(m_{0}, n_{0})$\;
	$(\alpha^{\text{old}},\nu^{\text{old}},\xi^{\text{old}},\omega^{\text{old}}) \gets (\alpha_{0},\nu_{0},\xi_{0},\omega_{0})$\;
	$i \gets 0$\;
	\While{$i < m_{0}$}{
		$\nu^{\text{new}} \gets \nu^{\text{old}} + \delta \Phi^{-1}[\Phi(-\nu^{\text{old}} / \delta) + u_{\nu} \Phi(\nu^{\text{old}} / \delta)]$\;
		$p_{\nu} \gets  \log\Phi(\nu^{\text{old}} / \delta) - \log\Phi(\nu^{\text{new}} / \delta)$\;
		$\omega^{\text{new}}  \gets \omega^{\text{old}} + \delta \Phi^{-1}[\Phi(-\omega^{\text{old}} / \delta) + u_{\omega} \Phi(\omega^{\text{old}} / \delta)]$\;
		$p_{\omega} \gets \log\Phi(\omega^{\text{old}} / \delta) - \log\Phi(\omega^{\text{new}} / \delta)$\;
		$\alpha^{\text{new}} \gets \alpha^{\text{old}} + \epsilon_{\alpha}$\;
		$\xi^{\text{new}} \gets \xi^{\text{old}} + \epsilon_{\xi}$\;
		$p_{\text{new}} \gets \log(\boldsymbol{\pi}(\alpha^{\text{new}},\nu^{\text{new}},\xi^{\text{new}},\omega^{\text{new}}|\mathbf{x}))$\;
		$p_{\text{old}} \gets \log(\boldsymbol{\pi}(\alpha^{\text{old}},\nu^{\text{old}},\xi^{\text{old}},\omega^{\text{old}}|\mathbf{x}))$\;
		$A \gets p_{\text{new}} - p_{\text{old}} + p_{\nu} + p_{\omega}$\;
		\If{$\log(u) \leq A$}{
			$(\alpha^{\text{old}},\nu^{\text{old}},\xi^{\text{old}},\omega^{\text{old}}) \gets (\alpha^{\text{new}},\nu^{\text{new}},\xi^{\text{new}},\omega^{\text{new}})$\;
		}
		$\Theta[i, :] \gets (\alpha^{\text{old}},\nu^{\text{old}},\xi^{\text{old}},\omega^{\text{old}})$\;
		$i \gets i + 1$\;
	}
	\Return{$\Theta[\lfloor m_{0}*b\rfloor+1:, :]$}\;
\end{algorithm}
In~\Cref{alg:mh_adjust}, the Metropolis-Hastings algorithm draws samples from the posterior distribution. The step size \( \delta \) is set to 0.5, and the burn-in rate \( b \) is set to 0.4. The algorithm returns the samples after the burn-in period. \(u_{\nu},u_{\omega}, u\) are random numbers drawn from the uniform distribution \( U(0,1) \). \(\epsilon_{\alpha}\) and \(\epsilon_{\xi}\) are drawn from the Normal distribution with mean zero and standard deviation \(\delta\). A brief explanation of~\Cref{alg:mh_adjust} is as follows:
\begin{itemize}
	\item Lines 6-9: Draw the new parameters \( \nu^{\text{new}} \) and \( \omega^{\text{new}} \) from the proposal distributions. As \( \nu \) and \( \omega \) are positive, the proposal distributions for \( \nu \) and \( \omega \) are the truncated normal distributions with mean \( \nu^{\text{old}} \), \( \omega^{\text{old}} \) and standard deviation \( \delta \), respectively.
	\item Lines 10-11: Draw the new parameters \( \alpha^{\text{new}} \) and \( \xi^{\text{new}} \) from the proposal distributions. The proposal distributions for \( \alpha \) and \( \xi \) are normal distributions with mean \( \alpha^{\text{old}} \), \( \xi^{\text{old}} \) and standard deviation \( \delta\), respectively.
	\item Lines 12-13: Compute the logarithm of the posterior distribution for the new and old parameters.
	\item Lines 14-19: Compute the acceptance rate \( A \) and update the parameters. The truncated proposal distributions do not have symmetric densities, so we have to adjust the acceptance rate \( A \) by adding the logarithm of the ratios, i.e., \( p_{\nu} \) and \( p_{\omega} \) to the acceptance rate \( A \) in line 14.
\end{itemize}

Steps 3, 4 and 5 employ the nested sampling method to determine the critical interval. One popular way is to use the following steps \(3^{\prime}\) and \(4^{\prime}\):
\begin{enumerate}[label={\arabic*$'$},start=3]
	\item Obtain the maximum posterior estimators of \( \alpha,\nu,\xi,\omega \) by maximising the posterior distribution \( \boldsymbol{\pi}(\alpha,\nu,\xi,\omega|\mathbf{x}) \), denoted as \( \hat{\alpha}_{\mathrm{MAP}},\hat{\nu}_{\mathrm{MAP}},\hat{\xi}_{\mathrm{MAP}},\hat{\omega}_{\mathrm{MAP}}\).
	\item Given the significance level \( \beta \), the credible interval of \( x^{*} \) given \( \mathbf{x} \) is \( [x^{\mathrm{MAP}}_{\beta/2},x^{\mathrm{MAP}}_{1-\beta/2}] \), where \( x^{\mathrm{MAP}}_{\beta/2} \) and \( x^{\mathrm{MAP}}_{1-\beta/2} \) are the \( \beta/2 \) and \( 1-\beta/2 \) quantiles of \( \boldsymbol{\pi}(\hat{\alpha}_{\mathrm{MAP}},\hat{\nu}_{\mathrm{MAP}},\hat{\xi}_{\mathrm{MAP}},\hat{\omega}_{\mathrm{MAP}}|\mathbf{x}) \), respectively. If \( x^{*} \) does not fall within this credible interval, then the null hypothesis \( \mathcal{H}_{0} \) can be rejected at the significance level \( \beta \).
\end{enumerate}
However, in practice, we found that the maximum posterior estimation is time-consuming, and the steps \(3^{\prime}\) and \(4^{\prime}\) do not perform better than the steps 3, 4 and 5, see the comparison study in~\Cref{subsec:Simulation_for_Bayesian_inference_procedures}. Therefore, we recommend using the nested sampling method to determine the critical interval.

Moreover, when \( \alpha\to +\infty \) or \( -\infty \), the skewed Student \( t \)  distribution degenerates to half Student \( t \) distribution. Then, the two-sided credible interval in Step 5 above should be slightly modified as \( [x_{(1)},x_{(\lceil B(1-\beta)\rceil)}] \) if \( \alpha \to +\infty \) or \( [x_{(\lfloor B\beta\rfloor)},x_{(B)}] \) if \( \alpha \to -\infty \).

\subsection{Jeffreys Priors}\label{sec:jeffery_priors}
In Bayesian probability, Jeffreys prior~\citep{jefferyprior1946} refers to a prior density function proportional to the square root of the determinant of the Fisher information matrix. In this section, we proposed an explicit Jeffery prior for the parameters of the skewed Student \( t \) distribution and compared it to existing priors.

In the literature, two types of priors are commonly used for \( \xi \) and \( \omega \) in~\Cref{eq:skewed_Student_t}: the partial information prior~\citep{sunReferencePriorsPartial1998,cohenBayesianStrategicClassification2025} and the independence Jeffreys prior~\citep{rubioInferenceTwoPieceLocationScale2014,fortuinPriorsBayesianDeep2022}. Both of these priors assume that the joint prior of \( \xi \) and \( \omega \), denoted as \( \boldsymbol{\pi}(\xi,\omega) \), is proportional to \( \omega^{-1} \). Consequently, the joint prior of \( \alpha,\nu,\xi,\omega \) can be expressed as follows:
\begin{equation*}
	\boldsymbol{\pi}(\alpha,\nu,\xi,\omega) \propto \omega^{-1}\boldsymbol{\pi}(\alpha,\nu).
\end{equation*}
The joint prior of \( \alpha \) and \( \nu \), \( \boldsymbol{\pi}(\alpha,\nu) \), may be specified by the Jeffreys prior~\citep{brancoObjectiveBayesianAnalysis2013} or the natural informative prior~\citep{detteNaturalNonInformative2018}.
~\citet{brancoObjectiveBayesianAnalysis2013} proposed the following prior for the general skewed Student \( t \) density function:
\begin{equation*}
	\boldsymbol{\pi}^{B}(\alpha,\nu,\xi,\omega) \propto \omega^{-1}\left(8 \alpha ^2+\pi ^2\right)^{-3/4}\boldsymbol{\pi}(\nu),
\end{equation*}
where \( \boldsymbol{\pi}(\nu) \) is proposed by~\cite{fonsecaObjectiveBayesianAnalysis2008} for the Student \( t \) distribution and given by
\begin{equation*}
	{\footnotesize \boldsymbol{\pi}(\nu)\propto \left( \frac{\nu}{\nu+3} \right)^{1/2}\left\{ \psi ^{(1)}\left( \frac{\nu}{2} \right)-\psi ^{(1)}\left( \frac{\nu+1}{2} \right)-\frac{2(\nu+3)}{\nu(\nu+1)^{2}}\right\}^{1/2}},
\end{equation*}
where \( \psi ^{(1)}(\cdot) \) is the trigamma function, i.e., the second derivative of \( \log(\Gamma(\cdot)) \). Here, \( \Gamma(\cdot) \) is the Gamma function. When \( \alpha\rightarrow +\infty \), the order of \( \boldsymbol{\pi}^{B}(\alpha,\nu,\xi,\omega) \) with respect to \( \alpha \) is  \( O(\alpha^{-3/2}) \)~\citep{brancoObjectiveBayesianAnalysis2013}.

~\citet{detteNaturalNonInformative2018} proposed the so-called natural informative prior of the parameter \( \alpha \) based on the total variation distance, which is given by
	{\small\begin{align*}
			& BTV(\alpha|\theta_{1},\theta_{2}) = \frac{1}{B(\theta_{1},\theta_{2})}\left( \frac{\tan^{-1}(\alpha)}{\pi}+\frac{1}{2} \right)^{\theta_{1}-1}\left( \frac{-\tan^{-1}(\alpha)}{\pi}+\frac{1}{2} \right)^{\theta_{2}-1}\frac{1}{\pi(1+\alpha^{2})} \\
		\end{align*}}
where \( B(\cdot,\cdot) \) represents the Beta function.
When \( \theta_{1}=1,\theta_{2}=1 \), the prior \( BTV(\alpha|1,1) \) reduces to the standard Cauchy prior. Similarly,~\citet{detteNaturalNonInformative2018} constructed the joint prior of \( \alpha,\nu,\xi, \omega \) based on \( BTV(\alpha|1,1) \) as
\begin{equation*}
	\boldsymbol{\pi}^{D}(\alpha,\nu,\xi,\omega) \propto \omega^{-1}\pi^{-1}\left(1+ \alpha ^2\right)^{-1}\boldsymbol{\pi}(\nu),
\end{equation*}
which is order of \( O(\alpha^{-2}) \) when \( \alpha\rightarrow +\infty\).

A notable concern with \( \boldsymbol{\pi}^{B}(\alpha,\nu,\xi,\omega) \) and \( \boldsymbol{\pi}^{D}(\alpha,\nu,\xi,\omega) \) is that the marginal prior of \( \nu \) is independent of \( \alpha \). However, given that the skewed Student \( t \) distribution is a generalisation of the Student \( t \) distribution, the marginal prior of \( \nu \) should ideally depend on \( \alpha \). We propose the Jeffreys prior for the skewed Student \( t \) density function, as detailed in~\Cref{thm:jeffery_prior_nu_alpha}.

\begin{theorem}\label{thm:jeffery_prior_nu_alpha}
	Let \( f(z|\alpha,\nu) \) be the skewed Student \( t \) density function, defined as follows:
	\begin{equation*}
		f(z|\alpha,\nu) \coloneqq 2t(z|\nu)T(w|\nu+1), \quad z\in\mathbb{R}, \alpha\in\mathbb{R}, \nu>0,
	\end{equation*}
	where \( r(z,\nu)=\sqrt{(\nu+1)/(\nu+z^2)} \) and \( w= \alpha z r(z,\nu)\). Given any \( \nu>0 \), let \( \boldsymbol{\pi}^{J}(\alpha,\nu) \) represent the joint Jeffreys prior for \( \alpha \) and \( \nu \). We have
	\begin{equation}\label{eq:jeffery_prior_nu_alpha}
		\boldsymbol{\pi}^{J}(\alpha,\nu) \propto\sqrt{I_{\alpha\alpha}I_{\nu\nu}-I_{\alpha\nu}^2},
	\end{equation}
	and the joint prior of \( \alpha,\nu,\xi,\omega \) is given by
	\begin{equation}\label{eq:joint_jeffery_prior_nu_alpha_xi_omega}
		\boldsymbol{\pi}^{J}(\alpha,\nu,\xi,\omega) \propto \omega^{-1}\boldsymbol{\pi}^{J}(\alpha,\nu),
	\end{equation}
	where
	\begin{equation*}\label{eq:the_fisher_information}\\
		\begin{aligned}
			I_{\alpha\alpha} & \approx \frac{\pi  \Gamma \left(\frac{\nu}{2}+1\right)^2 \left(\, _2F_1\left(\frac{1}{2},\nu+1;\frac{\nu+1}{2};-\frac{\alpha^2}{\sigma ^2_{\nu+1}}\right)-\, _2F_1\left(\frac{1}{2},\nu+2;\frac{\nu+1}{2};-\frac{\alpha^2}{\sigma ^2_{\nu+1}}\right)\right)}{\alpha^2 \Gamma \left(\frac{\nu+1}{2}\right)^2}, \\
			I_{\nu\nu} & =\mathbb{E}\left[ h^{2}(w)\frac{\alpha^{2} z^{2}(\nu +1)  }{4 \left(\nu +z^2\right)^{3}} \right]+\frac{1}{4}\mathbb{E}\left[ \left( \log \frac{\nu }{\nu +z^2} \right)^{2} \right]+\frac{1}{4}\mathbb{E}\left[ \left( \frac{z^2-1}{\nu +z^2} \right)^{2} \right] \\
			& \quad +d^{2}_{\nu}+d_{\nu}\mathbb{E}\left[  \log \frac{\nu }{\nu +z^2} \right]+\frac{1}{2}\mathbb{E}\left[ \left( \log \frac{\nu }{\nu +z^2} \right) \left( \frac{z^2-1}{\nu +z^2} \right)\right] \\
			I_{\alpha\nu} & = - \frac{\pi  \alpha  \Gamma \left(\frac{\nu }{2}+1\right)^2 }{8 \left(\alpha ^2+\sigma_{\nu+1} ^2\right)\Gamma \left(\frac{\nu +1}{2}\right) \Gamma \left(\frac{\nu +5}{2}\right)} \left((\nu +4)H_{5} -\frac{\left(\alpha ^2 (2 \nu +3)+(\nu +3) \sigma_{\nu+1} ^2\right) H_{6}}{\sigma_{\nu+1} ^2}\right),
		\end{aligned}
	\end{equation*}
	and
	\begin{equation*}\label{eq:expectations}\\
		\small
		\begin{aligned}
			& \mathbb{E}\left[  \log \frac{\nu }{\nu +z^2}  \right] =  \psi \left(\frac{\nu }{2}\right)-\psi \left(\frac{\nu +1}{2}\right),\quad \mathbb{E}\left[ \left( \log \frac{\nu }{\nu +z^2} \right) \left( \frac{z^2-1}{\nu +z^2} \right)\right]  = -\frac{1}{\nu ^2+\nu } \\
			& \mathbb{E}\left[ \left( \frac{z^2-1}{\nu +z^2} \right)^{2} \right]  = \frac{1}{\nu ^2+3 \nu }, \quad \mathbb{E}\left[ \left( \log \frac{\nu }{\nu +z^2} \right)^{2} \right] =  \left(\psi \left(\frac{\nu }{2}\right)-\psi \left(\frac{\nu +1}{2}\right)\right)^2 \psi ^{(1)}\left(\frac{\nu }{2}\right)-\psi ^{(1)}\left(\frac{\nu +1}{2}\right) \\
			& \mathbb{E}\left[ h^{2}(w)\frac{\alpha^{2} z^{2}(\nu +1)  }{4 \left(\nu +z^2\right)^{3}} \right]  = \frac{\pi ^{5/2} \Gamma \left(\frac{\nu }{2}+1\right) }{8 \nu ^2 \Gamma \left(\frac{\nu +5}{2}\right) B\left(\frac{\nu }{2},\frac{1}{2}\right) B\left(\frac{\nu +1}{2},\frac{1}{2}\right)^2}  \left((\nu +3)  H_{1}-(\nu +3) H_{2}-H_{3}+H_{4}\right). \\
		\end{aligned}
	\end{equation*}
	Furthermore, we define \( d_{\nu}\coloneqq-\frac{1}{2}\psi\left(\frac{\nu }{2}\right) +\frac{1}{2}\psi\left(\frac{\nu +2}{2}\right)+2c_{\nu}*g_{1}(\nu) \), where \( c_{\nu} \) is a constant that depends solely on \( \nu \). The hypergeometric functions, denoted as \( H_{1},H_{2},H_{3},H_{4},H_{5},H_{6} \), along with \( h^{2}(w) \) and \( g_{1}(\nu) \) are defined in the proof. \( _2F_1\left(a,b;c;d\right) \) represents the hypergeometric function. The term \( \sigma^{2}_{\nu+1} \) is defined in~\Cref{lem:hz_approximation}. The functions~\( \psi(\cdot) \) and \( \psi ^{(1)}(\cdot) \) represent the digamma and trigamma function respectively.
\end{theorem}

The proof of~\Cref{thm:jeffery_prior_nu_alpha} can be found in Appendix. Upon applying a Taylor expansion, we find that as \( \alpha\rightarrow 0 \), the order of \( \boldsymbol{\pi}^{J}(\alpha,\nu,\xi,\omega) \) with respect to \( \alpha \) is \( O(1+\alpha^{2}) \). Conversely, as \( \alpha\rightarrow +\infty \), the order becomes \( O(\alpha^{-3/2}) \), aligning with the findings of~\citet{brancoObjectiveBayesianAnalysis2013}. The computation of \( \boldsymbol{\pi}^{J}(\alpha,\nu,\xi,\omega) \) relies solely on the basic functions provided by the Python packages \textit{scipy} and \textit{numpy}. As such, the formulas in~\Cref{thm:jeffery_prior_nu_alpha} can be directly implemented to define a function for computing the Jeffreys prior. Additionally, we have developed a Python package, \texttt{skewt-scipy} (\href{https://pypi.org/project/skewt-scipy/}{https://pypi.org/project/skewt-scipy/}), for the skewed Student \( t \) distribution.

For the Bayesian estimation of \( \nu \), we anticipate that \( \boldsymbol{\pi}^{J}(\alpha,\nu,\xi,\omega) \) will outperform \( \boldsymbol{\pi}^{B}(\alpha,\nu,\xi,\omega) \). Because \( \boldsymbol{\pi}^{J}(\alpha,\nu,\xi,\omega) \) depends on \( \alpha \), unlike \( \boldsymbol{\pi}(\nu) \) in \( \boldsymbol{\pi}^{B}(\alpha,\nu,\xi,\omega) \) is independent of \( \alpha \). This is verified by the comparison in~\Cref{subsec:Jeffreys_Prior}. The Jeffreys prior \( \boldsymbol{\pi}^{J}(\alpha,\nu,\xi,\omega) \) is a suitable choice for the skewed Student \( t \) density function, see~\Cref{tab:mad_comparison}.~\Cref{tab:mad_comparison} also shows that the BIGPAST based on the Jeffreys prior \( \boldsymbol{\pi}^{J}(\alpha,\nu,\xi,\omega) \) is robust to detecting the departures from normality and outperforms the existing approaches based on the assumption that the data come from a Normal distribution.

\section{Simulation}\label{sec:Simulation}
\subsection{Jeffreys Prior Comparison Study}\label{subsec:Jeffreys_Prior}
In this section, we conduct a comparative analysis of our proposed prior \( \boldsymbol{\pi}^{J}(\alpha,\nu,\xi,\omega) \) against the priors \( \boldsymbol{\pi}^{B}(\alpha,\nu,\xi,\omega) \) and \( \boldsymbol{\pi}^{D}(\alpha,\nu,\xi,\omega) \). We evaluate these priors under six different parameter settings for \( (\alpha, \nu) \): \( (-1, 1) \), \( (-10,10) \), \( (-30,30) \), \( (-50,50) \), \( (-1, 50) \), and \( (-50,1) \). For each setting, the location and scale parameters are fixed at \( \xi=-2, \omega=\sqrt{2} \). To simplify the notation, we use \( \boldsymbol{\pi}^{J} \), \( \boldsymbol{\pi}^{B} \), and \( \boldsymbol{\pi}^{D} \) as shorthand for \( \boldsymbol{\pi}^{J}(\alpha,\nu,\xi,\omega) \), \( \boldsymbol{\pi}^{B}(\alpha,\nu,\xi,\omega) \), and \( \boldsymbol{\pi}^{D}(\alpha,\nu,\xi,\omega) \), respectively. For each parameter setting, we generate \( N=1000 \) samples, each with a sample size of \( n=500 \), from the skewed Student \( t \) density function. We then numerically minimise the negative logarithm of the posterior distribution to obtain the maximum posterior estimators of \( \alpha,\nu,\xi,\omega \). The performance of different priors is evaluated using the mean absolute deviation (MAD) from the true parameter. Alternative metric distances may also be applicable in this context; however, a detailed discussion of these alternatives is beyond the scope of this paper. The MAD is calculated as \( 1/N\sum_{i=1}^{N}{\lvert \hat{\theta}_{i}-\theta \rvert} \), where \( \theta \) is the true parameter and \( \hat{\theta}_{i} \) is the \( i \)-th maximum posterior estimator of \( \theta \). Our comparative study also includes the maximum likelihood estimators (MLE). The results are presented in~\Cref{tab:mad_comparison}.

As demonstrated in \Cref{tab:mad_comparison}, when \( |\alpha| \) and \( \nu \) are less than 30, the performance of \( \boldsymbol{\pi}^{J} \) aligns closely with that of \( \boldsymbol{\pi}^{B} \) and \( \boldsymbol{\pi}^{D} \) in terms of estimating \( \alpha \) and \( \nu \). However, for larger values of \( |\alpha| \) and \( \nu \), \( \boldsymbol{\pi}^{J} \) outperforms both \( \boldsymbol{\pi}^{B} \) and \( \boldsymbol{\pi}^{D} \). Furthermore, \( \boldsymbol{\pi}^{J} \), \( \boldsymbol{\pi}^{B} \), and \( \boldsymbol{\pi}^{D} \) all outperform the maximum likelihood estimators (MLE) for larger \( |\alpha| \) and \( \nu \). In terms of estimating the location \( \xi \) and scale \( \omega \), the mean absolute deviations (MADs) of all priors are comparable to the MLE.\ These results underscore the efficacy of \( \boldsymbol{\pi}^{J} \) for the skewed Student \( t \) density function, particularly when \( |\alpha| \) and \( \nu \) exceed 30.

\begin{table*}[htbp]
	\small\sf\centering
	\caption{Comparison of the mean absolute deviation (MAD) among \( \boldsymbol{\pi}^{J} \) with \( c_{\nu}=0 \) and \( c_{\nu}=1 \), \( \boldsymbol{\pi}^{B} \), \( \boldsymbol{\pi}^{D} \), and the Uniform prior. The estimation with a Uniform prior distribution is generally equivalent to the maximum likelihood estimation (MLE). Each entry in the table represents the MAD, calculated based on 1000 replications.}\label{tab:mad_comparison} %
	\begin{threeparttable}
		\begin{tabular}{cccccccc}
			\toprule
			\multirow{2}{*}{Parameters} & \multirow{2}{*}{Priors}& \multicolumn{6}{c}{\( (\alpha,\nu) \)} \\
			\cmidrule{3-8}
			&  & \( (-1, 1) \) & \( (-10,10) \) & \( (-30,30) \) & \( (-50,50) \) & \( (-1, 50) \) & \( (-50,1) \) \\
			\midrule
			\multirow{5}{*}{\( \alpha \)} & \( \boldsymbol{\pi}^{J},c_{\nu}=0 \) & 0.156 & 1.989 & 9.402 & 11.839 & 0.471 & 8.415 \\
			& \( \boldsymbol{\pi}^{J},c_{\nu}=1 \) & 0.162 & 2.009 & 9.105 & 11.072 & 0.457 & 7.703 \\
			& \( \boldsymbol{\pi}^{B} \) & 0.142 & 1.901 & 9.399 & 14.267 & 0.497 & 11.489 \\
			& \( \boldsymbol{\pi}^{D} \) & 0.142 & 1.913 & 9.122 & 14.362 & 0.532 & 11.807 \\
			& Uniform & 0.146 & 2.451 & 236.974 & 375.183 & 0.385 & 612.510 \\
			\midrule
			\multirow{5}{*}{\( \nu \)} & \( \boldsymbol{\pi}^{J},c_{\nu}=0 \) & 0.079 & 2.612 & 6.130 & 6.656 & 14.321 & 0.064 \\
			& \( \boldsymbol{\pi}^{J},c_{\nu}=1 \) & 0.085 & 2.418 & 5.474 & 6.210 & 8.961 & 0.063 \\
			& \( \boldsymbol{\pi}^{B} \) & 0.064 & 2.642 & 15.143 & 20.924 & 35.187 & 0.061 \\
			& \( \boldsymbol{\pi}^{D} \) & 0.063 & 2.671 & 15.247 & 21.587 & 35.319 & 0.061 \\
			& Uniform & 0.064 & 13.619 & 82.848 & 50.030 & 27.068 & 0.062 \\
			\midrule
			\multirow{5}{*}{\( \xi \)} & \( \boldsymbol{\pi}^{J},c_{\nu}=0 \) & 0.136 & 0.035 & 0.017 & 0.012 & 0.344 & 0.017 \\
			& \( \boldsymbol{\pi}^{J},c_{\nu}=1 \) & 0.148 & 0.035 & 0.016 & 0.011 & 0.333 & 0.013 \\
			& \( \boldsymbol{\pi}^{B} \) & 0.104 & 0.028 & 0.018 & 0.013 & 0.373 & 0.012 \\
			& \( \boldsymbol{\pi}^{D} \) & 0.105 & 0.029 & 0.019 & 0.014 & 0.401 & 0.013 \\
			& Uniform & 0.105 & 0.026 & 0.016 & 0.010 & 0.261 & 0.011 \\
			\midrule
			\multirow{5}{*}{\( \omega \)} & \( \boldsymbol{\pi}^{J},c_{\nu}=0 \) & 0.120 & 0.071 & 0.049 & 0.041 & 0.156 & 0.088 \\
			& \( \boldsymbol{\pi}^{J},c_{\nu}=1 \) & 0.128 & 0.067 & 0.046 & 0.039 & 0.143 & 0.085 \\
			& \( \boldsymbol{\pi}^{B} \) & 0.099 & 0.076 & 0.076 & 0.064 & 0.199 & 0.085 \\
			& \( \boldsymbol{\pi}^{D} \) & 0.099 & 0.078 & 0.079 & 0.068 & 0.211 & 0.086 \\
			& Uniform & 0.099 & 0.070 & 0.047 & 0.040 & 0.109 & 0.083 \\
			\bottomrule
		\end{tabular}
	\end{threeparttable}
\end{table*}

\subsection{Comparison with Existing Approaches}\label{subsec:comparison_with_existing_approaches}

To evaluate the performance of BIGPAST in comparison to the \( z \)-score, \( t \)-score~\citep{crawfordComparingIndividualTest1998}, Crawford-Garthwaite Bayesian approach~\citep{crawfordComparisonSingleCase2007}, and Anderson-Darling non-parametric approach~\citep{stephens1987} for abnormality detection in a single-subject against a control group, we adopt the ``\textit{severe skew (\( \gamma_{1}=-0.7 \))}'' settings from Section 3.1.1 in~\citet{crawfordTestingDeficitSinglecase2006}. The parameter \( \gamma_{1} \) is a measure of skewness in~\citet{crawfordTestingDeficitSinglecase2006}. This setting is equivalent to  parameter settings: \( \alpha=-3.23 \), \( \nu=7 \), \( \xi=0 \), and \( \omega=1 \) in the definition of our skewed Student \( t \) distribution. The single-subject and control group observations are generated from \( f(x|-3.23,7,0,1) \). We consider sample sizes of 50, 100, 200, and 400 for the control group.
The simulation code in this section can be found in~\href{https://github.com/Jieli12/BIGPAST}{GitHub repository: BIGPAST}. Given a significance level of 0.05, a one-sided test (less) and \textbf{\(\mathbf{c:d=100:0}\)}, we record the abnormality detection results using the \( z \)-score, \( t \)-score, hyperbolic arcsine transformation (\(\log(x+\sqrt{x^{2}+1}), x\in \mathbb{R}\)), Crawford-Garthwaite Bayesian method, Anderson-Darling method, and BIGPAST.\ Other transformations for \(x \in \mathbb{R}\) are applicable; however, determining the optimal transformation is beyond the scope of this paper. Therefore, we will only consider the hyperbolic arcsine transformation in this study. This procedure is repeated one million times as~\citet{crawfordTestingDeficitSinglecase2006}. The results, presented in~\Cref{tab:fpr_acc_less_comparison_all_negative}, indicate that BIGPAST outperforms the other methods in terms of false positive rate (FPR) and accuracy (i.e., ACC=(TP+TN)/(TP+TN+FP+FN).) when all single-subject observations are from the ground truth distribution.

\begin{table}[htbp]
	\footnotesize\sf\centering
	\caption{The table presents the simulation results for false positive rate (FPR) and accuracy (ACC), derived from single-subject observations that are exclusively negative, with a one-sided test (less). Each entry in the table is computed based on one million test outcomes. The abbreviations `CG', `CG-HA' and `AD' denote Crawford-Garthwaite Bayesian method, Crawford-Garthwaite Bayesian method based on hyperbolic arcsine transformation, and Anderson-Darling methods, respectively.}\label{tab:fpr_acc_less_comparison_all_negative}
	\begin{tabular}{cccccccc}
		\toprule
		& \( n \) & \( z \) & \( t \) & CG-HA & CG & AD & BIGPAST \\
		\midrule
		\multirow{4}{*}{FPR} & 50 & 0.0700 & 0.0664 & 0.0544 & 0.0664 & 0.0733 & 0.0519 \\
		& 100 & 0.0696 & 0.0678 & 0.0550 & 0.0678 & 0.0566 & 0.0537 \\
		& 200 & 0.0666 & 0.0657 & 0.0567 & 0.0657 & 0.0576 & 0.0526 \\
		& 400 & 0.0679 & 0.0674 & 0.0554 & 0.0674 & 0.0562 & 0.0521 \\
		\cmidrule{2-8}
		\multirow{4}{*}{ACC} & 50 & 0.9300 & 0.9336 & 0.9456 & 0.9336 & 0.9267 & 0.9481 \\
		& 100 & 0.9304 & 0.9322 & 0.9450 & 0.9322 & 0.9434 & 0.9463 \\
		& 200 & 0.9334 & 0.9343 & 0.9433 & 0.9343 & 0.9424 & 0.9474 \\
		& 400 & 0.9321 & 0.9326 & 0.9446 & 0.9326 & 0.9438 & 0.9479 \\
		\bottomrule
	\end{tabular}
\end{table}

Given that all single-subject observations are from ground truth distribution, implying a negative underlying result, the true positive rate and false negative rate (Type II error) are undefined. To address this, we modify the generation of single-subject observations such that 50\% yield negative results and the remaining 50\% yield positive results, i.e., \(c:d=50:50\). Even in the case of \(c:d=100:0\), ~\Cref{tab:fpr_acc_less_comparison_all_negative} demonstrates that BIGPAST outperforms the other methods. To balance the trade-off between the false positive rate (FPR) and the true positive rate (TPR), we will focus on the case of \(c:d=50:50\) for the remainder of this section. Additionally, results for the case of \(c:d=80:20\) are provided in Appendix, see~\Cref{tab:fpr_acc_less_comparison_mixed_two_sided82}.

The observations are generated using the method described in~\Cref{subsec:Simulation_for_Bayesian_inference_procedures}.
The true positive rate and false positive rate for the \( z \) score, \( t \) score, hyperbolic arcsine transformation, Crawford-Garthwaite Bayesian method, Anderson-Darling method, and BIGPAST, pertaining to a two-sided test, are presented in~\Cref{tab:fpr_acc_less_comparison_mixed_two_sided}. Similar simulations have been conducted for the one-sided test (greater) and one-sided test (less), with the results detailed in Appendix, see~\Cref{tab:fpr_acc_less_comparison_mixed_less,tab:fpr_acc_less_comparison_mixed_greater}.

From~\Cref{tab:fpr_acc_less_comparison_mixed_two_sided}, we can conclude that BIGPAST outperforms the \( t \)-score and Crawford-Garthwaite methods in terms of false positive rate (FPR) and accuracy (ACC) when the single-subject observations are mixed. As Normal distribution is a special case of skewed Student \( t \) distribution (\( \alpha=0, \nu\to+\infty \)), the performance of BIGPAST is still as good as those of \( t \)-score and Crawford-Garthwaite methods when the underlying distribution is the Normal distribution, see~\Cref{tab:fpr_acc_less_comparison_mixed_two-sided_normal} in Appendix. The results are not surprising because the errors in this context originate from two primary sources: model misspecification and inherent randomness. Notably, BIGPAST can effectively mitigate the errors arising from model misspecification; see~\Cref{fig:model_misspecification} in the next section.
\begin{table}[htbp]
	\footnotesize\sf\centering
	\caption{The simulation results of false positive rate (FPR), true positive rate (TPR) and accuracy (ACC), derived from single-subject observations that consist of 50\% positive and 50\% negative, are presented under a two-sided test. Each cell in the resulting table is calculated based on one million test outcomes. `CG', `CG-HA' and `AD' denote the Crawford-Garthwaite Bayesian method, the Crawford-Garthwaite Bayesian method based on hyperbolic arcsine transformation and the Anderson-Darling methods, respectively.}\label{tab:fpr_acc_less_comparison_mixed_two_sided}
	\begin{tabular}{cccccccc}
		\toprule
		& \( n \) & \( z \) & \( t \) & CG-HA & CG & AD & BIGPAST \\
		\midrule
		\multirow{4}{*}{FPR} & 50 & 0.1320 & 0.1182 & 0.0450 & 0.1182 & 0.1107 & 0.0428 \\
		& 100 & 0.1269 & 0.1199 & 0.0569 & 0.1199 & 0.0782 & 0.0349 \\
		& 200 & 0.1207 & 0.1166 & 0.0333 & 0.1167 & 0.0644 & 0.0334 \\
		& 400 & 0.1236 & 0.1215 & 0.0338 & 0.1215 & 0.0368 & 0.0187 \\
		\cmidrule{2-8}
		\multirow{4}{*}{TPR} & 50 & 0.6908 & 0.6502 & 0.8595 & 0.6502 & 0.9101 & 0.8327 \\
		& 100 & 0.6597 & 0.6415 & 0.8917 & 0.6416 & 0.9356 & 0.8992 \\
		& 200 & 0.6420 & 0.6312 & 0.9196 & 0.6311 & 0.9655 & 0.9552 \\
		& 400 & 0.6316 & 0.6264 & 0.9229 & 0.6264 & 0.9660 & 0.9413 \\
		\cmidrule{2-8}
		\multirow{4}{*}{ACC} & 50 & 0.7794 & 0.7660 & 0.9072 & 0.7660 & 0.8997 & 0.8949 \\
		& 100 & 0.7664 & 0.7608 & 0.9174 & 0.7609 & 0.9287 & 0.9322 \\
		& 200 & 0.7606 & 0.7573 & 0.9432 & 0.7572 & 0.9505 & 0.9609 \\
		& 400 & 0.7540 & 0.7525 & 0.9445 & 0.7524 & 0.9646 & 0.9613 \\
		\bottomrule
	\end{tabular}
\end{table}
\subsection{Model Misspecification Error}\label{subsec:Model_misspecification_error}
\begin{figure*}[!ht]
	\centering
	\begin{subfigure}[bt]{0.48\textwidth}
		\centering
		\includegraphics[width=\textwidth]{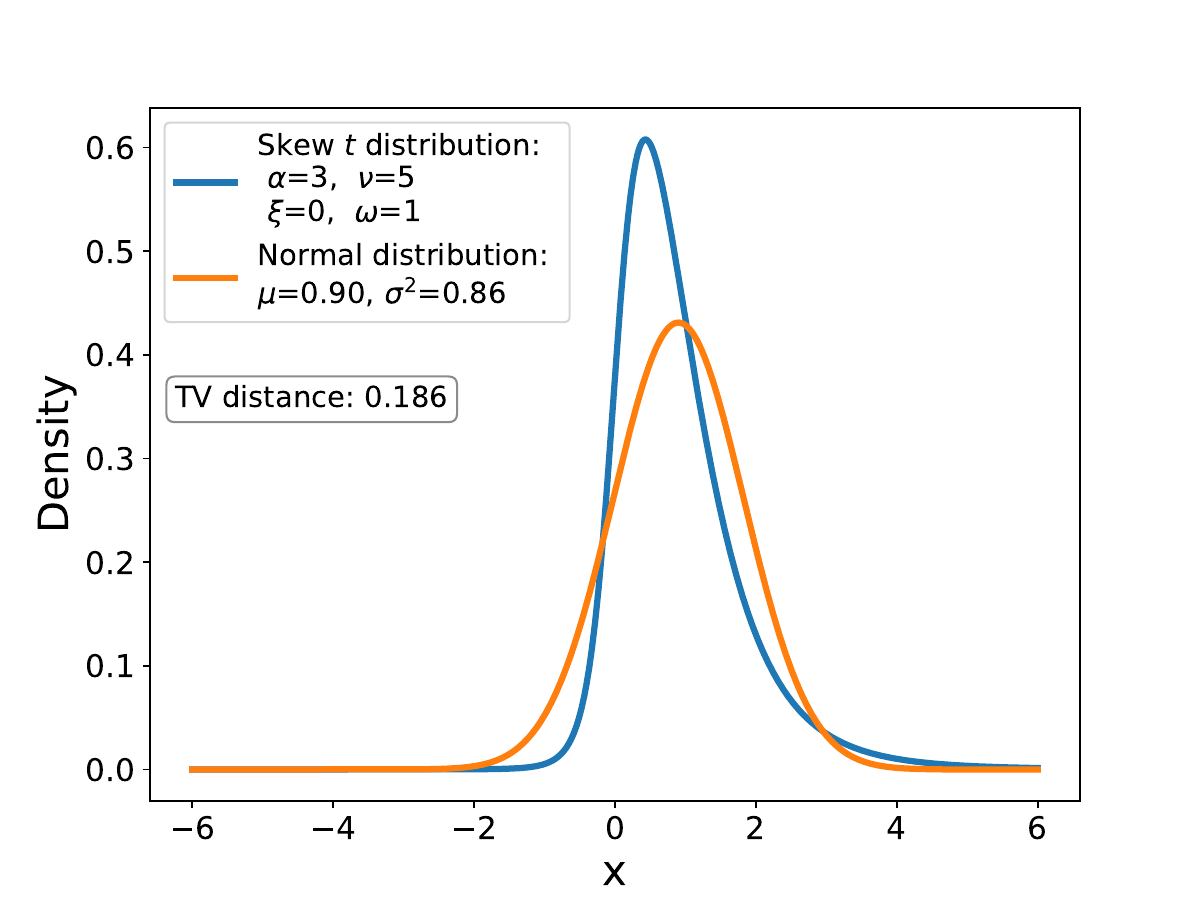}
		\caption{The density}\label{fig:figs/alpha3df5.pdf}
	\end{subfigure}
	\begin{subfigure}[bt]{0.48\textwidth}
		\centering
		\includegraphics[width=\textwidth]{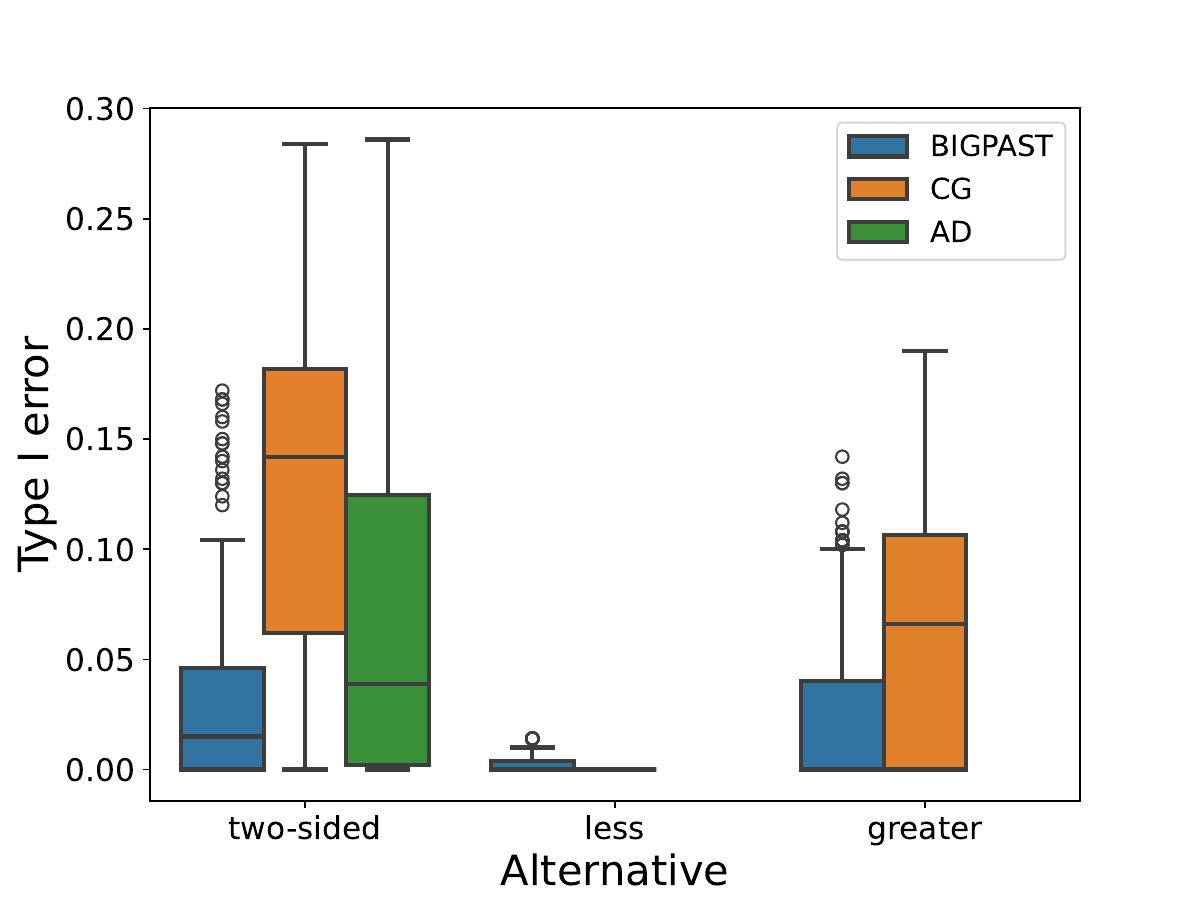}
		\caption{The Type I error}\label{fig:figs/TypeIErroralpha3df5.pdf}
	\end{subfigure}
	\begin{subfigure}[bt]{0.48\textwidth}
		\centering
		\includegraphics[width=\textwidth]{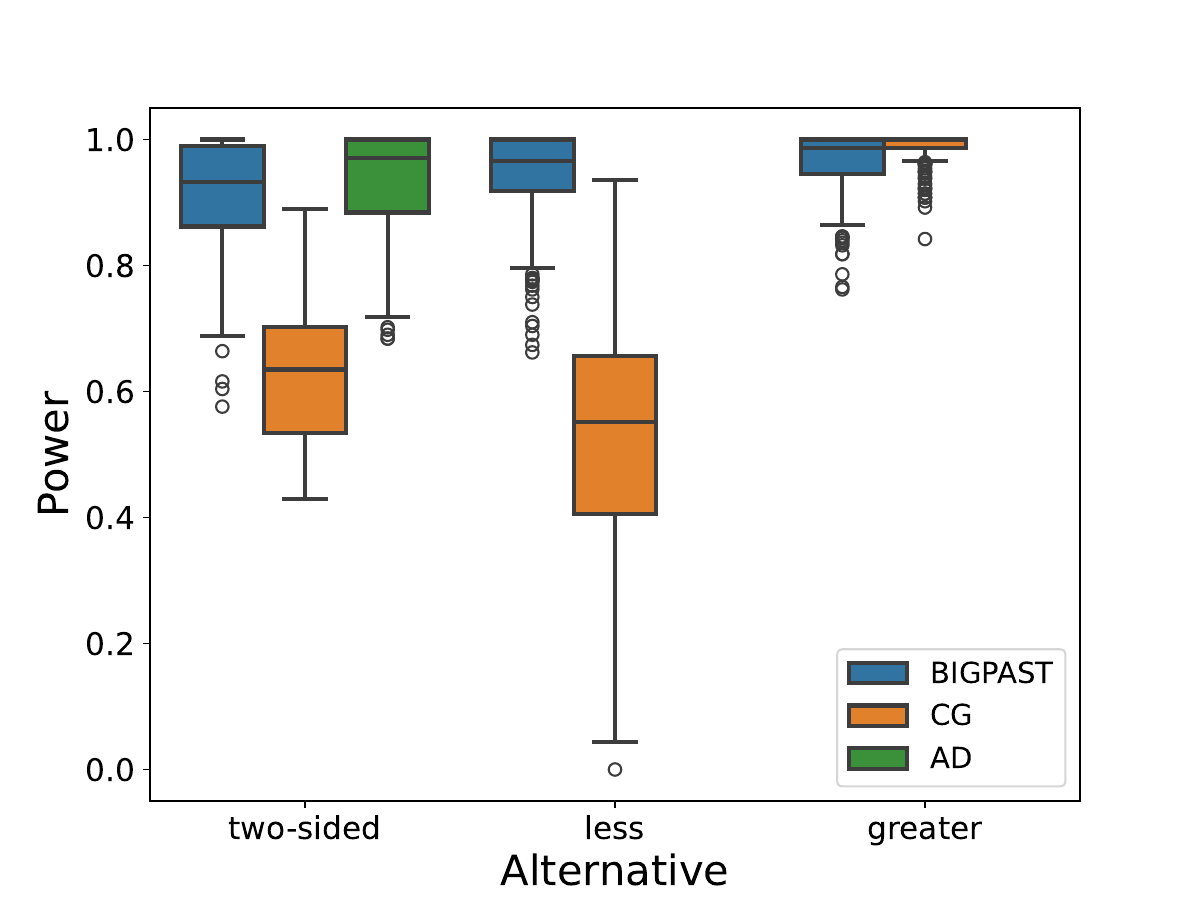}
		\caption{The power}\label{fig:figs/Poweralpha3df5.pdf}
	\end{subfigure}
	\begin{subfigure}[bt]{0.48\textwidth}
		\centering
		\includegraphics[width=\textwidth]{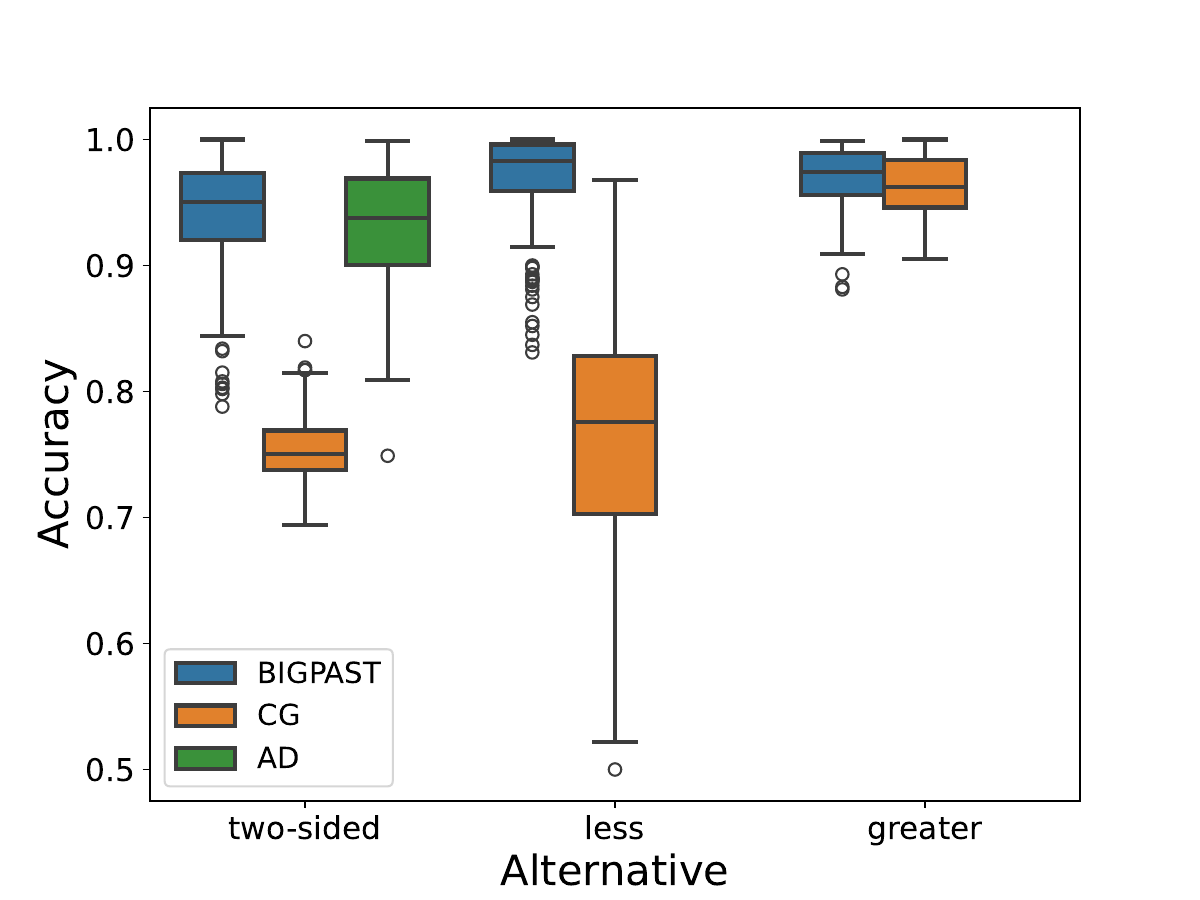}
		\caption{The accuracy}\label{fig:figs/Accuracyalpha3df5.pdf}
	\end{subfigure}
	\caption{The comparison results of BIGPAST, CG and AD approaches under three alternative hypotheses: `two-sided', `less' and `greater' respectively when \( \alpha=3 \) and \( \nu=5 \).~\Cref{fig:figs/alpha3df5.pdf} shows the densities of skewed Student \( t \) and Normal distributions, the \( \mu \) and \( \sigma^{2} \) of Normal distribution are equal to the mean and variance of skewed Student \( t \) distribution. In fact, the Normal distribution is the theoretical assumption of the CG test when the sample comes from the skewed Student \( t \) distribution. TV distance is short for total variation distance. Each box plot summarises over 200 independent replications.}\label{fig:alpha3df5}
\end{figure*}

In this section, we conduct a comparative analysis of the BIGPAST, CG, and AD methodologies under varying alternative hypotheses and parameter configurations. The predetermined significance level is 0.05. The skewness parameter \( \alpha \) and degrees of freedom \( \nu \) are assigned values from the sets \( (1, 10) \), \( (1, 5) \), \( (2, 5) \), and \( (3, 5) \), respectively. The location and scale of the skewed Student \( t \) distribution are set to 0 and 1, respectively. Other sample settings include a sample size \( n=100 \), \( c_{\nu}=1 \), and a Metropolis-Hastings sampling size of 2000. The comparison procedures are as follows: Firstly, \( N=200 \) independent samples are randomly generated from the skewed Student \( t \) distribution with parameters \( \alpha \) and \( \nu \). Secondly, \( m=1000 \) single-subjects are randomly drawn, and the BIGPAST, CG, and AD tests are conducted on each of the \( N \) samples. Thirdly, for each of \( N \) samples, the Type I error (false positive rate), power (true positive rate), and accuracy are calculated based on the  \( m=1000 \) single-subjects. Finally, the Type I error, power, and accuracy are reported in the form of box plots for the four skewed Student \( t \) parameter settings, as shown in~\Cref{fig:alpha1df10,fig:alpha1df5,fig:alpha2df5,fig:alpha3df5}.

Under the mechanism of these parameter settings, the model is misspecified for the CG test. The theoretical assumption of CG test is the Normal distribution, the mean and variance can be directly computed from the skewed Student \( t \) distribution by \( \mu=\delta b_{\nu},\nu>1 \) and \( \sigma^{2}=\nu/(\nu-2)-(b_{\nu}\delta)^2, \nu>2 \), where
\begin{equation*}
	b_{\nu}=\frac{\sqrt{\nu}\Gamma\left( \frac{1}{2}(\nu-1) \right)}{\sqrt{\pi}\Gamma\left( \frac{\nu}{2} \right)},\quad \nu>1; \quad \delta=\frac{\alpha}{\sqrt{1+\alpha^{2}}}.
\end{equation*}

We utilise the total variation distance to quantify the disparity between skewed Student \( t \) (for BIGPAST) and Normal (for CG) distributions. The total variation distances of the aforementioned four settings are: 0.057, 0.111, 0.155 and 0.186, respectively, as illustrated in~\Cref{fig:figs/alpha1df10.pdf,fig:figs/alpha1df5.pdf,fig:figs/alpha2df5.pdf,fig:figs/alpha3df5.pdf}. In the context of a two-sided hypothesis test, the divergence between the BIGPAST and CG methodologies becomes larger as the total variation distance increases, as shown in~\Cref{fig:alpha1df10,fig:alpha1df5,fig:alpha2df5,fig:alpha3df5}. The result of the AD approach demonstrates robustness for different total variation distances but performs significantly worse than BIGPAST in terms of Type I error, as seen in~\Cref{fig:figs/TypeIErroralpha3df5.pdf}. The AD approach also performs slightly worse than BIGPAST in terms of accuracy, as depicted in~\Cref{fig:figs/Accuracyalpha3df5.pdf}.

\begin{figure*}[ht]
	\centering
	\begin{subfigure}[bt]{0.49\textwidth}
		\centering
		\includegraphics[width=\textwidth]{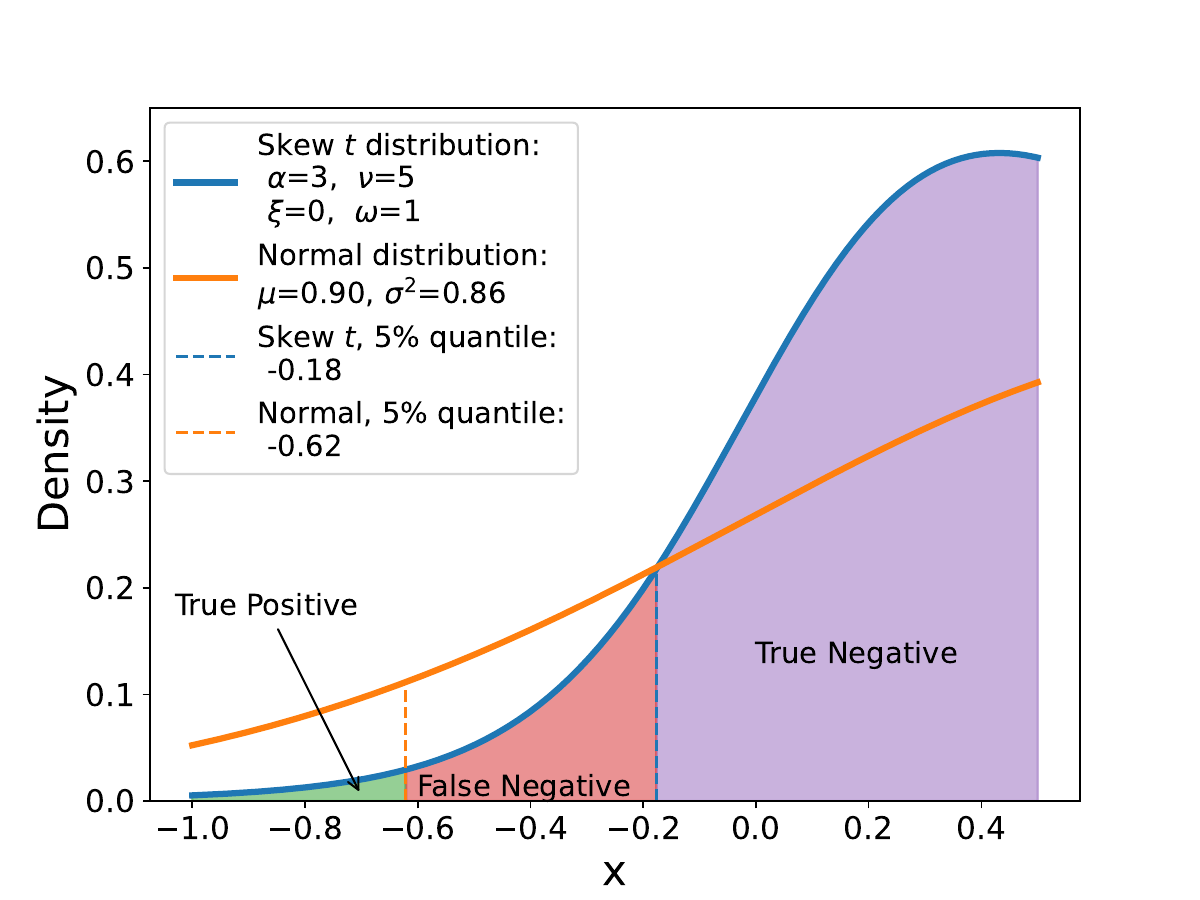}
		\caption{Alternative: `less'}\label{fig:figs/alpha3df5light_tail.pdf}
	\end{subfigure}
	\begin{subfigure}[bt]{0.49\textwidth}
		\centering
		\includegraphics[width=\textwidth]{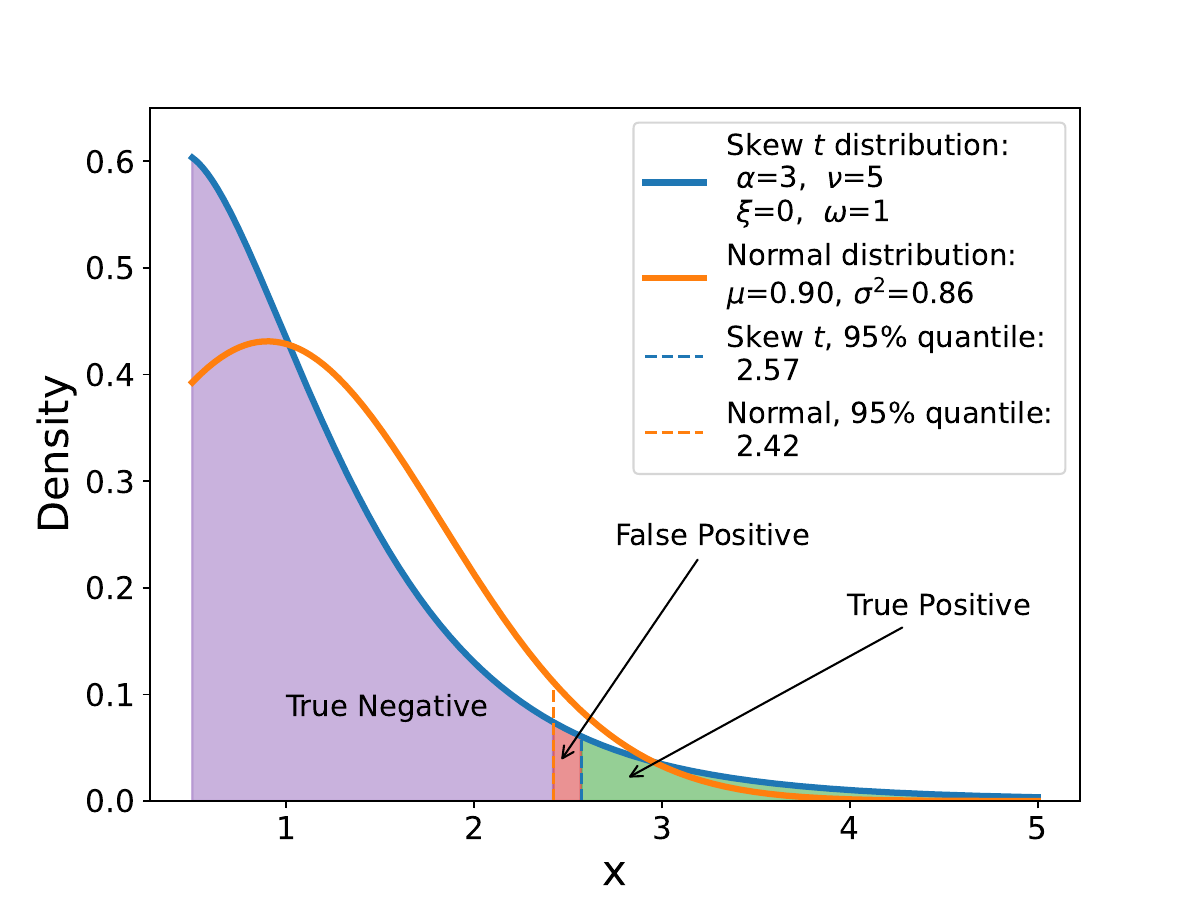}
		\caption{Alternative: `greater'}\label{fig:figs/alpha3df5heavy_tail.pdf}
	\end{subfigure}
	\caption{Model misspecification error on the light and heavy tails for one-sided test.}\label{fig:model_misspecification}
\end{figure*}

The complexity of the one-sided test (either `greater' or `less') compared to the two-sided test due to the skewed Student \( t \) distribution increases due to the skewed Student \( t \) distribution having light and heavy tails. The model misspecification error impacts the tails of the skewed Student \( t \) distribution differently. When \( \alpha>0 \) (resp.\ \( \alpha<0 \)), the skewed Student \( t \) distribution has a right (resp.\ left) heavy tail. In this paragraph, we explore the scenario where the direction of the alternative hypothesis is `greater', i.e., one-sided test (greater). If we implement the CG test, the 95\% quantile of the Normal distribution is less than the 95\% quantile of the skewed Student \( t \) distribution. Consequently, the Type I error of the CG test exceeds that of the BIGPAST test, leading to a higher power for the CG test, as shown in~\Cref{fig:figs/TypeIErroralpha3df5.pdf,fig:figs/Poweralpha3df5.pdf}. This can be explained by the light-red region between the 95\% quantile of skewed Student \( t \) distribution and the 95\% quantile of Normal distribution, as depicted in~\Cref{fig:figs/alpha3df5heavy_tail.pdf}. Theoretically, all single-case observations from this region will be classified as positive by the CG test and as negative by the BIGPAST test, leading us to refer to this region as the false positive region. The error generated by the false positive region arises from model misspecification rather than sampling randomness, as illustrated in~\Cref{fig:model_misspecification}.

Conversely, for the light tail of the skewed Student \( t \) distribution, the 5\% quantile of the Normal distribution is less than the 5\% quantile of the skewed Student \( t \) distribution. Given that the alternative is `less' and \( \alpha>0 \),  the region between the 5\% quantile of skewed Student \( t \) distribution and the 5\% quantile of the Normal distribution is referred to as the false negative region other than the false positive region. Because all the single-case observations from this region will be classified as negative by the CG test but as positive by the BIGPAST test. Consequently, the theoretical Type I error is zero, excluding the random error of sampling, as depicted in~\Cref{fig:figs/TypeIErroralpha3df5.pdf}. The power of the CG test, i.e., the actual positive rate, is lower than that of the BIGPAST test, as illustrated in~\Cref{fig:figs/Poweralpha3df5.pdf}.

For the left skewed Student \( t \) distribution (\( \alpha<0 \)), we can obtain similar results just by reflecting the densities in line \( x=0 \). This will not be detailed here. From the preceding discussion, we can conclude that (1) The difference between the BIGPAST and CG methodologies becomes more significant in terms of Type I error, power, and accuracy as the total variation distance increases; (2) Assuming a skewed Student \( t \) distribution, the BIGPAST approach outperforms the CG approach in terms of accuracy due to the presence of model misspecification error; (3) Given the asymmetry between the light and heavy tails of the skewed Student \( t \) distribution, we recommend using a two-sided test instead of a one-sided test. This is because the combinations of a one-sided test (`less' or `greater') and the sign of \( \alpha \) significantly influence the evaluation of detection.

For the left skewed Student \( t \) distribution (\( \alpha<0 \)), we can obtain similar results just by reflecting the densities in line \( x=0 \). This will not be detailed here. From the preceding discussion, we can conclude that (1) The difference between the BIGPAST and CG methodologies becomes more significant in terms of Type I error, power, and accuracy as the total variation distance increases; (2) Assuming a skewed Student \( t \) distribution, the BIGPAST approach outperforms the CG approach in terms of accuracy due to the presence of model misspecification error; (3) Given the asymmetry between the light and heavy tails of the skewed Student \( t \) distribution, we recommend using a two-sided test instead of a one-sided test. This is because the combinations of a one-sided test (`less' or `greater') and the sign of \( \alpha \) significantly influence the evaluation of detection.
\subsection{Comparison Study with Other Frameworks}\label{subsec:Simulation_for_Bayesian_inference_procedures}
This section is dedicated to assessing the performance of the proposed BIGPAST methodology and existing approaches when a control group is present. The first approach is a variant of BIGPAST based on the maximum a posteriori (MAP) estimation displayed in steps \(3^{\prime}\) and \(4^{\prime}\).

The second approach is a non-Bayesian approach that relies on maximum likelihood estimation (MLE). This approach assumes that the control group follows a skewed Student \( t \) distribution. The MLE procedures can be outlined as follows:
\begin{enumerate}[label={\arabic*$'$}]
	\item Obtain the maximum likelihood estimators of \( \alpha,\nu,\xi,\omega \) by maximising the likelihood function \( l(\mathbf{x}|\alpha,\nu,\xi,\omega) \). These estimators are denoted as \( \hat{\alpha}_{\mathrm{MLE}},\hat{\nu}_{\mathrm{MLE}},\hat{\xi}_{\mathrm{MLE}},\hat{\omega}_{\mathrm{MLE}}\).
	\item Given the significance level \( \beta \), the confidence interval of \( x^{*} \) given \( \mathbf{x} \) is \( [x^{\mathrm{MLE}}_{\beta/2},x^{\mathrm{MLE}}_{1-\beta/2}] \), where \( x^{\mathrm{MLE}}_{\beta/2} \) and \( x^{\mathrm{MLE}}_{1-\beta/2} \) are the \( \beta/2 \) and \( 1-\beta/2 \) quantiles of \( f_{c}(\hat{\alpha}_{\mathrm{MLE}},\hat{\nu}_{\mathrm{MLE}},\hat{\xi}_{\mathrm{MLE}},\hat{\omega}_{\mathrm{MLE}}) \), respectively. If \( x^{*} \) does not fall within this confidence interval, then the null hypothesis \( \mathcal{H}_{0} \) can be rejected at the significance level \( \beta \).
\end{enumerate}

The third approach is a non-parametric (NP) approach. Given group data \( \mathbf{x} \) and a significance level \( \beta \), we can derive the empirical confidence interval \( [x^{\mathrm{NP}}_{(\lfloor n*\beta/2\rfloor )},x^{\mathrm{NP}}_{(\lceil n*(1-\beta/2) \rceil)}] \) where \( x^{\mathrm{NP}}_{(\lfloor n*\beta/2\rfloor )} \) and \( x^{\mathrm{NP}}_{(\lceil n*(1-\beta/2) \rceil)}  \) are the empirical quantiles of \( \mathbf{x} \) at \( \beta/2 \) and \( 1-\beta/2 \), respectively. If \( x^{*} \) does not fall within this empirical confidence interval, then the null hypothesis \( \mathcal{H}_{0} \) can be rejected at the significance level \( \beta \). The NP approach does not rely on any distributional assumptions. Given that our BIGPAST methodology employs the Metropolis-Hastings sampler in Step 3, we will refer to the four approaches above: MH, MAP, MLE, and NP, respectively.

We now outline the data generation mechanism for a single-subject and control group. Given the parameters \(\alpha, \nu, \xi, \omega \), we randomly draw \( N \) independent control groups \( \mathbf{x}_{1},\mathbf{x}_{2},\ldots,\mathbf{x}_{N} \) from \( f_{c}(\alpha,\nu,\xi,\omega) \). The following is a straightforward method for generating single-subject data. Define \( S_{1}\coloneqq (-\infty,q_{2.5\%}]\cup [q_{97.5\%},\infty) \) and \( S_{2}\coloneqq (q_{2.5\%},q_{5\%}]\cup [q_{95\%},q_{97.5\%}) \) where \( q_{2.5\%},q_{5\%},q_{95\%}, q_{97.5\%} \) are 0.025, 0.05, 0.95, 0.975 quantiles of \( f_{c}(\alpha,\nu,\xi,\omega) \). We draw \( k_1 \) and \( k_{2} \) random copies of single-subjects \( x^{*}_{1}\ldots,x^{*}_{k_1},x^{*}_{k_1+1},\ldots, x^{*}_{k_1+k_2}\) respectively from the uniform distributions defined on \( S_{1} \) and \( S_{2} \). Let \( K=k_1+k_2 \). At the significance level of 0.05, the ground truth for the first \( k_{1} \) single-subjects is negative, while the ground truth for the remaining \( k_{2} \) single-subjects is positive. In the simulation, we first apply the MH, MAP, MLE, and NP approaches to \( \mathbf{x}_{i}, i=1,2,\ldots, N \) to obtain the four types of intervals. Next, we conduct the hypothesis tests by comparing \( \mathbf{x}^{*}=( x^{*}_{1},x^{*}_{2},\ldots,x^{*}_{K}) \) with each type of intervals. Finally, we compute the false discovery rate (FDR) and accuracy (ACC) defined by
\begin{equation*}
	\mathrm{FDR}_{i}^{\mathcal{M}}\coloneqq \frac{\mathrm{FP}_{i}^{\mathcal{M}}}{\mathrm{TP}_{i}^{\mathcal{M}}+\mathrm{FP}_{i}^{\mathcal{M}}},\quad \mathrm{ACC}_{i}^{\mathcal{M}}\coloneqq \frac{\mathrm{TP}_{i}^{\mathcal{M}}+\mathrm{TN}_{i}^{\mathcal{M}}}{K},
\end{equation*}
where \(i=1,2,\ldots,N\), \( \mathrm{TP} \), \( \mathrm{FP} \), and \( \mathrm{TN} \) represent the number of true positive, false positive and true negative, respectively.\ \( \mathcal{M} \) could be one of MH, MAP, MLE, NP.\

\begin{figure*}[htbp]
	\centering
	\begin{subfigure}[bt]{0.9\textwidth}
		\centering
		\includegraphics[width=\textwidth]{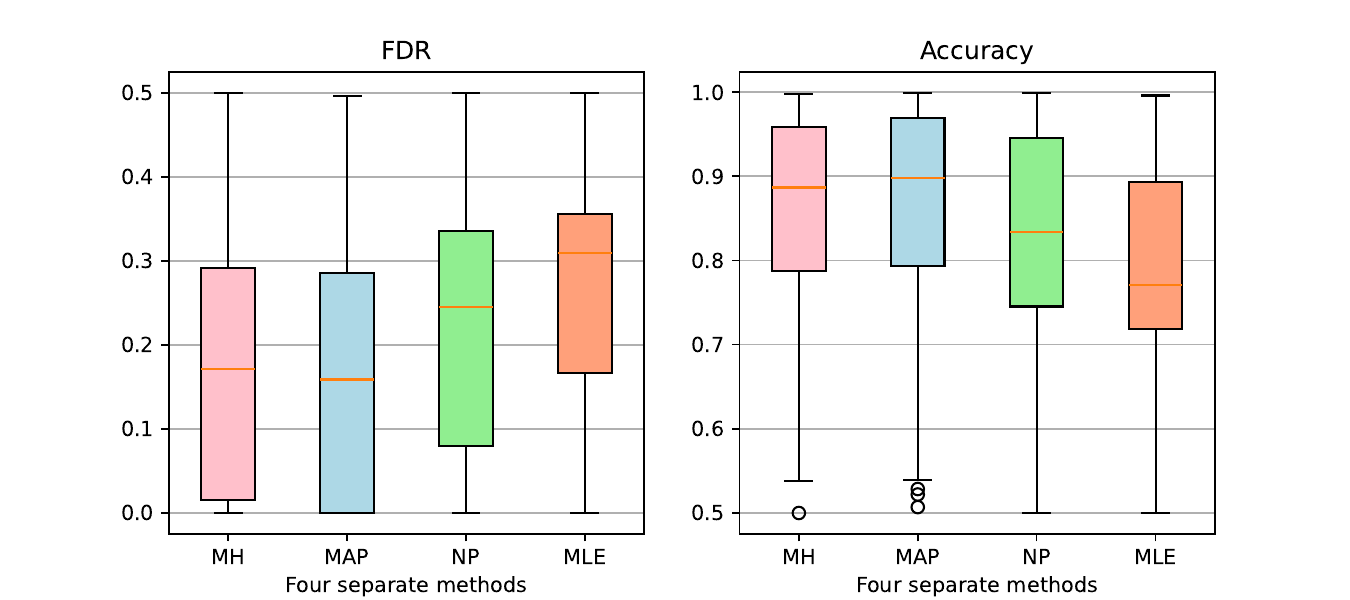}
		\caption{\( \alpha=0,\nu=3 \)}\label{fig:figs/Result_n100_alpha0_df3.pdf}
	\end{subfigure}
	\begin{subfigure}[bt]{0.9\textwidth}
		\centering
		\includegraphics[width=\textwidth]{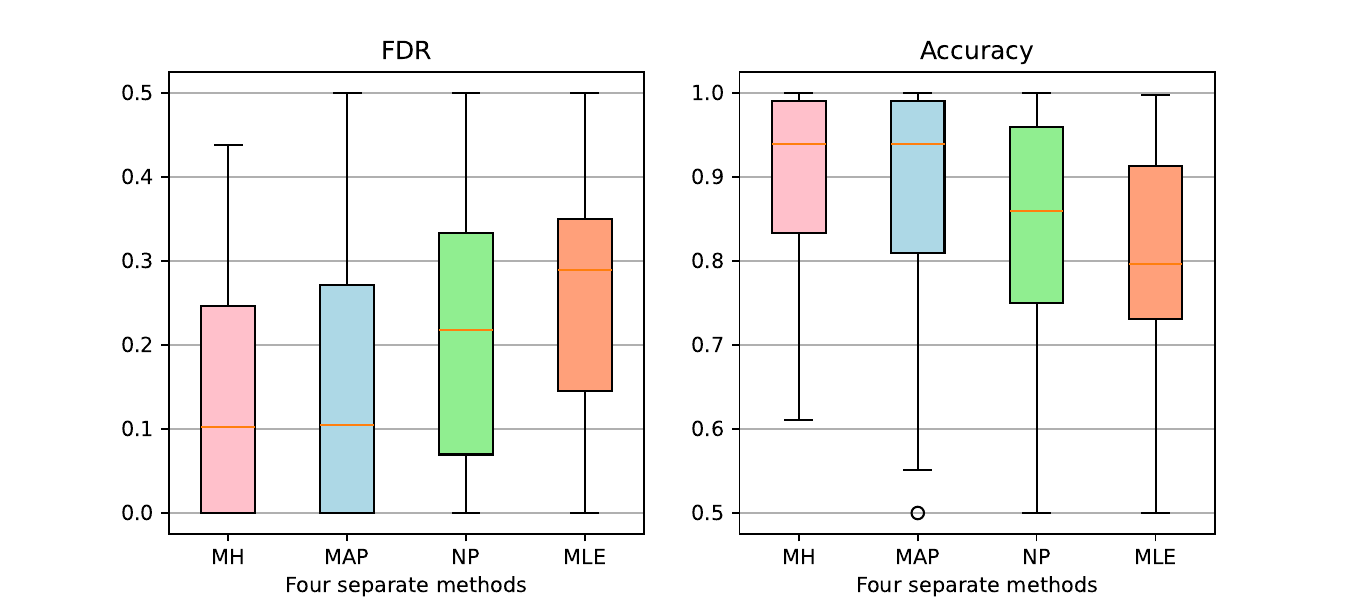}
		\caption{\( \alpha=5,\nu=5 \)}\label{fig:figs/Result_n100_alpha5_df5.pdf}
	\end{subfigure}
	\begin{subfigure}[bt]{0.9\textwidth}
		\centering
		\includegraphics[width=\textwidth]{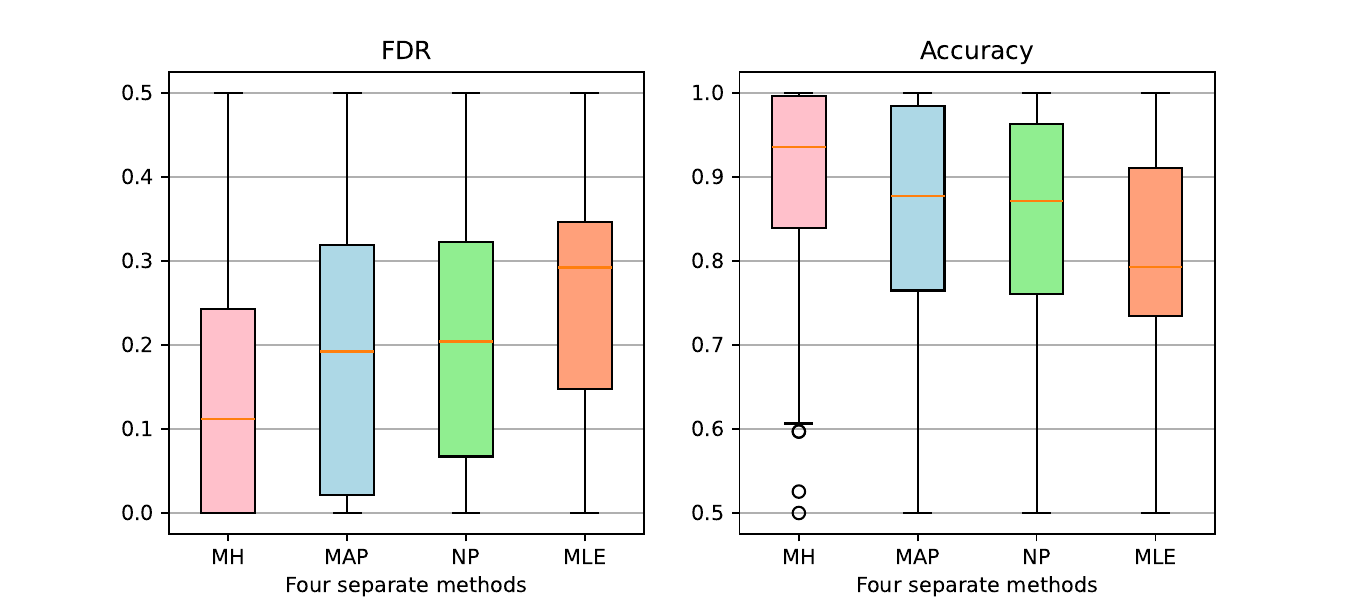}
		\caption{\( \alpha=50,\nu=50 \)}\label{fig:figs/Result_n100_alpha50_df50.pdf}
	\end{subfigure}
	\caption{The box plots for false discovery rate (FDR) and accuracy (ACC) of the MH, MAP, MLE, and NP approaches for three distinct skewness parameters: low (\( \alpha=0 \)), medium (\( \alpha=5 \)), and high (\( \alpha=50 \)). In this context, MH refers to the original BIGPAST method, MAP denotes the Bayesian inference framework based on maximum a posteriori estimation, MLE represents the method based on maximum likelihood estimation, and NP signifies the non-parametric method.}\label{fig:fdr_acc_boxplot}
\end{figure*}

The parameters settings are as follows: we consider ten combinations of \( (\alpha,\nu) \): (0,3), (1,3), (3,3), (5,5), (10,10), (20,20), (30,30), (50,50), (5,50) and (50,5) with fixed \( \xi=-2 \) and \( \omega=2 \). In MH approach, we let \( m=5000 \), \( s=100 \) and \( c_{\nu}=1 \). The other parameter settings are \( N=400 \), \( n=100 \), \( k_1=k_2=1000\), \( \beta=0.05 \). The results are presented in~\Cref{tab:fdr_acc_mean_table} and~\Cref{fig:fdr_acc_boxplot}.

\Cref{tab:fdr_acc_mean_table} displays the FDR and ACC of the four approaches.
The MH and MAP methods consistently outperform the NP and MLE methods in terms of FDR and ACC.\ When \( {\left\lvert \alpha \right\rvert}\leq3 \) and \( \nu\leq 3 \), both the FDR and ACC of the MAP approach surpass those of the MH approach. This is expected, as the MAP approach
provides more accurate estimation of
the underlying parameters of the skewed Student \( t \) distribution when \( {\left\lvert \alpha \right\rvert} \) and \( \nu \) are small, as is shown in column 3 of~\Cref{tab:mad_comparison}. When \( {\left\lvert \alpha \right\rvert} \) or \( \nu \) is large,
the FDR and ACC of the MH approach perform better than those of the MAP approach.
This can be attributed to the randomness of parameters in the Metropolis-Hastings sampler. However, the MAP approach fails to estimate the underlying parameters when \( \alpha \) or \( \nu \) is large, as is shown in column 4 of~\Cref{tab:mad_comparison}, and lacks randomness of parameters. The box plots in~\Cref{fig:fdr_acc_boxplot} clearly illustrate the comparison between the MH and MAP approaches. The results demonstrate that the FDR of MH (i.e. BIGPAST) is generally lower than others, and the ACC of MH (i.e. BIGPAST) is generally greater than other methods in conducting hypothesis tests for single-subject data.

\begin{table}[htbp]
	\centering
	\caption{The average false discovery rate (FDR) and accuracy (ACC) for the MH, MAP, MLE, and NP methods were calculated over 400 repetitions. In this context, MH refers to the original BIGPAST method, MAP denotes the Bayesian inference framework based on maximum a posteriori estimation, MLE represents the method based on maximum likelihood estimation, and NP signifies the non-parametric method.}\label{tab:fdr_acc_mean_table}
	\begin{tabular}{ccccccccc}
		\toprule
		\multirow{2}{*}{\( (\alpha,\nu) \)}& \multicolumn{4}{c}{FDR}   & \multicolumn{4}{c}{ACC} \\
		\cline{2-5} \cline{6-9}
		& MH & MAP & NP & MLE & MH & MAP & NP & MLE \\
		\midrule
		(0,3) & 0.174 & 0.165 & 0.216 & 0.266 & 0.865 & 0.872 & 0.833 & 0.787 \\
		(1,3) & 0.158 & 0.145 & 0.201 & 0.257 & 0.877 & 0.888 & 0.845 & 0.797 \\
		(3,3) & 0.142 & 0.129 & 0.197 & 0.254 & 0.891 & 0.901 & 0.849 & 0.798 \\
		(5,5) & 0.131 & 0.136 & 0.206 & 0.255 & 0.905 & 0.899 & 0.846 & 0.803 \\
		(10,10) & 0.118 & 0.145 & 0.200 & 0.259 & 0.915 & 0.894 & 0.851 & 0.797 \\
		(20,20) & 0.132 & 0.157 & 0.190 & 0.255 & 0.907 & 0.886 & 0.861 & 0.803 \\
		(30,30) & 0.132 & 0.171 & 0.188 & 0.261 & 0.907 & 0.875 & 0.862 & 0.796 \\
		(50,50) & 0.134 & 0.184 & 0.193 & 0.254 & 0.906 & 0.864 & 0.859 & 0.801 \\
		(5,50) & 0.144 & 0.158 & 0.185 & 0.261 & 0.898 & 0.885 & 0.865 & 0.796 \\
		(50,5) & 0.139 & 0.165 & 0.218 & 0.249 & 0.899 & 0.876 & 0.834 & 0.805 \\
		\bottomrule
	\end{tabular}
\end{table}
\section{Real Data Analysis}\label{sec:Real_data_analysis}
This dataset is derived from a mild traumatic brain injury (mTBI) study conducted by~\href{https://www.innovision-ip.co.uk}{Innovision IP Ltd}. It comprises a functional time series of Magnetoencephalography (MEG) data and an anatomical T1-weighted Magnetic Resonance Imaging (MRI) scan for 103 healthy subjects and one patient with mild traumatic brain injury. The MEG data, recorded from 306 sensors, was pre-processed using the `MNE-Python' package~\citep{GramfortEtAl2013a}. The MEG data were analysed to reconstruct the source generators using a minimum norm inverse method. The MRI data were pre-processed with the `FreeSurfer' package~\citep{hanReliabilityMRIderivedMeasurements2006}. The standard FreeSurfer pipeline (detailed on their website) was followed and includes motion correction, intensity normalisation, skull-stripping, normalisation to template space and segmentation of the cortex and subcortical structures using the Desikan-Killiany atlas. This results in 68 regions covering the entire cortex. For each cortical region, we calculate the average of source powers within this region. This process yields a control group dataset with dimensions \( 103\times 68\times100 \) and a patient dataset with dimensions \( 68\times100 \). To simplify, we treat the epochs as repeated observations and take averages over them to reduce the data's complexity. Consequently, the control group dataset has dimensions \( 103\times 68 \), and the patient dataset is reduced to a vector of length \( 68 \).

For each brain region of the Desikan-Killiany cortical atlas, we apply the BIGPAST method to the control group and patient dataset to test whether the patient significantly differs from the control group. This real data analysis shows that the single-subject has mTBI.\ However, in practice, we can calculate the observation of a new client and conduct the hypothesis test using BIGPAST to determine whether the client significantly differs from the control group.

We further explore the impact of waveband frequency on the comparison between a control group and a single-subject. The wavebands are arranged in ascending order of frequency: delta (1-4 Hz), theta (4-8 Hz), alpha (8-12 Hz), beta (12-30 Hz), and gamma (30-45 Hz).
The subsequent results also include those of the Crawford-Garthwaite method, denoted as CG for brevity. The prespecified significance level is 0.05. The test results of Crawford-Garthwaite (CG) and BIGPAST are presented in~\Cref{tab:result_CG_BIGPAST_no_log_transformation}, where the bracket \textit{(Normal, mTBI)} in the third row represents that the result of CG is Normal, but the result of BIGPAST is mTBI.\ The numbers in cells are the indices of human brain regions; for example, number 1 in the ``delta'' column represents the caudal anterior cingulate in the left brain hemisphere. For this brain region, CG's result is Normal, while the result of BIGPAST is mTBI. We did not consider the false discovery rate adjustment here as (1) steps 3 and 4 in BIGPAST employ the nested sampling technique, and the p-values are hard to obtain. (2) The aim of~\Cref{tab:result_CG_BIGPAST_no_log_transformation} is to provide a comparison between the CG and BIGPAST methods, and the FDR adjustment is not necessary for this purpose.
\begin{table*}[htbp]
	\small\sf\centering

	\caption{The test results of the Crawford-Garthwaite (CG) based on Gaussian assumption and BIGPAST approaches based on the skewed Student \( t \) assumption. The numbers in cells corresponding to specific brain cortical regions are 1:  lh-caudal anterior cingulate, 2: lh-caudal middle frontal, 4: lh-entorhinal, 6: lh-fusiform, 10: lh-isthums cingulate, 12: lh-lateral orbitofrontal, 14: lh-medial orbitofrontal, 17: lh-parahippocampal, 22: lh-postcentral, 24: lh-precentral, 26: lh-rostral anterior cingulate, 35: rh-caudal anterior cingulate, 38: rh-entorhinal, 40: rh-fusiform, 47: rh-lingual, 50: rh-paracentral, 51: rh-parahippocampal, 58: rh-precentral, 59: rh-precuneus, 62: rh-superior frontal. The prefixes `lh-' and `rh-' indicate the left and right hemispheres of the brain, respectively. `NA' stands for `not available', and `Others' represents the remaining brain cortical indices not listed in the column but can be found in the indices in~\Cref{tab:brain_region_index}.}\label{tab:result_CG_BIGPAST_no_log_transformation}
	\begin{threeparttable}
		\begin{tabular}{cccccc}
			\toprule
			\multirow{2}{*}{The Results of (CG, BIGPAST)} & \multicolumn{5}{c}{Wave bands} \\
			\cmidrule{2-6}
			& delta & theta & alpha & beta & gamma \\
			\midrule
			(mTBI, mTBI) & \begin{tabular}{@{}c@{}}6,17,\\40,51\end{tabular}  &\begin{tabular}{@{}c@{}}4,6,\\17,48\end{tabular}  & \begin{tabular}{@{}c@{}}6,14,\\17,40\end{tabular} & \begin{tabular}{@{}c@{}}6,12,17,\\38,40\end{tabular} & 17 \\
			\midrule
			(Normal, mTBI) & \begin{tabular}{@{}c@{}}1,10,16,26,\\ 35,50,59,62\end{tabular} & \begin{tabular}{@{}c@{}}1,2,24,\\35,62\end{tabular} & \begin{tabular}{@{}c@{}}2,24,35,\\58,59\end{tabular} & \begin{tabular}{@{}c@{}}1,2,10,22,\\24,35,59\end{tabular} & 1,2,35 \\
			\midrule
			(mTBI, Normal) &\begin{tabular}{@{}c@{}}NA\end{tabular}  & 14,40 & NA &\begin{tabular}{@{}c@{}}14,47\end{tabular}  &6,40 \\
			\midrule
			(Normal, Normal) & Others & Others & Others & Others & Others \\
			\bottomrule
		\end{tabular}
	\end{threeparttable}
\end{table*}

\begin{figure*}[!ht]
	\centering
	\begin{subfigure}[bt]{0.4\textwidth}
		\centering
		\includegraphics[width=\textwidth]{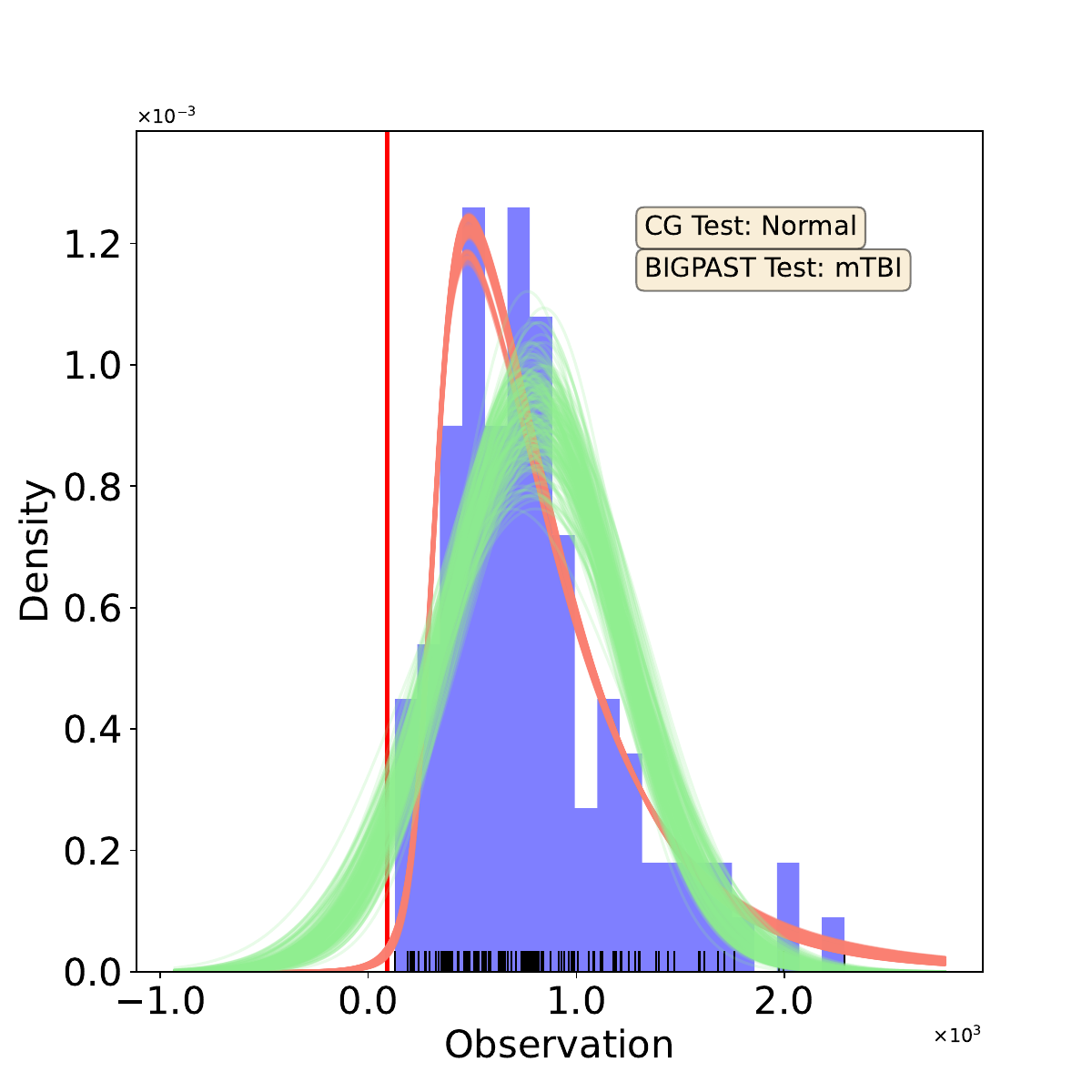}
		\caption{Delta Band, \( \bar{\hat{\alpha}}\approx3.95 \), Left Caudal Anterior Cingulate}\label{fig:figs/delta1ctx-lh-caudalanteriorcingulate_nolog50two_sided.pdf}
	\end{subfigure}
	\begin{subfigure}[bt]{0.4\textwidth}
		\centering
		\includegraphics[width=\textwidth]{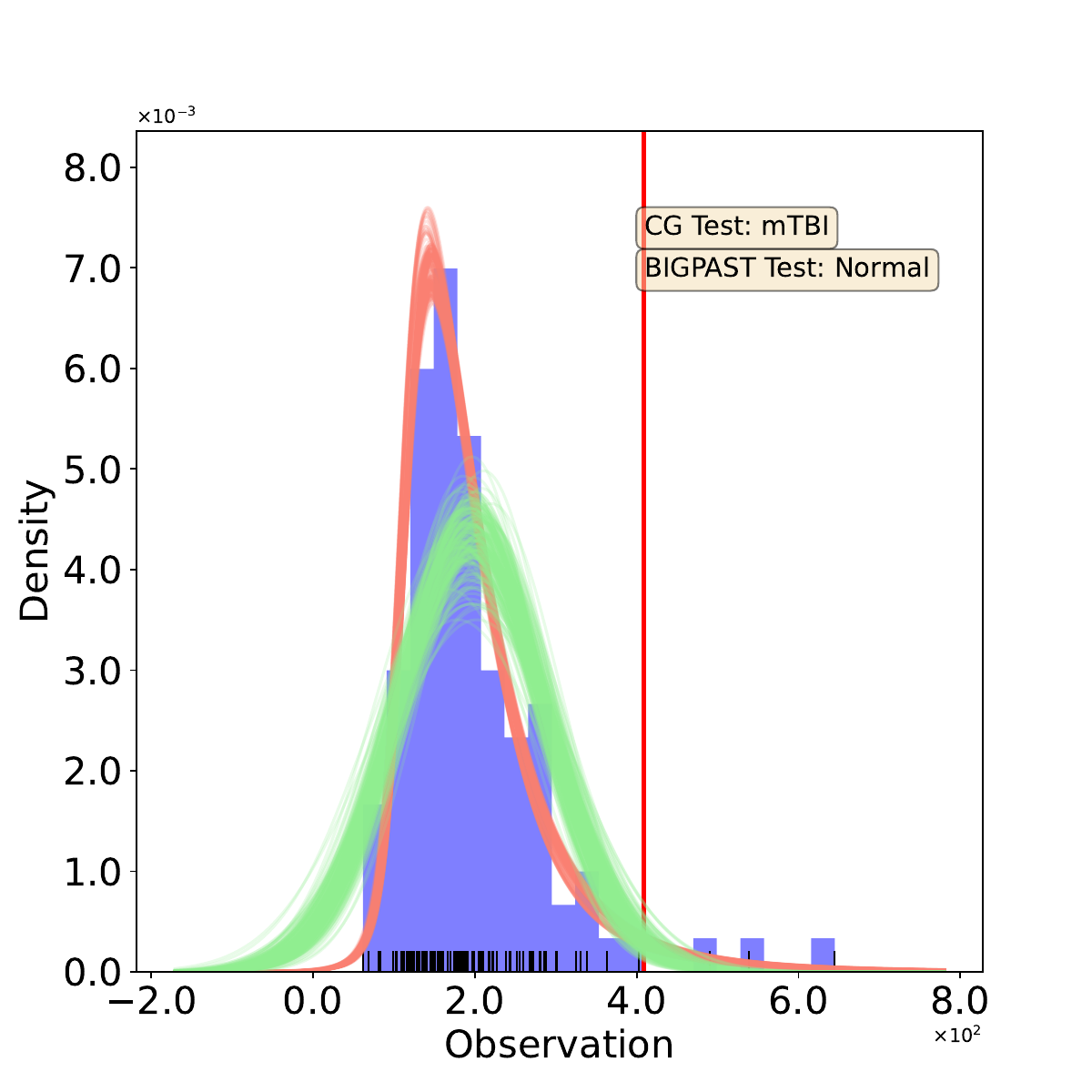}
		\caption{Gamma Band, \( \bar{\hat{\alpha}}\approx2.69 \), Right Fusiform\\ \quad}\label{fig:figs/gamma40ctx-rh-fusiform_nolog50two_sided.pdf}
	\end{subfigure}
	\caption{The comparison results between the Crawford-Garthwaite (CG) and BIGPAST approaches in terms of density estimations. The light-red lines represent the skewed Student \( t \) densities, with parameters drawn from the BIGPAST approach's posterior distribution. The light green lines depict the Normal densities, with parameters drawn from the CG approach's posterior distribution. The CG test is based on 10,000 densities, while the BIGPAST test is based on 5,000 densities (the first 5,000 densities are discarded for burn-in). For clarity, each panel only displays 200 densities drawn from the CG and BIGPAST posterior distribution. The red vertical line on the \( x \) axis represents the observation of a single-subject (i.e., a patient with mTBI). A histogram is used to demonstrate the goodness of fit for both the CG and BIGPAST approaches.}\label{fig:left_right_skew_advantage_no_log_transformation}
\end{figure*}

The null hypothesis in this real data analysis is two-sided, with one heavy tail and one light tail in the skewed Student \( t \) distribution. The cortical region indices in row 3 of~\Cref{tab:result_CG_BIGPAST_no_log_transformation} exhibit a positive skewness parameter, i.e., \( \alpha>0 \). Consequently, the light tail is on the left, and the heavy tail is on the right. For all cortical region indices in row 3 of~\Cref{tab:result_CG_BIGPAST_no_log_transformation}, the single-subject observation is located at the light tail of the skewed Student \( t \) distribution. This positioning suggests a high probability of rejecting the null hypothesis, see~\Cref{fig:figs/delta1ctx-lh-caudalanteriorcingulate_nolog50two_sided.pdf}. Similarly, for cortical regions 6 (lh-fusion, gamma band), 14 (lh-medial orbitofrontal, theta and beta bands), 40 (rh-fusiform, theta and gamma bands), and 47 (rh-lingual, beta band), the single-subject observation is located at the heavy tail of the skewed Student \( t \) distribution, as shown in~\Cref{fig:figs/gamma40ctx-rh-fusiform_nolog50two_sided.pdf}. The skewed Student \( t \) assumption is more appropriate for these cortical regions than the assumption of Normal distribution. The BIGPAST approach outperforms the CG approach in detecting the single-subject observation at the heavy tail of the skewed Student \( t \) distribution, as demonstrated in~\Cref{fig:left_right_skew_advantage_no_log_transformation}. Because the BIGPAST approach corrects the model misspecification error at both tails compared to the CG approach. By combining rows 2 and 3 of~\Cref{tab:result_CG_BIGPAST_no_log_transformation} and~\Cref{fig:left_right_skew_advantage_no_log_transformation}, we conclude that the BIGPAST approach is more reliable than the CG approach in hypothesis tests for single-subject data when the control group data exhibits skewness.

\section{Conclusion}\label{sec:discussion}
In this paper, we proposed a novel Bayesian inference framework for the abnormality detection on a single-subject versus a control group. Under the assumption of a Normal distribution, the \( t \)-score~\citep{crawfordComparingIndividualTest1998} is capable of managing scenarios where the sample size is fewer than 30~\citep{machin2018sample}. However, in neuroscience, the number of epochs often exceeds 100, sometimes even reaching 1000, and the distributions, whether over epochs or subjects, exhibit skewness. Under these conditions, methods predicated on the assumption of normality cease to be effective. Our proposed methodology, BIGPAST, leverages the skewed Student \( t \) distribution, effectively capturing the inherent asymmetry of the data.

Our proposed BIGPAST methodology outperforms the existing Crawford-Garthwaite (CG) approach in terms of accuracy, power, and Type I error. Particularly when the control group data exhibits skewness, BIGPAST provides a more reliable solution for hypothesis testing on single-subject data. Furthermore, BIGPAST demonstrates superior robustness to model misspecification errors compared to the CG approach. The performance of the BIGPAST and CG methodologies is significantly influenced by the total variation distance between the skewed Student \( t \) distribution and the Normal distribution. When this distance is large, the BIGPAST approach exhibits superior accuracy compared to the CG method. The skewed Student \( t \) distribution's asymmetry, characterised by its light and heavy tails, further emphasises the necessity of employing a two-sided test over a one-sided test.

The BIGPAST framework can be readily adapted to accommodate other distributions, such as Sub-normal, Sub-Weibull, or Extreme Value distributions. The key requirement is to identify an appropriate prior distribution for the parameters of the chosen distribution. A further extension of BIGPAST could encompass multivariate distributions, thereby enabling the application of the BIGPAST methodology to high-dimensional data in future.
\section*{Data and Code Availability}

The calculation of skewed Student \( t \) distribution is implemented in the Python package \href{https://pypi.org/project/skewt-scipy/}{skewt-scipy}. The code for simulation studies can be found in~\href{https://github.com/Jieli12/BIGPAST}{GitHub repository: BIGPAST}. The real data comes from the \href{https://cam-can.mrc-cbu.cam.ac.uk/dataset/}{Cambridge Centre for Ageing and Neuroscience (Cam-CAN) dataset}~\citep{cam-can2014}.

\section*{Competing interests}
The authors from the University of Kent, J.L., J.Z.,  and P.L. have no competing interests.\ G.G. and S.C. are employees of Innovision IP Ltd, which provides commercial reports on individuals who may have had a head injury.

\section*{Author contributions statement}

JL:\ Conceptualisation, Methodology, Programming and Software,  Investigation, and Writing—Original Draft. GG:\ Conceptualisation, Supervision, Methodology, Resources, Investigation,  Project Administration and Writing—Review \& Editing. SC:\ Methodology and Writing- Review \& Editing. PL:\ Supervision, Methodology, Project Administration and Writing—Review \& Editing. JZ:\ Conceptualisation, Supervision, Methodology,  Investigation, Project Administration and Writing—Review \& Editing.
\section*{Funding}
This research received financial support jointly from UKRI through Innovate UK and Innovision IP Ltd. (Grant No: 10053865).
\section*{Acknowledgments}
This work was supported by the High-Performance Computing Cluster at the University of Kent.

\small\bibliographystyle{chicago}
\bibliography{main.bib}
\clearpage
\newpage
\appendix
\section{Appendix}
\begin{lemma}\label{lem:hz_approximation}
	Let \( t(z|\nu)\) and \( T(z|\nu)\) be the probability density function and cumulative distribution function of the Student \( t \)-distribution, respectively, for \(\nu\geq 1 \), then \( t(z|\nu)[T(z|\nu)]^{-1/2} [1-T(z|\nu)]^{-1/2}/\pi\) can be approximated by \( t(z/\sigma_{\nu}|\nu)/\sigma_{\nu} \), where \( \sigma_{\nu}=1.5536 \) if \( \nu>2400 \). If \( 1\leq \nu\leq 2400 \), then
	\begin{equation}\label{eq:polyfit}\\
		\begin{aligned}
			\sigma_{\nu} & = 0.00000543(\log(\nu))^{7}- 0.00016303(\log(\nu))^{6} +0.00199613(\log(\nu))^{5} \\
			& \quad- 0.01285016(\log(\nu))^{4} +0.04631303(\log(\nu))^{3}- 0.08761023(\log(\nu))^{2} \\
			& \quad+  0.05036188\log(\nu)+  1.62021189
		\end{aligned}
	\end{equation}
\end{lemma}
\begin{proof}
	Firstly, when \( 1\leq \nu\leq 2400 \),  we can use \( t(z/\sigma_{\nu}|\nu)/\sigma_{\nu} \) to approximate \( t(z|\nu)[T(z|\nu)]^{-1/2} [1-T(z|\nu)]^{-1/2}/\pi \) by the similar discussion in~\citet{bayesBayesianInferenceSkewness2007}. Apparently, the optimal scale \( \sigma_{\nu} \) depends on \( \nu \). To obtain~\Cref{eq:polyfit}, for simplicity, let \( f\coloneqq t(z|\nu)[T(z|\nu)]^{-1/2} [1-T(z|\nu)]^{-1/2}/\pi \) and \( g\coloneqq t(z/\sigma_{\nu}|\nu)/\sigma_{\nu} \). Define the total variational distance between \( f \) and \( g \) as
	\begin{equation}\label{eq:TV_distance}
		TV(\sigma_{\nu})\coloneqq TV(f,g) = \frac{1}{2}\int_{-\infty}^{\infty} {\left\lvert f-g \right\rvert}  \,\mathrm{d}z
	\end{equation}
	Let \begin{equation*}
		\nu_{i}=
		\begin{cases}
			0.5+0.5i       & i=1,2,\ldots,199      \\
			100+50*(i-199) & i=200,201,\ldots,245,
		\end{cases}
	\end{equation*}
	given \( \nu_{i} \), we use grid search method to minimise \( KL(\sigma(\nu_{i})) \) and obtain the optimal scale \( \hat{\sigma}(\nu_{i}) \). Let \( \sigma_{j,\nu_{i}} =1+0.0002j\), \( j=0,1,\ldots,5000 \), then \(\hat{\sigma}_{\nu_{i}}=\argmin_{\sigma_{j,\nu_{i}}}KL(\sigma_{j,\nu_{i}}) \) where \( KL(\sigma_{j,\nu_{i}}) \) is evaluated using the Monte Carlo method. Finally, we carry out the polynomial fit with polynomial degree \( p=2,3,\ldots,7 \) based on the optimal scale \( \hat{\sigma}(\nu_{i}) \) and \( \log(\nu_{i}) \), \( i=1,2,\ldots,245 \).~\Cref{eq:polyfit} is the best polynomial fit with polynomial degree \( p=7 \) based on Bayes information criterion, see~\Cref{fig:figs/polyfit.png}. Secondly, when \( \nu>2400 \), \( t(z|\nu)[T(z|\nu)]^{-1/2} [1-T(z|\nu)]^{-1/2}/\pi\) can be approximated by \( \phi(z)[\Phi(z)]^{-1/2} [1-\Phi(z)]^{-1/2}/\pi \). According to~\citet{bayesBayesianInferenceSkewness2007}, \( \phi(z)[\Phi(z)]^{-1/2} [1-\Phi(z)]^{-1/2}/\pi \) can be further approximated by \( N(0,\sigma^{2}_{\nu}) \) where the optimal scale \( \sigma_{\nu} \) chosen by total variational distance is 1.5536. Note that~\citet{bayesBayesianInferenceSkewness2007} used the optimal scale \( \sigma_{\nu} \) chosen by the Kullback-Leibler divergence (defined in~\Cref{eq:KL_divergence}), the optimal scale \( \sigma_{\nu} \) is 1.54 for Normal distribution. However, our simulation result shows that the total variational distance criterion performs better than the Kullback-Leibler divergence criterion for Student \( t \) with a small degree of freedom, see~\Cref{fig:figs/comparison_KL_AV.eps}. Therefore, we employ the criterion of total variational distance to choose the optimal scale \( \sigma_{\nu} \) in \( t(z/\sigma_{\nu}|\nu)/\sigma_{\nu} \) to approximate \( t(z|\nu)[T(z|\nu)]^{-1/2} [1-T(z|\nu)]^{-1/2}/\pi \) throughout this paper.~\Cref{fig:Density_f_solid_curve)} shows the density \( t(z|\nu)[T(z|\nu)]^{-1/2} [1-T(z|\nu)]^{-1/2}/\pi \) and its approximation \( t(z/\sigma_{\nu}|\nu)/\sigma_{\nu} \) with \( \sigma_{\nu} \) in~\Cref{eq:polyfit} for \( \nu=1,5,10,100 \).
	\begin{figure}[ht]
		\centering
		\includegraphics[width=0.7\textwidth]{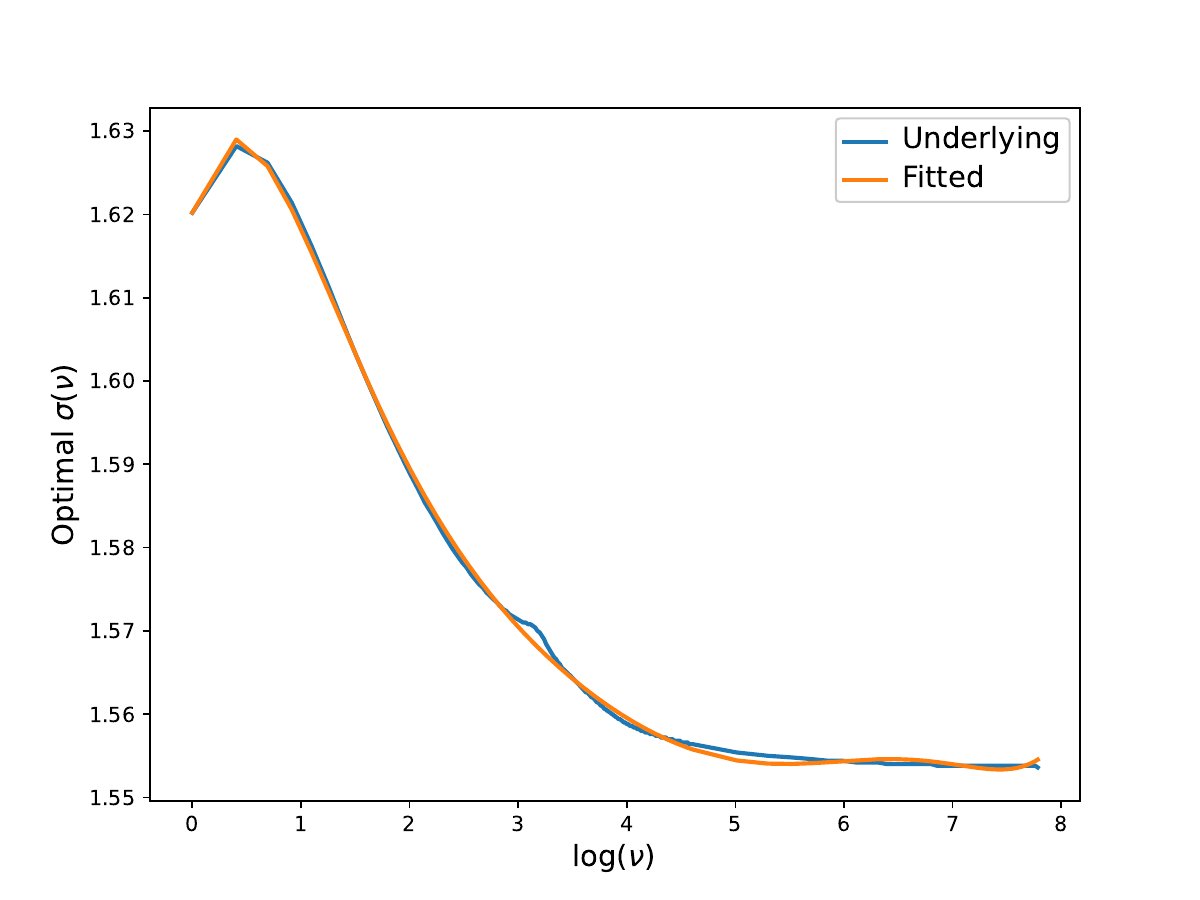}
		\caption{Polynomial fitted curve}\label{fig:figs/polyfit.png}
	\end{figure}
	\begin{figure}[ht]
		\centering
		\includegraphics[width=0.7\textwidth]{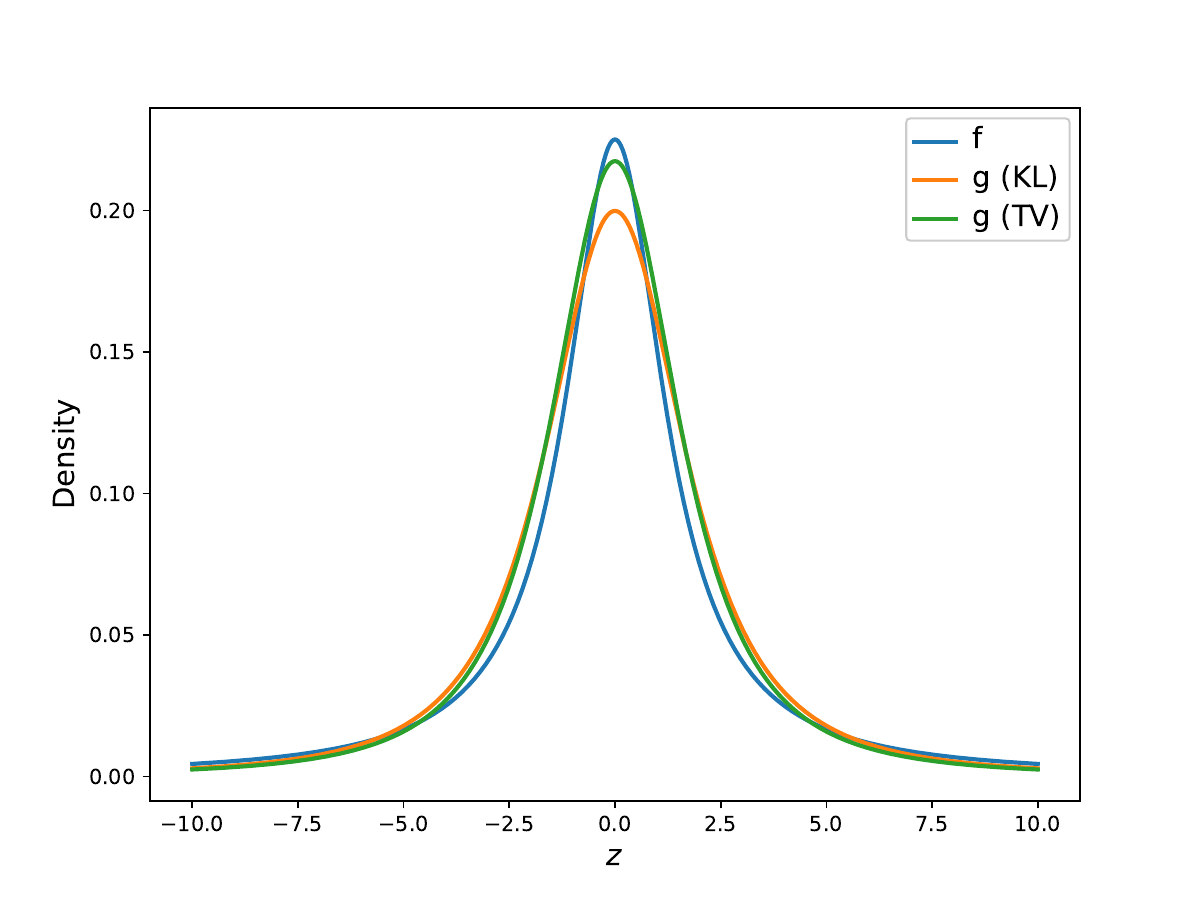}
		\caption{Comparison of approximation based on Kullback-Leibler divergence and total variational distance when the degree of freedom \( \nu=2 \). The blue line represents the density of \( f \). The green line represents the density of approximation \( g \) based on total variational distance, and the orange line is the density of approximation \( g \) based on Kullback-Leibler divergence.}\label{fig:figs/comparison_KL_AV.eps}
	\end{figure}

	\begin{equation}\label{eq:KL_divergence}
		KL(\sigma_{\nu})\coloneqq KL(f,g) = \int_{-\infty}^{\infty} f \log\left(\frac{f}{g}\right)dz+\int_{-\infty}^{\infty} g \log\left(\frac{g}{f}\right)dz.
	\end{equation}
	\begin{figure}[ht]
		\centering
		\begin{subfigure}[bt]{0.45\textwidth}
			\centering
			\includegraphics[width=\textwidth]{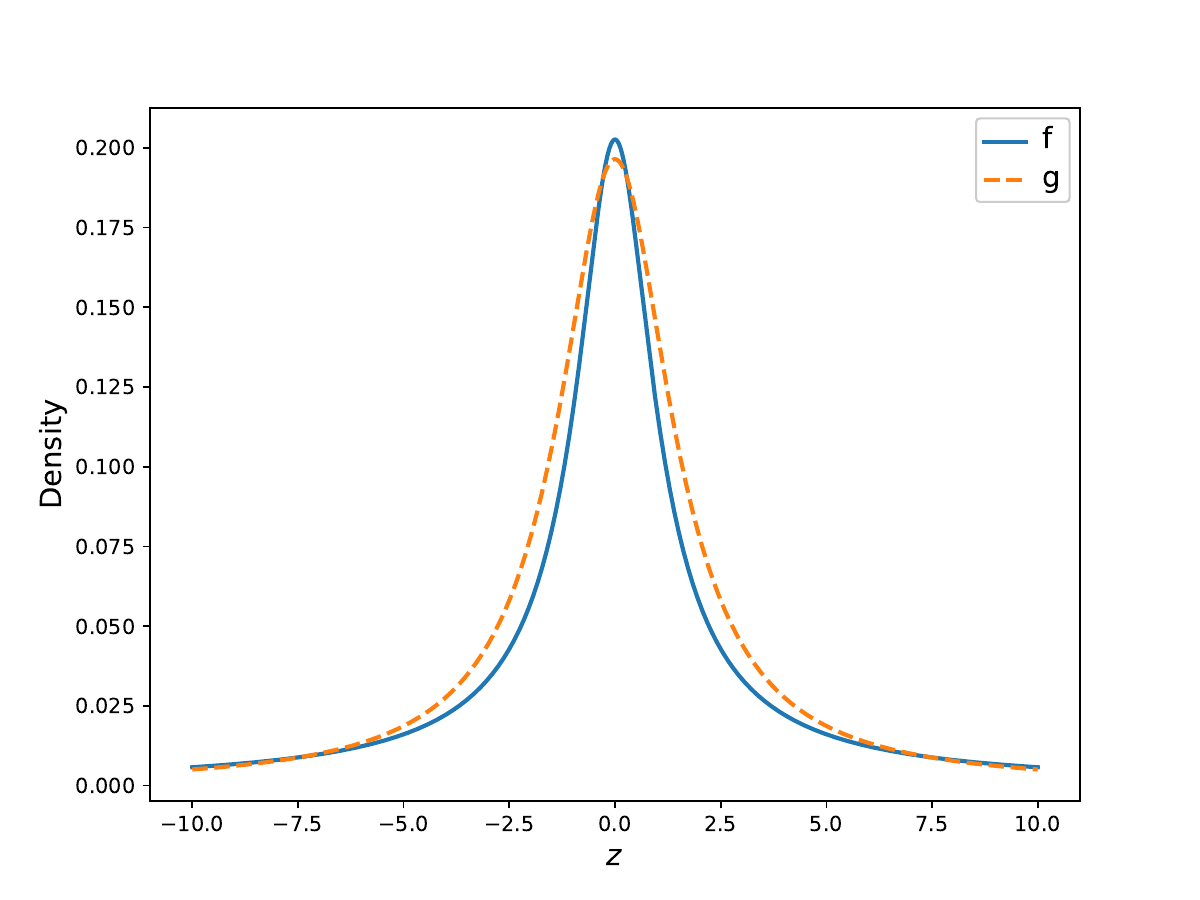}
			\caption{\( \nu=1 \)}\label{fig:figs/comparison_TV1.pdf}
		\end{subfigure}
		\begin{subfigure}[bt]{0.45\textwidth}
			\centering
			\includegraphics[width=\textwidth]{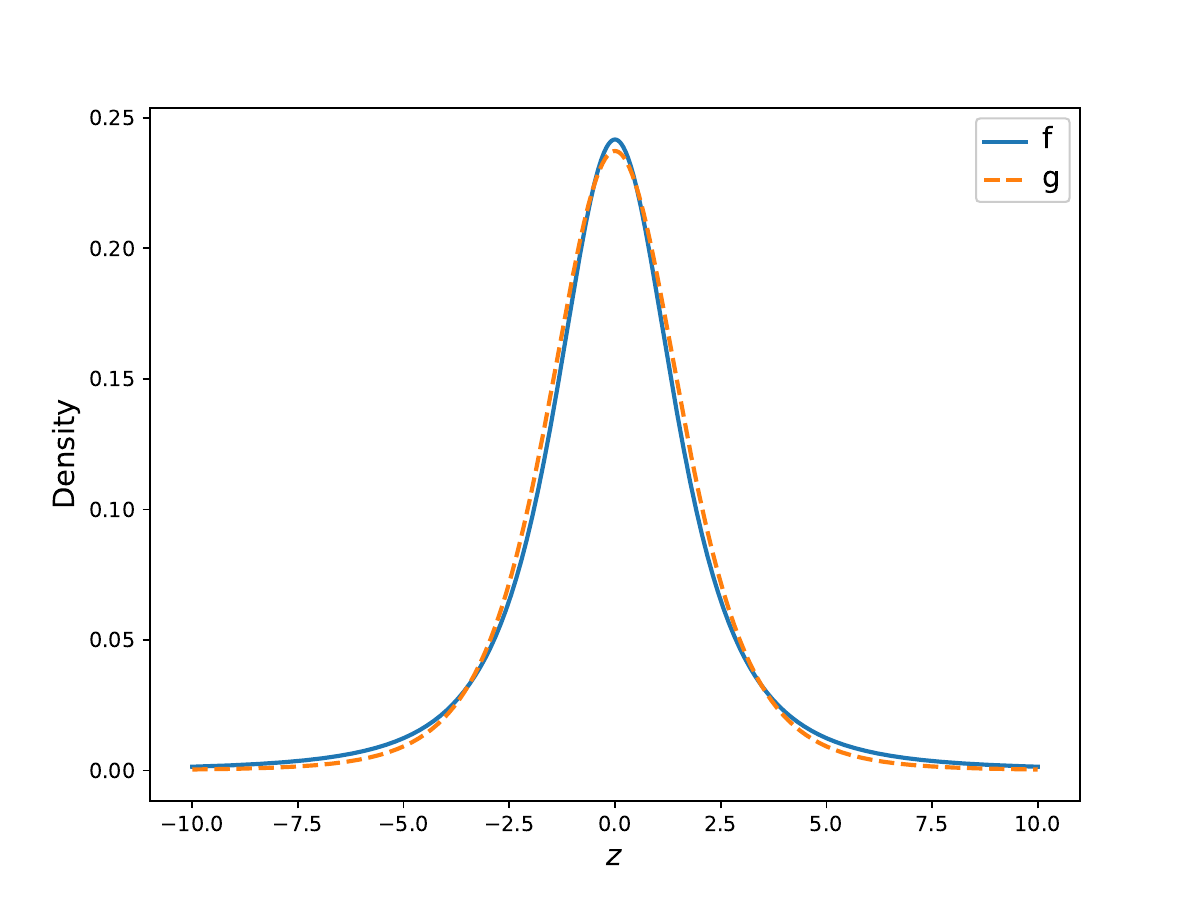}
			\caption{\( \nu=5 \)}\label{fig:figs/comparison_TV5.pdf}
		\end{subfigure}
		\begin{subfigure}[bt]{0.45\textwidth}
			\centering
			\includegraphics[width=\textwidth]{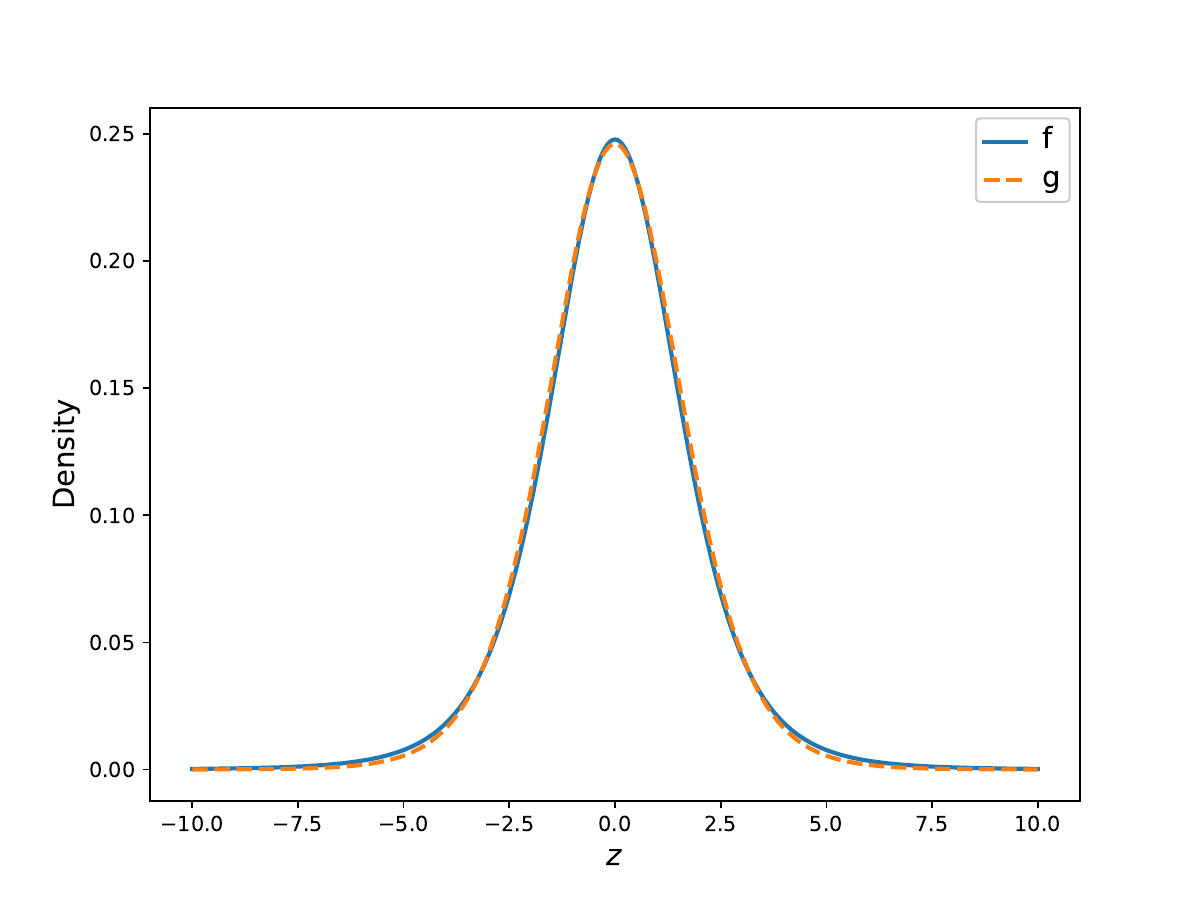}
			\caption{\( \nu=10 \)}\label{fig:figs/comparison_TV10.pdf}
		\end{subfigure}
		\begin{subfigure}[bt]{0.45\textwidth}
			\centering
			\includegraphics[width=\textwidth]{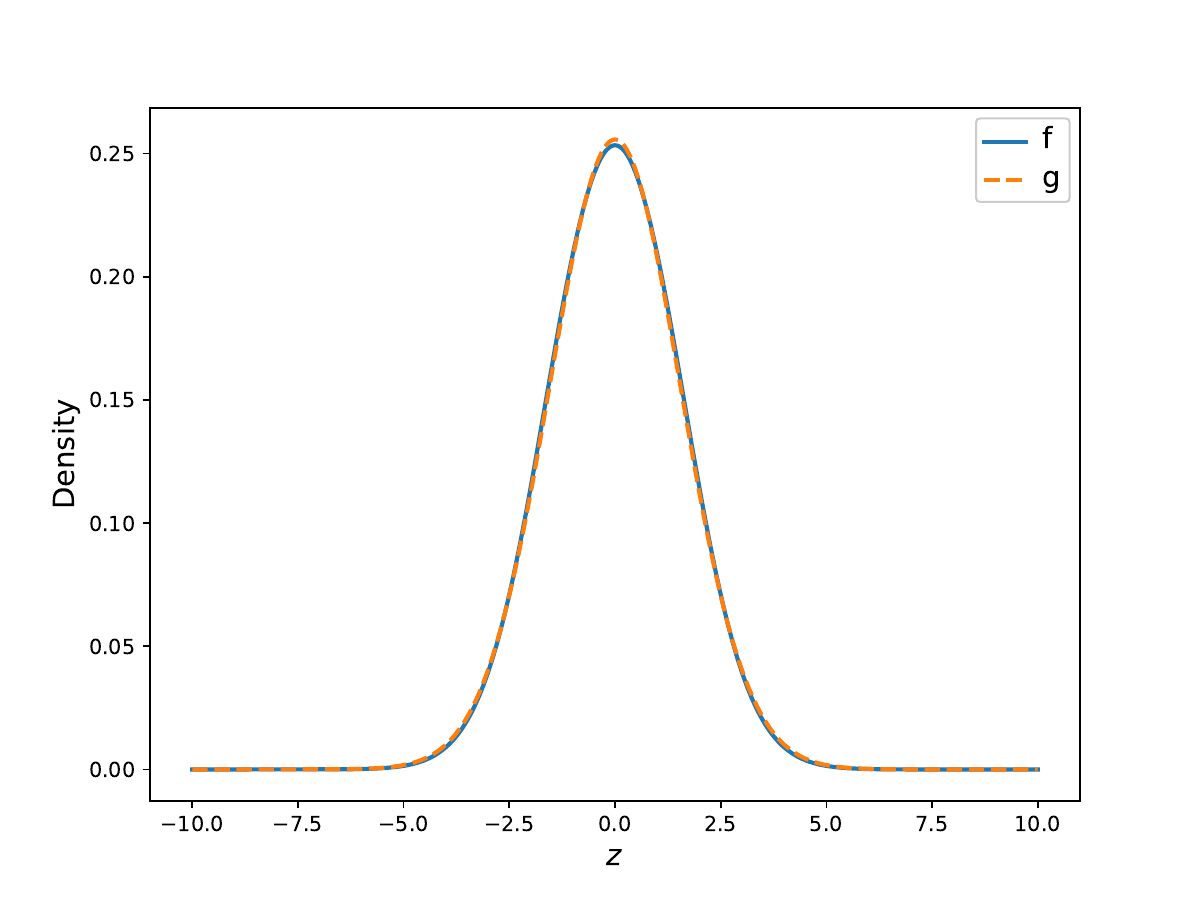}
			\caption{\( \nu=100 \)}\label{fig:figs/comparison_TV100.pdf}
		\end{subfigure}
		\caption{Density \( f \) (solid blue curve) and its approximation \( g \) with \( \sigma_{\nu} \) (dashed orange curve) for \( \nu=1,5,10,100 \).}\label{fig:Density_f_solid_curve)}
	\end{figure}
\end{proof}

\begin{lemma}\label{lem:jeffery_prior}
	Let \( f(z|\alpha,\nu) \) be the skewed Student \( t \) density function, defined as follows:
	\begin{equation*}
		f(z|\alpha,\nu) \coloneqq 2t(z|\nu)T(\alpha z r(z,\nu)|\nu+1), \quad z\in\mathbb{R}, \alpha\in\mathbb{R}, \nu>0,
	\end{equation*}
	where \( r(z,\nu)=\sqrt{(\nu+1)/(\nu+z^2)} \). Given any \( \nu>0 \), let \( \boldsymbol{\pi}(\alpha|\nu) \) represent the Jeffreys prior for \( \alpha \), we have
	\begin{equation}\label{eq:jeffery_prior}
		\boldsymbol{\pi}(\alpha|\nu) \propto \frac{\sqrt{\pi}  \Gamma \left(\frac{v}{2}+1\right) \sqrt{_2F_1\left(\frac{1}{2},v+1;\frac{v+1}{2};-\frac{\alpha^2}{\sigma^2_{\nu+1}}\right)-\, _2F_1\left(\frac{1}{2},v+2;\frac{v+1}{2};-\frac{\alpha^2}{\sigma ^2_{\nu+1}}\right)} }{{\left\lvert \alpha \right\rvert} \Gamma \left(\frac{v+1}{2}\right)},
	\end{equation}
	where \( _2F_1\left(a,b;c;d\right) \) represents the hypergeometric function, \( \sigma_{\nu+1} \) is defined in~\Cref{lem:hz_approximation}, \( \Gamma(\cdot) \) is the Gamma function.
\end{lemma}
\begin{proof}
	By definition of Jeffreys prior, we have \( \boldsymbol{\pi}(\alpha|\nu)\propto \sqrt{I_{\alpha\alpha}} \), where \( I_{\alpha\alpha} \) is the Fisher information of \( \alpha \) in the skewed Student \( t \) density function \( f(z|\alpha,\nu) \). According to~\citet[Proposition 2]{brancoObjectiveBayesianAnalysis2013}, the Fisher information of \( \alpha \) in \( f(z|\alpha,\nu) \) is given by
	\begin{equation}\label{eq:fisher_information}
		I_{\alpha\alpha} = 2\int_{0}^{+\infty} \frac{z^2 r(z,\nu)^{2} t(z|\nu)t^{2}(\alpha z r(z,\nu)|\nu+1)}{T(\alpha z r(z,\nu)|\nu+1)\left[ 1-T(\alpha z r(z,\nu)|\nu+1) \right]}\,\mathrm{d}z.
	\end{equation}
	Let \( m= \alpha z r(z,\nu)\), by the change of variable \( z=\frac{\sqrt{m^2 \nu}}{\sqrt{\alpha^2 \nu+\alpha^2-m^2}} \),~\Cref{eq:fisher_information} can be rewritten as
	\begin{equation}\label{eq:fisher_information_change_variable}
		I_{\alpha\alpha} = 2\int_{0}^{\alpha\sqrt{\nu+1}} \frac{m^2 t\left(\frac{\sqrt{m^2 \nu}}{\sqrt{\alpha^2 \nu+\alpha^2-m^2}}\Big|\nu\right)t^{2}(m|\nu+1)}{T(m|\nu+1)\left[ 1-T(m|\nu+1) \right]}\frac{\sqrt{\nu} (\nu+1)}{ \left(\alpha^2 (\nu+1)-m^2\right)^{3/2}}\,\mathrm{d}m.
	\end{equation}
	By \textcolor{darkblue}{Lemma}~\ref{lem:hz_approximation}, we have \( t(m|\nu+1)[T(m|\nu+1)]^{-1/2} [1-T(m|\nu+1)]^{-1/2}/\pi \approx t(m/\sigma_{\nu+1}|\nu+1)/\sigma_{\nu+1} \). Substituting this approximation into~\Cref{eq:fisher_information_change_variable}, we obtain
	\begin{equation}\label{eq:fisher_information_by_m}
		I_{\alpha\alpha} \approx 2\int_{0}^{\alpha\sqrt{\nu+1}} \frac{m^2\pi^{2} \left(\frac{1}{\frac{m^2}{\alpha^2 \nu+\alpha^2-m^2}+1}\right)^{\frac{\nu+1}{2}} \left(\frac{\nu+1}{\frac{m^2}{\sigma ^2_{\nu+1}}+\nu+1}\right)^{\nu+2}}{\left(\alpha^2 (\nu+1)-m^2\right)^{3/2}\sigma ^2_{\nu+1} B\left(\frac{\nu}{2},\frac{1}{2}\right) B\left(\frac{\nu+1}{2},\frac{1}{2}\right)^2}\,\mathrm{d}m,
	\end{equation}
	where \( B(\cdot,\cdot) \) is the Beta function.
	Furthermore, let \( x=m^2 \), by change of variable,~\Cref{eq:fisher_information_by_m} can be rewritten as
	\begin{equation}\label{eq:fisher_information_by_x}
		I_{\alpha\alpha} \approx 2\int_{0}^{\alpha^{2}(\nu+1)}\frac{\pi ^2 \sqrt{x} \left(\frac{1}{\frac{x}{\alpha^2 (\nu+1)-x}+1}\right)^{\frac{\nu+1}{2}} \left(\frac{\nu+1}{\nu+\frac{x}{\sigma ^2_{\nu+1}}+1}\right)^{\nu+2}}{\sigma ^2_{\nu+1} B\left(\frac{\nu}{2},\frac{1}{2}\right) B\left(\frac{\nu+1}{2},\frac{1}{2}\right)^2 \left(\alpha^2 (v+1)-x\right)^{3/2}} \,\mathrm{d}x.
	\end{equation}
	By integration, we obtain the following expression:
	\begin{equation*}
		I_{\alpha\alpha} \approx \frac{\pi  \Gamma \left(\frac{\nu}{2}+1\right)^2 \left(\, _2F_1\left(\frac{1}{2},\nu+1;\frac{\nu+1}{2};-\frac{\alpha^2}{\sigma ^2_{\nu+1}}\right)-\, _2F_1\left(\frac{1}{2},\nu+2;\frac{\nu+1}{2};-\frac{\alpha^2}{\sigma ^2_{\nu+1}}\right)\right)}{\alpha^2 \Gamma \left(\frac{\nu+1}{2}\right)^2},
	\end{equation*}
	which completes the proof.
\end{proof}
\paragraph*{The Proof of~\Cref{thm:jeffery_prior_nu_alpha}}
\begin{proof}
	The log-likelihood function of \( f(z|\alpha,\nu) \), say \( \ell(\alpha,\nu) \), can be expressed as
	\begin{equation}\label{eq:log-likelihood_skew_t}
		\ell(\alpha,\nu) = \log\left(\frac{2}{\sqrt{\pi}}\right) + \frac{\nu +1}{2}\log\left( \frac{\nu }{\nu +z^2} \right) -\frac{1}{2}\log(\nu) +\log\left( \Gamma\left( \frac{\nu +1}{2} \right) \right)- \log\left( \Gamma\left( \frac{\nu}{2} \right) \right)+ \log\left( T(w|\nu+1) \right).
	\end{equation}
	The derivative of \( \ell(\alpha,\nu) \) with respect to \( \nu \) is given by
	\begin{equation}\label{eq:log-likelihood_skew_t_Derivative_v}
		\frac{\partial  \ell(\alpha,\nu)}{\partial \nu} = \frac{1}{2}\log \left(\frac{\nu }{\nu +z^2}\right)-\frac{1}{2}\psi\left(\frac{\nu }{2}\right) +\frac{1}{2}\psi\left(\frac{\nu +1}{2}\right)+  \frac{z^2-1}{2 \left(\nu +z^2\right)} + \frac{1}{T(w|\nu+1)}\frac{\partial T(w|\nu+1) }{\partial \nu},
	\end{equation}
	where \( \psi(x)\coloneqq \mathrm{d}\log\Gamma(x)/\mathrm{d} x \) is the digamma function. The derivative of \( T(w|\nu+1) \) with respect to the parameter \( \nu \) can be expressed by
	\begin{equation}\label{eq:Derivative_T_w_nu}\\
		\begin{aligned}
			\frac{\partial T(w|\nu+1) }{\partial \nu} & = t(w|\nu+1)\frac{\partial w }{\partial \nu}+\int_{-\infty}^{w} \frac{\partial t(x|\nu+1) }{\partial \nu}\,\mathrm{d}x \\
			& = t(w|\nu+1) \frac{\alpha z \left(z^2-1\right)}{2 \sqrt{\nu +1} \left(\nu +z^2\right)^{3/2}}+\frac{1}{2}\left[-\psi\left(\frac{\nu +1}{2}\right) +\psi\left(\frac{\nu }{2}+1\right) \right]T(w|\nu+1) \\
			& \quad+\int_{-\infty}^{w} \frac{1}{2}\left[ \log \left(\frac{\nu +1}{\nu +x^2+1}\right)+\frac{x^2-1}{\left(\nu +x^2+1\right)} \right]t(x|\nu+1)\,\mathrm{d}x \\
			& = t(w|\nu+1) \frac{\alpha z \left(z^2-1\right)}{2 \sqrt{\nu +1} \left(\nu +z^2\right)^{3/2}}+\frac{1}{2}\left[ -\psi\left(\frac{\nu +1}{2}\right) +\psi\left(\frac{\nu }{2}+1\right) \right]T(w|\nu+1) \\
			& \quad+g(\nu,w)-\frac{w}{2(\nu+1)}t(w|\nu+1) \\
			& = -t(w|\nu+1) \frac{\alpha z\sqrt{\nu +1}  }{2 \left(\nu +z^2\right)^{3/2}}+\frac{1}{2}\left[ -\psi\left(\frac{\nu +1}{2}\right) +\psi\left(\frac{\nu }{2}+1\right) \right]T(w|\nu+1) \\
			& \quad+g(\nu,w)
		\end{aligned}
	\end{equation}
	where \( g(\nu,w)\coloneqq \int_{-\infty}^{w} \frac{1}{2}\log \left(\frac{\nu +1}{\nu +x^2+1}\right)t(x|\nu+1)\,\mathrm{d}x \). Substituting~\Cref{eq:Derivative_T_w_nu} into~\Cref{eq:log-likelihood_skew_t_Derivative_v}, we obtain
	\begin{equation}\label{eq:log-likelihood_skew_t_Derivative_v_inter}\\
		\begin{aligned}
			\frac{\partial  \ell(\alpha,\nu)}{\partial \nu} & = \frac{1}{2}\log \left(\frac{\nu }{\nu +z^2}\right)-\frac{1}{2}\psi\left(\frac{\nu }{2}\right) +\frac{1}{2}\psi\left(\frac{\nu +2}{2}\right)+  \frac{z^2-1}{2 \left(\nu +z^2\right)} \\
			& \quad -h(w)\frac{\alpha z\sqrt{\nu +1}  }{2 \left(\nu +z^2\right)^{3/2}}+\frac{g(\nu,w)}{T(w|\nu+1)},
		\end{aligned}
	\end{equation}
	where \( h(w)\coloneqq t(w|\nu+1)/T(w|\nu+1) \). Next, we focus on the calculation of \( g(\nu,w) \). Note that \( \log\frac{\nu+1}{\nu+x^2+1} \) and \( t(x|\nu+1) \) are symmetric with respect to \( x \), therefore for \( w>0 \), we only consider
	\begin{equation}\label{eq:g_nu_calculation_part}\\
		\begin{aligned}
			g_{0}(\nu,w) & = \int_{0}^{w} \frac{1}{2}\log \left(\frac{\nu +1}{\nu +x^2+1}\right)t(x|\nu+1)\,\mathrm{d}x \\
			& = \frac{1}{2}\int_{0}^{w} \log \left(\frac{\nu +1}{\nu +x^2+1}\right)\frac{\left(\frac{\nu +1}{\nu +x^2+1}\right)^{\frac{\nu +2}{2}}}{\sqrt{\nu +1} B\left(\frac{\nu +1}{2},\frac{1}{2}\right)}\,\mathrm{d}x \\
			& = \frac{1}{4}\int_{0}^{w^{2}} \frac{1}{\sqrt{u}}\log \left(\frac{\nu +1}{\nu +u+1}\right)\frac{\left(\frac{\nu +1}{\nu +u+1}\right)^{\frac{\nu +2}{2}}}{\sqrt{\nu +1} B\left(\frac{\nu +1}{2},\frac{1}{2}\right)}\,\mathrm{d}u \\
			& = \frac{1}{4B\left(\frac{\nu +1}{2},\frac{1}{2}\right)}\int_{\frac{\nu+1}{\nu+w^{2}+1}}^{1} \frac{y^{\frac{\nu-1}{2}}}{\sqrt{1-y}}\log \left(y\right)\,\mathrm{d}y \\
			& = +\frac{\sqrt{1+w^2+\nu}}{(1+\nu)^{2}}t(w|\nu+1)\, _3F_2\left(\frac{1}{2},\frac{\nu }{2}+\frac{1}{2},\frac{\nu }{2}+\frac{1}{2};\frac{\nu }{2}+\frac{3}{2},\frac{\nu }{2}+\frac{3}{2};\frac{\nu +1}{\nu +w^2+1}\right) \\
			& \quad -\frac{\sqrt{1+w^2+\nu}}{2(1+\nu)}\log \left(\frac{\nu +1}{\nu +w^2+1}\right) t(w|\nu+1)\, _2F_1\left(\frac{1}{2},\frac{\nu +1}{2};\frac{\nu +3}{2};\frac{\nu +1}{\nu +w^2+1}\right) \\
			& \quad -\frac{1}{4}\left[ \psi \left(\frac{\nu }{2}+1\right)-\psi \left(\frac{\nu +1}{2}\right) \right].
		\end{aligned}
	\end{equation}
	Similarly, we have
	\begin{equation}\label{eq:g_nu_calculation_all}
		g_{1}(\nu) = \int_{0}^{+\infty} \frac{1}{2}\log \left(\frac{\nu +1}{\nu +x^2+1}\right)t(x|\nu+1)\,\mathrm{d}x=-\frac{1}{4}\left[ \psi \left(\frac{\nu }{2}+1\right)-\psi \left(\frac{\nu +1}{2}\right) \right].
	\end{equation}
	Finally, we obtain \( g(\nu,w) \) as follows:
	\begin{equation}\label{eq:g_nu_calculation_final}
		g(\nu,w)=
		\begin{cases}
			g_{0}(\nu,w)+g_{1}(\nu) & \text{if }w\geq 0  \\
			g_{1}(\nu)-g_{0}(\nu,w) & \text{otherwise }.
		\end{cases}
	\end{equation}
	By take the partial derivative of~\Cref{eq:log-likelihood_skew_t_Derivative_v_inter} with respect to \( \alpha \), we have
	\begin{equation}\label{eq:g_nu_calculation_final_Derivative_alpha}
		\frac{\partial^{2} \ell(\alpha,\nu)}{\partial \nu\partial \alpha}=-\frac{\alpha z^{2}(\nu+1)}{2(\nu+z^{2})^{2}}h^{\prime}(w)-\frac{z  \sqrt{\nu +1}}{2 \left(\nu +z^2\right)^{3/2}}h(w)+zh(w)\sqrt{\frac{\nu +1}{\nu +z^2}}\left[ \frac{1}{2}\log \left(\frac{\nu +1}{\nu +w^2+1}\right)-\frac{g(\nu,w)}{T(w|\nu+1)}\right],
	\end{equation}
	where
	\begin{equation*}
		h^{\prime}(w)=-\frac{w(2+\nu)}{1+w^2+\nu}h(w)-h^{2}(w).
	\end{equation*}
	Because~\Cref{eq:g_nu_calculation_final_Derivative_alpha} includes the terms with \( z^{1} \) which is odd, therefore the Fisher Information of~\Cref{eq:g_nu_calculation_final_Derivative_alpha} can be expressed as
	\begin{equation}\label{eq:Fisher_Information_nu_alpha}
		I_{\alpha\nu} = -\mathbb{E}\left[ h^{2}(w)\frac{\alpha z^{2}(\nu+1)}{2(\nu+z^{2})^{2}} \right]+\mathbb{E}\left[ zh(w)\sqrt{\frac{\nu +1}{\nu +z^2}} \frac{g(\nu,w)}{T(w|\nu+1)}\right]
	\end{equation}
	By numeric results, we found that
	\begin{equation*}
		\frac{g(\nu,w)}{T(w|\nu+1)}=2c_{\nu}*g_{1}(\nu),
	\end{equation*}
	where \( c_{\nu} \) is a constant only depending on \( \nu \), see~\Cref{fig:c_alpha_for_different_nu}.
	\begin{figure}[ht]
		\centering
		\begin{subfigure}[bt]{0.45\textwidth}
			\centering
			\includegraphics[width=\textwidth]{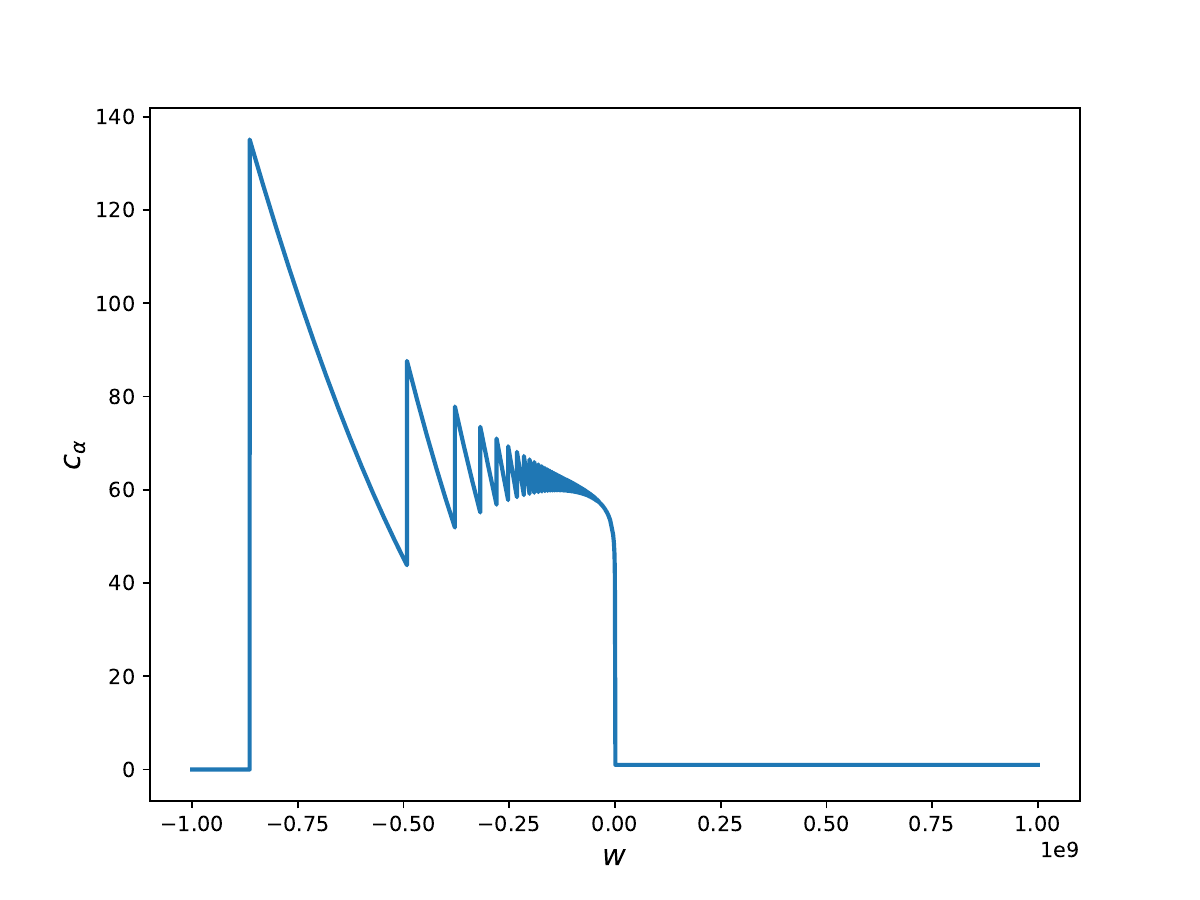}
			\caption{\( \nu=1 \)}\label{fig:figs/c_alpha_nu1.pdf}
		\end{subfigure}
		\begin{subfigure}[bt]{0.45\textwidth}
			\centering
			\includegraphics[width=\textwidth]{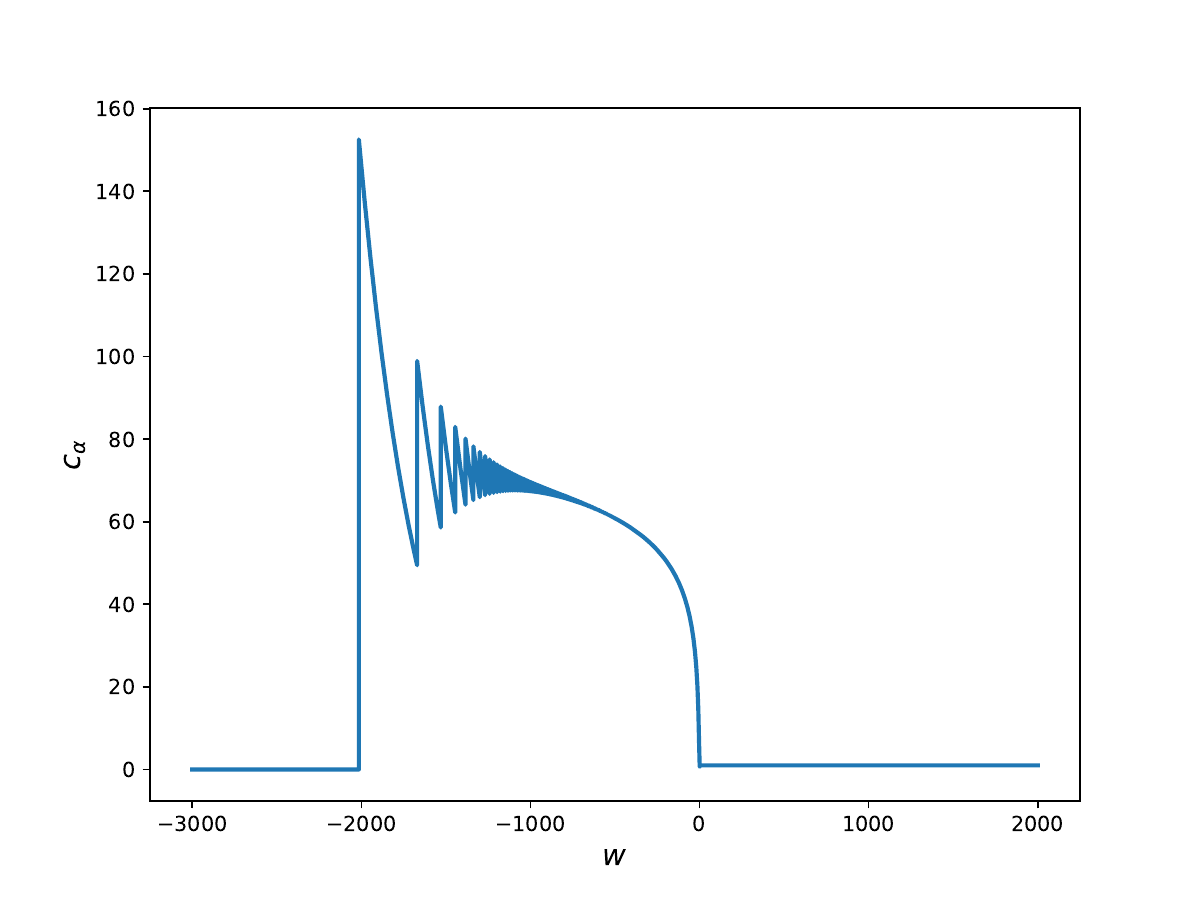}
			\caption{\( \nu=5 \)}\label{fig:figs/c_alpha_nu5.pdf}
		\end{subfigure}
		\begin{subfigure}[bt]{0.45\textwidth}
			\centering
			\includegraphics[width=\textwidth]{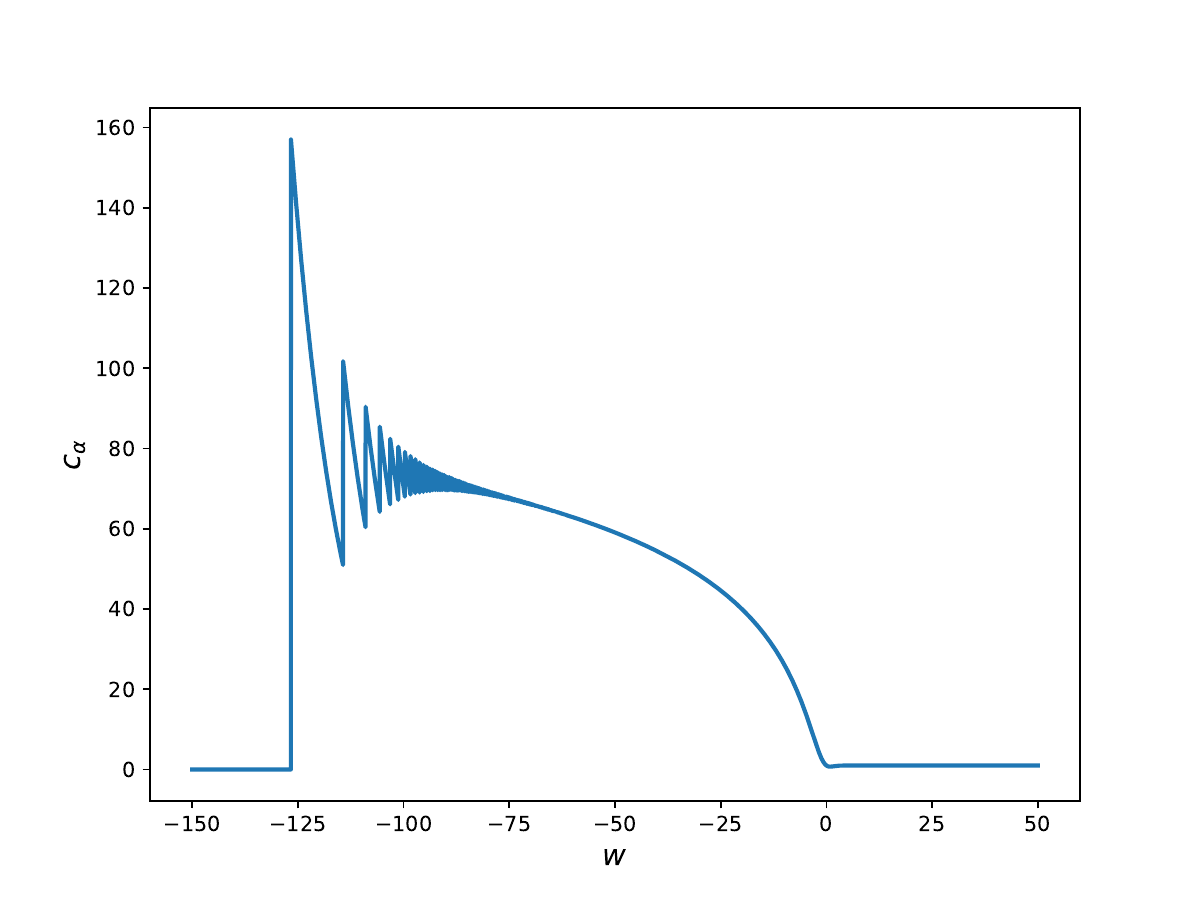}
			\caption{\( \nu=10 \)}\label{fig:figs/c_alpha_nu10.pdf}
		\end{subfigure}
		\begin{subfigure}[bt]{0.45\textwidth}
			\centering
			\includegraphics[width=\textwidth]{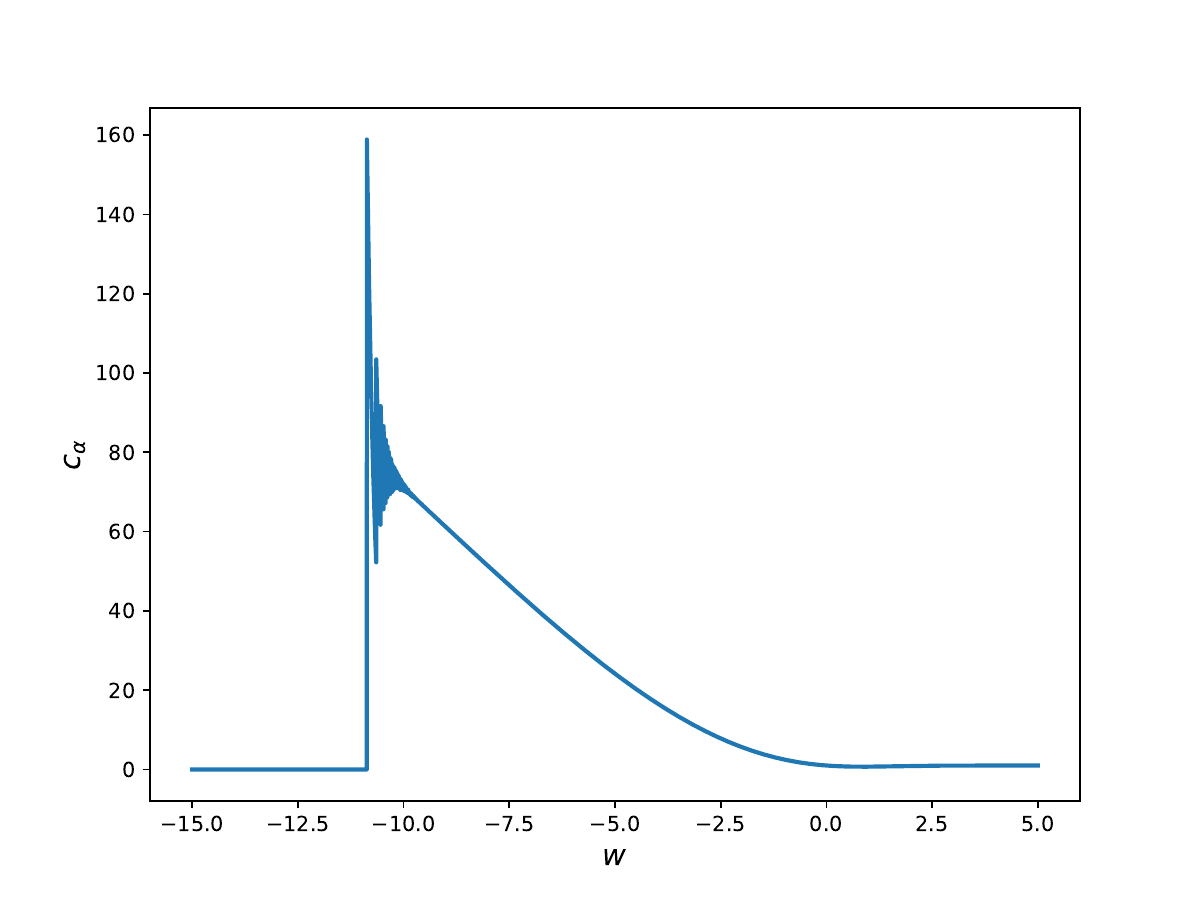}
			\caption{\( \nu=100 \)}\label{fig:figs/c_alpha_nu100.pdf}
		\end{subfigure}
		\caption{The \(c_{\nu}\) for \( \nu=1,5,10,100 \). When \( w\rightarrow -\infty \), \( c_{\nu}\rightarrow 0 \); When \( w\rightarrow \infty \), \( c_{\nu}\rightarrow 1 \). Specially, if \( w=0 \), \( c_{\nu}=1/2 \). In the above figures, \( c_{\nu}<160 \).}\label{fig:c_alpha_for_different_nu}
	\end{figure}
	Therefore,~\Cref{eq:Fisher_Information_nu_alpha} can be written as
	\begin{equation*}
		I_{\alpha\nu} = -\mathbb{E}\left[ h^{2}(w)\frac{\alpha z^{2}(\nu+1)}{2(\nu+z^{2})^{2}} \right],
	\end{equation*}
	and~\Cref{eq:log-likelihood_skew_t_Derivative_v_inter} can be simplified as
	\begin{equation*}
		\frac{\partial  \ell(\alpha,\nu)}{\partial \nu} = \frac{1}{2}\log \left(\frac{\nu }{\nu +z^2}\right)+  \frac{z^2-1}{2 \left(\nu +z^2\right)} -h(w)\frac{\alpha z\sqrt{\nu +1}  }{2 \left(\nu +z^2\right)^{3/2}}+d_{\nu},
	\end{equation*}
	where \( d_{\nu}\coloneqq-\frac{1}{2}\psi\left(\frac{\nu }{2}\right) +\frac{1}{2}\psi\left(\frac{\nu +2}{2}\right)+2c_{\nu}*g_{1}(\nu) \). Hence, the Fisher information with respect to \( \nu \) can be expressed as
	\begin{equation}\label{eq:fisher_info_nu}\\
		\begin{aligned}
			I_{\nu\nu} & = \mathbb{E}\left[ \left( \frac{\partial  \ell(\alpha,\nu)}{\partial \nu} \right)^{2} \right] \\
			& = \mathbb{E}\left[ h^{2}(w)\frac{\alpha^{2} z^{2}(\nu +1)  }{4 \left(\nu +z^2\right)^{3}} \right]+\frac{1}{4}\mathbb{E}\left[ \left( \log \frac{\nu }{\nu +z^2} \right)^{2} \right] +\frac{1}{4}\mathbb{E}\left[ \left( \frac{z^2-1}{\nu +z^2} \right)^{2} \right] +d^{2}_{\nu} \\
			& \qquad+\frac{1}{2}\mathbb{E}\left[ \left( \log \frac{\nu }{\nu +z^2} \right) \left( \frac{z^2-1}{\nu +z^2} \right)\right]+d_{\nu}\mathbb{E}\left[  \log \frac{\nu }{\nu +z^2}  \right]+d_{\nu}\mathbb{E}\left[  \frac{z^2-1}{\nu +z^2} \right].
		\end{aligned}
	\end{equation}
	After some integrations, we obtain
	\begin{equation*}
		\begin{aligned}
			\mathbb{E}\left[  \frac{z^2-1}{\nu +z^2} \right] & = 0 \\
			\mathbb{E}\left[  \log \frac{\nu }{\nu +z^2}  \right] & =  \psi \left(\frac{\nu }{2}\right)-\psi \left(\frac{\nu +1}{2}\right) \\
			\mathbb{E}\left[ \left( \log \frac{\nu }{\nu +z^2} \right) \left( \frac{z^2-1}{\nu +z^2} \right)\right] & = -\frac{1}{\nu ^2+\nu } \\
			\mathbb{E}\left[ \left( \frac{z^2-1}{\nu +z^2} \right)^{2} \right] & = \frac{1}{\nu ^2+3 \nu } \\
			\mathbb{E}\left[ \left( \log \frac{\nu }{\nu +z^2} \right)^{2} \right] & =  \left(\psi \left(\frac{\nu }{2}\right)-\psi \left(\frac{\nu +1}{2}\right)\right)^2+\psi ^{(1)}\left(\frac{\nu }{2}\right)-\psi ^{(1)}\left(\frac{\nu +1}{2}\right) \\
			\mathbb{E}\left[ h^{2}(w)\frac{\alpha^{2} z^{2}(\nu +1)  }{4 \left(\nu +z^2\right)^{3}} \right] & = \frac{\pi ^{5/2} \Gamma \left(\frac{\nu }{2}+1\right) \left((\nu +3)  H_{1}-(\nu +3) H_{2}-H_{3}+H_{4}\right)}{8 \nu ^2 \Gamma \left(\frac{\nu +5}{2}\right) B\left(\frac{\nu }{2},\frac{1}{2}\right) B\left(\frac{\nu +1}{2},\frac{1}{2}\right)^2} \\
			\mathbb{E}\left[ h^{2}(w)\frac{\alpha z^{2}(\nu+1)}{2(\nu+z^{2})^{2}} \right] & = \frac{\pi  \alpha  \Gamma \left(\frac{\nu }{2}+1\right)^2 \left((\nu +4)H_{5} -\frac{\left(\alpha ^2 (2 \nu +3)+(\nu +3) \sigma_{\nu+1} ^2\right) H_{6}}{\sigma_{\nu+1} ^2}\right)}{8 \left(\alpha ^2+\sigma_{\nu+1} ^2\right)\Gamma \left(\frac{\nu +1}{2}\right) \Gamma \left(\frac{\nu +5}{2}\right)},
		\end{aligned}
	\end{equation*}
	where
	\begin{align*}
		H_{1} & =\, _2F_1\left(\frac{1}{2},\nu +1;\frac{\nu +3}{2};-\frac{\alpha ^2}{\sigma_{\nu+1} ^2}\right) \\
		H_{2} & =  \, _2F_1\left(\frac{1}{2},\nu +2;\frac{\nu +3}{2};-\frac{\alpha ^2}{\sigma_{\nu+1} ^2}\right) \\
		H_{3} & =\, _2F_1\left(\frac{3}{2},\nu +1;\frac{\nu +5}{2};-\frac{\alpha ^2}{\sigma_{\nu+1} ^2}\right) \\
		H_{4} & =  \, _2F_1\left(\frac{3}{2},\nu +2;\frac{\nu +5}{2};-\frac{\alpha ^2}{\sigma_{\nu+1} ^2}\right) \\
		H_{5} & =  \, _2F_1\left(-\frac{1}{2},\nu +2;\frac{\nu +5}{2};-\frac{\alpha ^2}{\sigma_{\nu+1} ^2}\right) \\
		H_{6} & =  \, _2F_1\left(\frac{1}{2},\nu +2;\frac{\nu +5}{2};-\frac{\alpha ^2}{\sigma_{\nu+1} ^2}\right).
	\end{align*}
	Furthermore, the detailed proof of \(I_{\alpha\alpha}\) is in \textcolor{darkblue}{Lemma}~\ref{lem:jeffery_prior} which completes the proof.
\end{proof}
\subsection*{Goodness-of-fit for Skewed Student \textit{t} Distribution}\label{subsec:Goodness of fit for skew distribution}

For simplicity, let \( x_{1},x_{2},\ldots,x_{n} \) be the observations of \( n \) control subjects. We denote \( F(x|\alpha,\nu,\xi,\omega) \) as the cumulative function of this distribution where The parameters \( \alpha,\nu,\xi,\omega \) are unknown.  The definition of skewed Student \( t \) distribution is defined in~\Cref{sec:jeffery_priors}. To test the null hypothesis \( H_{0}: x_{1},x_{2},\ldots,x_{n} \) are random samples from the skewed Student \( t \) distribution, we propose the following testing process:

\begin{enumerate}[label=(\alph*)]
	\item Estimate the parameters \( \hat{\alpha},\hat{\nu},\hat{\xi},\hat{\omega} \) by maximising the likelihood function.
	\item Compute the empirical distribution function \( v_{i}= F(x|\hat{\alpha},\hat{\nu},\hat{\xi},\hat{\omega}) \) for \( i=1,2,\ldots,n \).
	\item Compute \( y_{i}=\Phi^{-1}(v_{i}) \), where \( \Phi \) is the standard Normal cumulative distribution function.
	\item Let \(u_i=\Phi((y_i-\bar{y})/s_y)\), where \(\bar{y}=n^{-1}\sum_{i=1}^n y_i\) and \(s_y^2=(n-1)^{-1}\sum_{i=1}^n(y_i-\bar{y})^2\). Let \(u_{(1)},\ldots,u_{(n)}\) be the order statistics of \(u_{1},\ldots,u_{n}\).
	\item Calculate the Anderson statistic \( A^{2} \) based on \(u_{(1)},\ldots,u_{(n)}\), i.e.,
	\begin{equation*}
		A^2:=-n-n^{-1}\sum_i^n[(2i-1)\log(u_{(i)})+(2n+1-2i)\log(1-u_{i})].
	\end{equation*}
	Modify \( A^{2} \) into \(A^{*}:=A^2(1+0.75/n+2.25/n^2)\). If \(A^{*}\) exceeds the critical value at a given significance level, reject \( H_{0} \).
\end{enumerate}
More details on the testing procedures and critical points can be found in~\citet{chenGeneralPurposeApproximate1995}. If we accept the null hypothesis \( H_{0} \) at some significance level, we can proceed with comparing a single-subject and the control group described in the next section. If \( H_{0} \) is rejected, one can employ a non-parametric method to compare a single-subject with the control group. However, the discussion of the non-parametric method is beyond the scope of this paper. We plan to extend this approach in future work.
\subsection*{Other results}
\begin{figure}[ht]
	\centering
	\begin{subfigure}[bt]{0.45\textwidth}
		\centering
		\includegraphics[width=\textwidth]{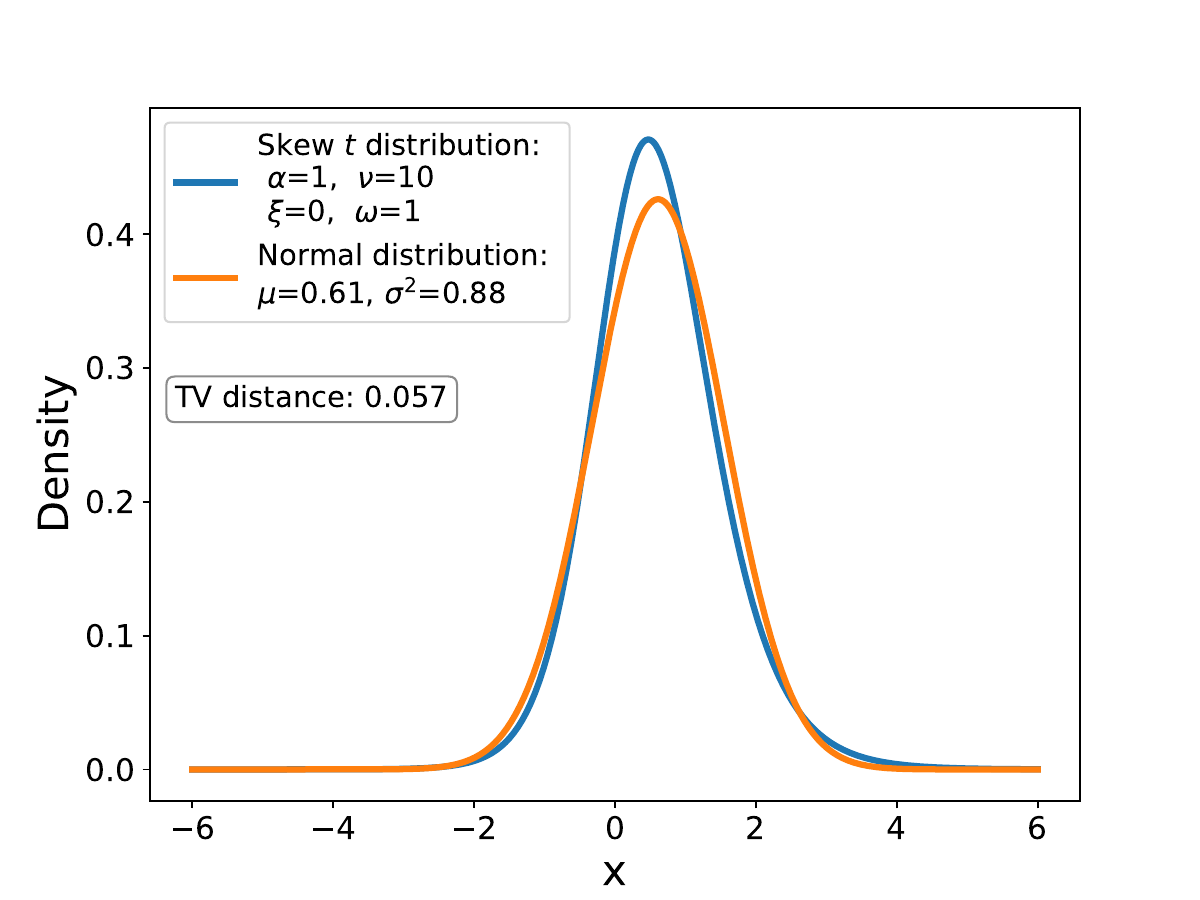}
		\caption{The density}\label{fig:figs/alpha1df10.pdf}
	\end{subfigure}
	\begin{subfigure}[bt]{0.45\textwidth}
		\centering
		\includegraphics[width=\textwidth]{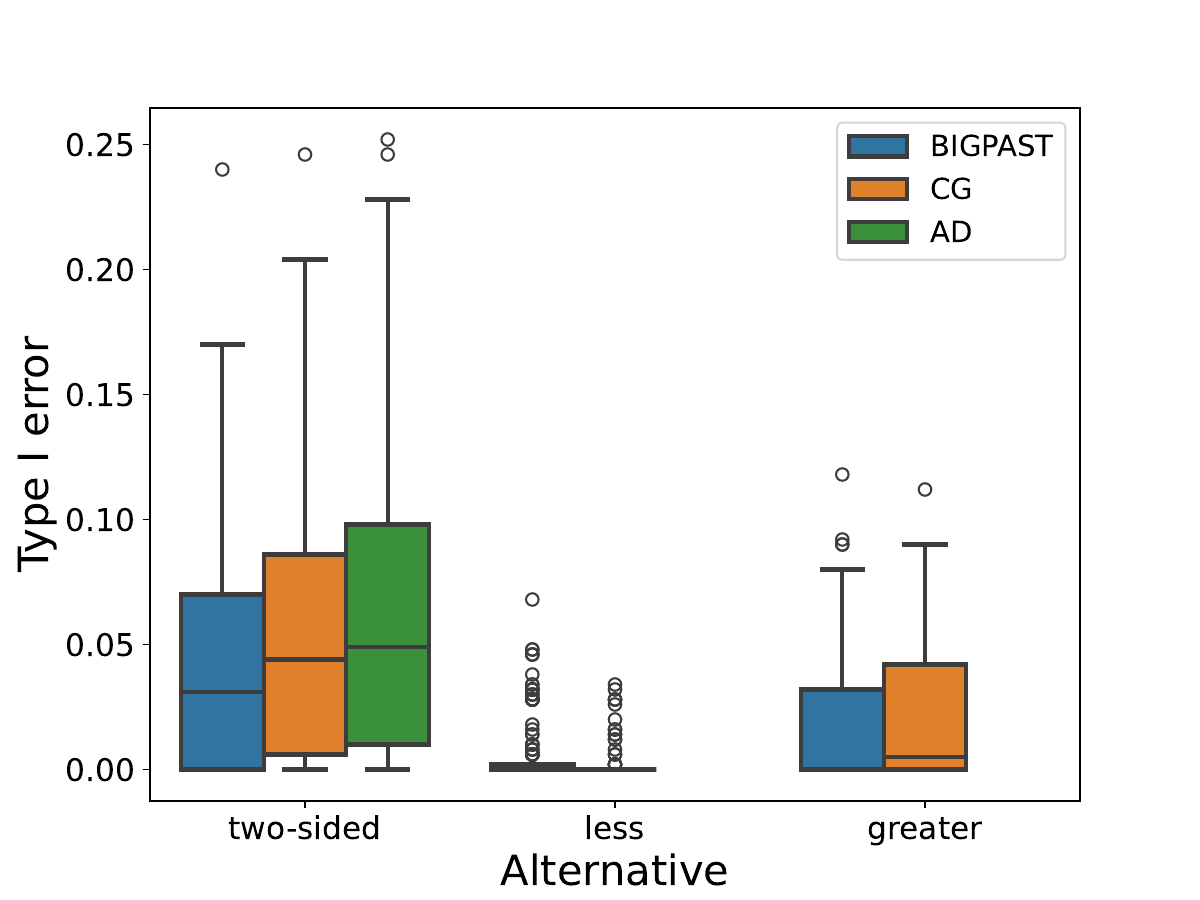}
		\caption{The type I error}\label{fig:figs/TypeIErroralpha1df10.pdf}
	\end{subfigure}
	\begin{subfigure}[bt]{0.45\textwidth}
		\centering
		\includegraphics[width=\textwidth]{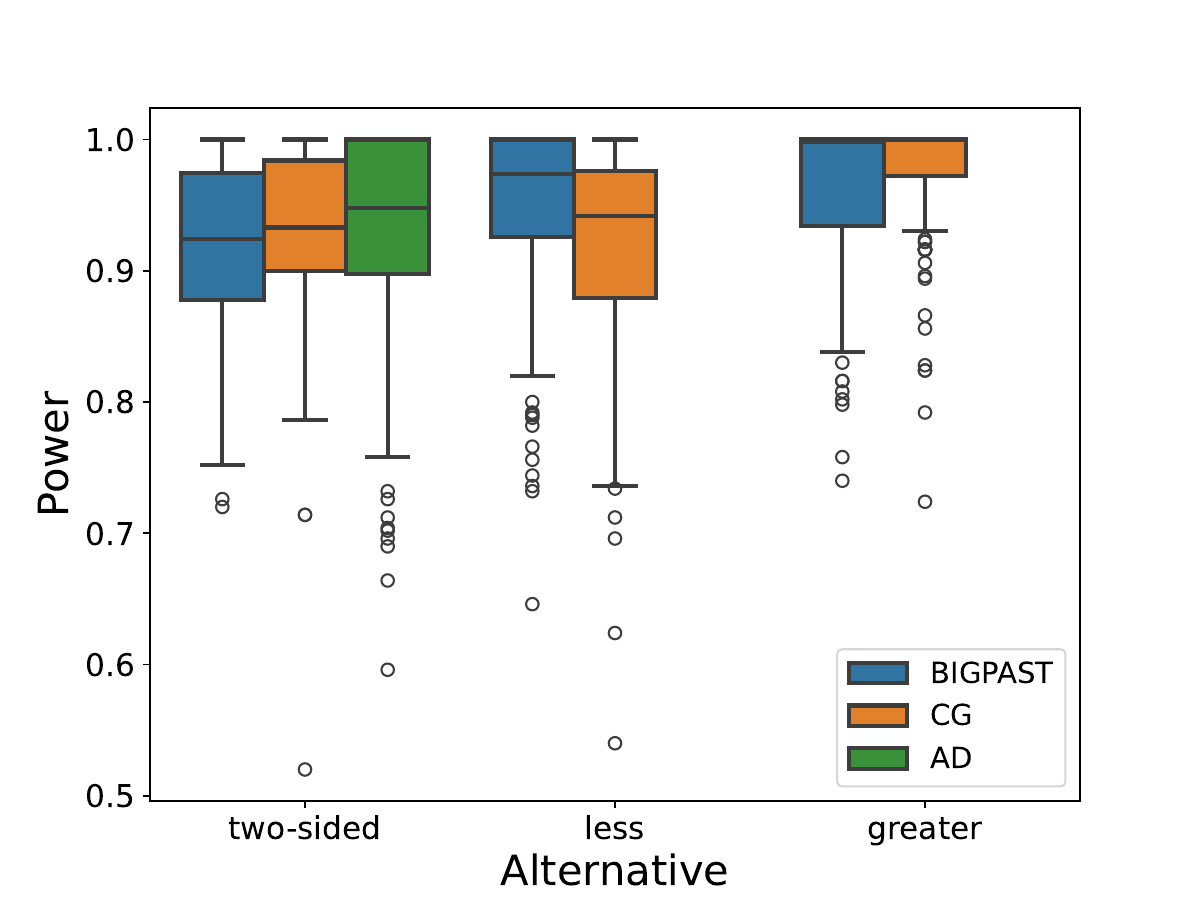}
		\caption{The power}\label{fig:figs/Poweralpha1df10.pdf}
	\end{subfigure}
	\begin{subfigure}[bt]{0.45\textwidth}
		\centering
		\includegraphics[width=\textwidth]{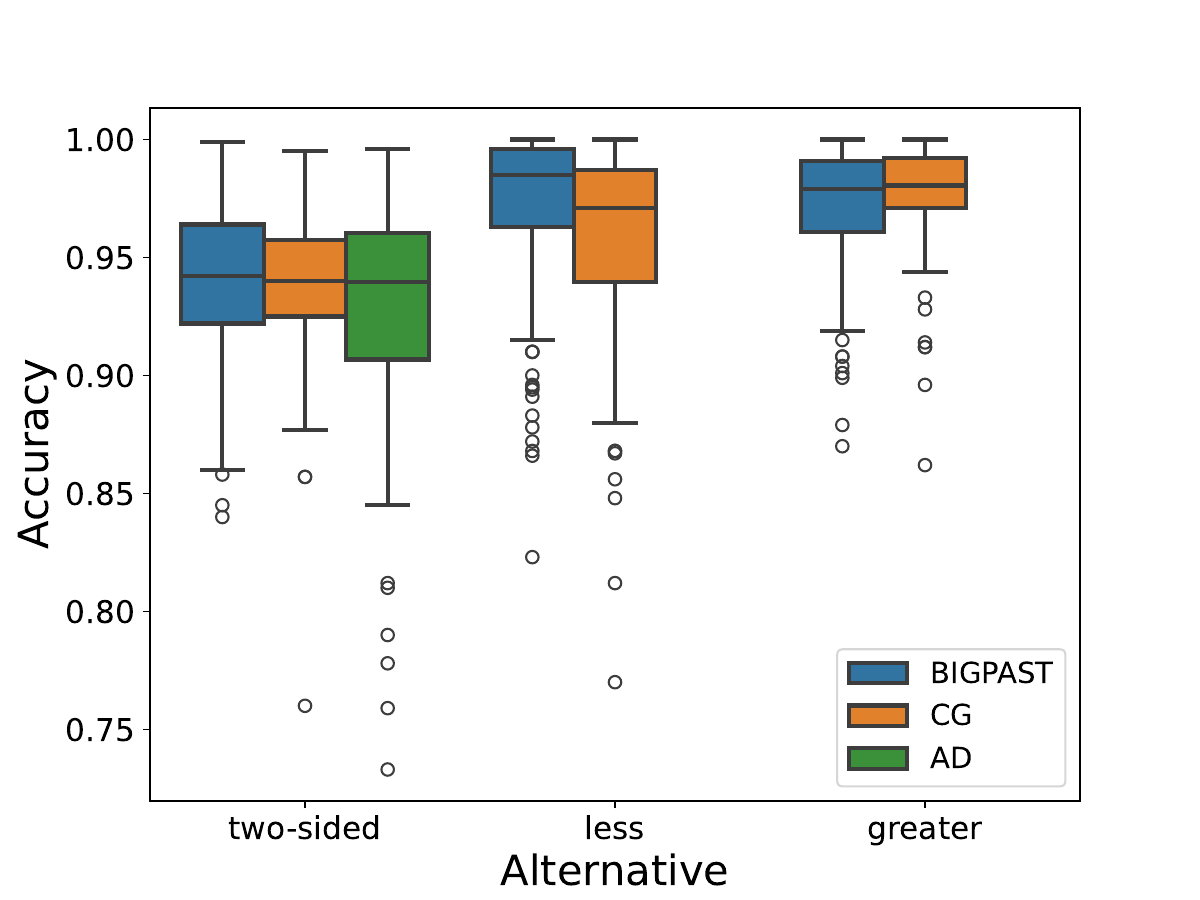}
		\caption{The accuracy}\label{fig:figs/Accuracyalpha1df10.pdf}
	\end{subfigure}
	\caption{The comparison results of BIGPAST, CG and AD approaches under three alternative hypotheses: `two-sided', `less' and `greater' respectively when \( \alpha=1 \) and \( \nu=10 \).~~\Cref{fig:figs/alpha1df10.pdf} shows the densities of skewed Student \( t \) and Normal distributions, the \( \mu \) and \( \sigma^{2} \) of Normal distribution are equal to the mean and variance of skewed Student \( t \) distribution. The Normal distribution is the theoretical assumption of the CG test when the sample comes from the skewed Student \( t \) distribution. TV distance is short for total variation distance. Each box plot summarises over 200 independent replications.}\label{fig:alpha1df10}
\end{figure}

\begin{figure}[ht]
	\centering
	\begin{subfigure}[bt]{0.45\textwidth}
		\centering
		\includegraphics[width=\textwidth]{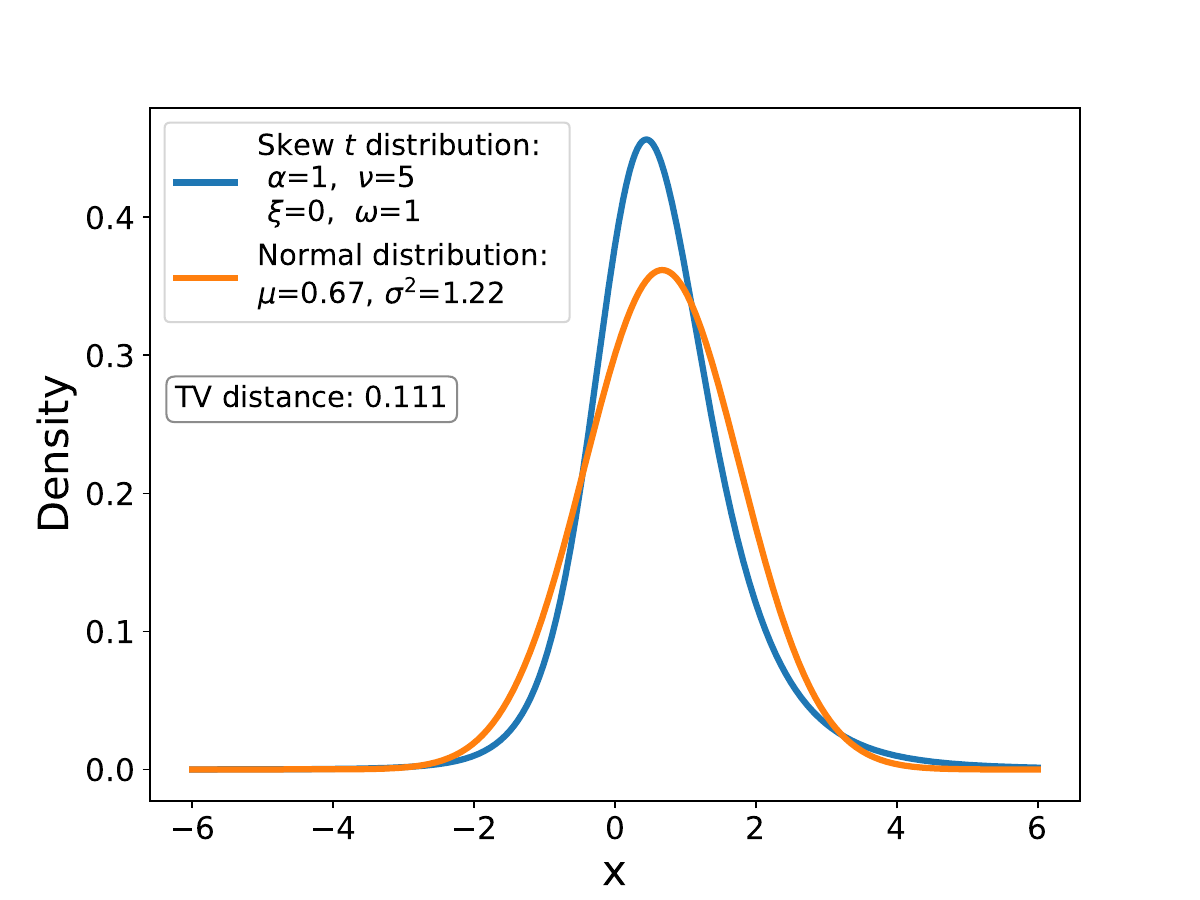}
		\caption{The density}\label{fig:figs/alpha1df5.pdf}
	\end{subfigure}
	\begin{subfigure}[bt]{0.45\textwidth}
		\centering
		\includegraphics[width=\textwidth]{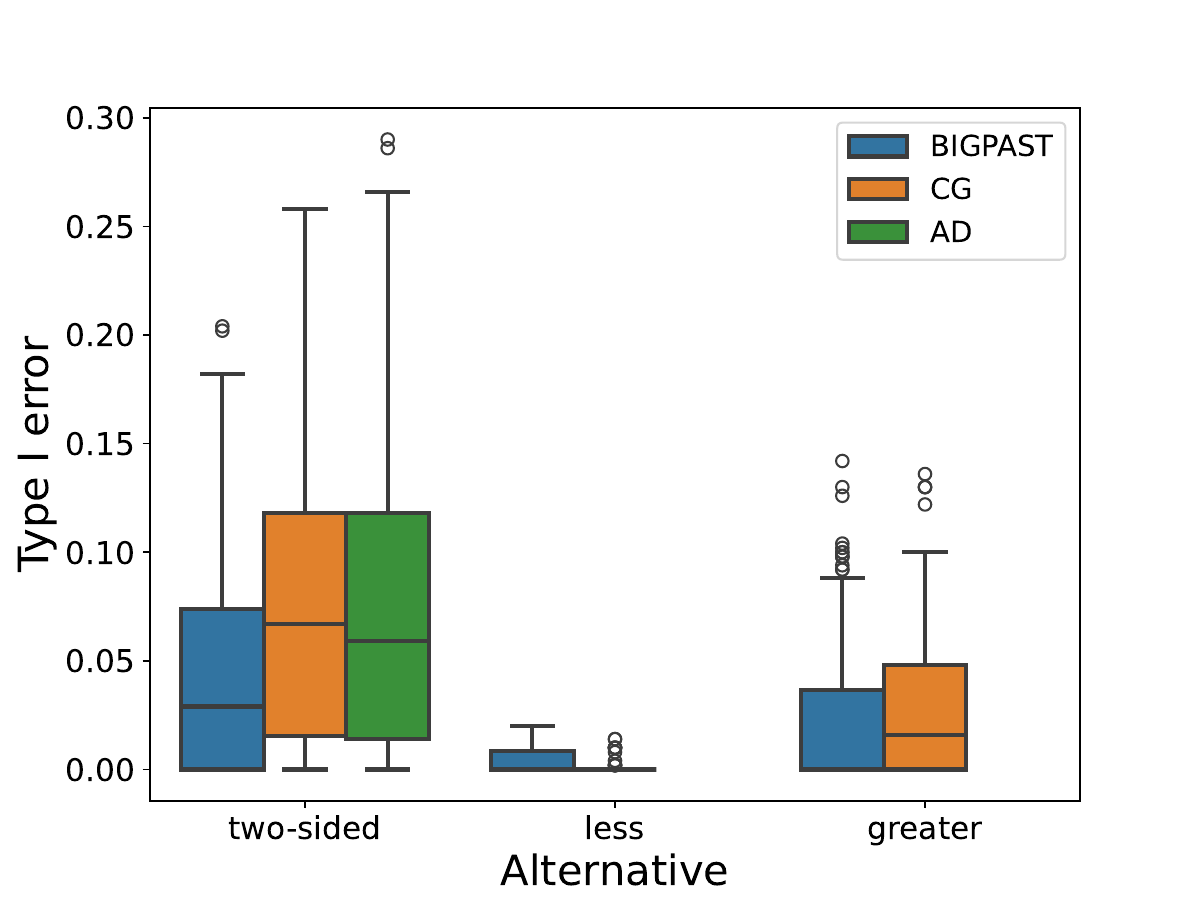}
		\caption{The type I error}\label{fig:figs/TypeIErroralpha1df5.pdf}
	\end{subfigure}
	\begin{subfigure}[bt]{0.45\textwidth}
		\centering
		\includegraphics[width=\textwidth]{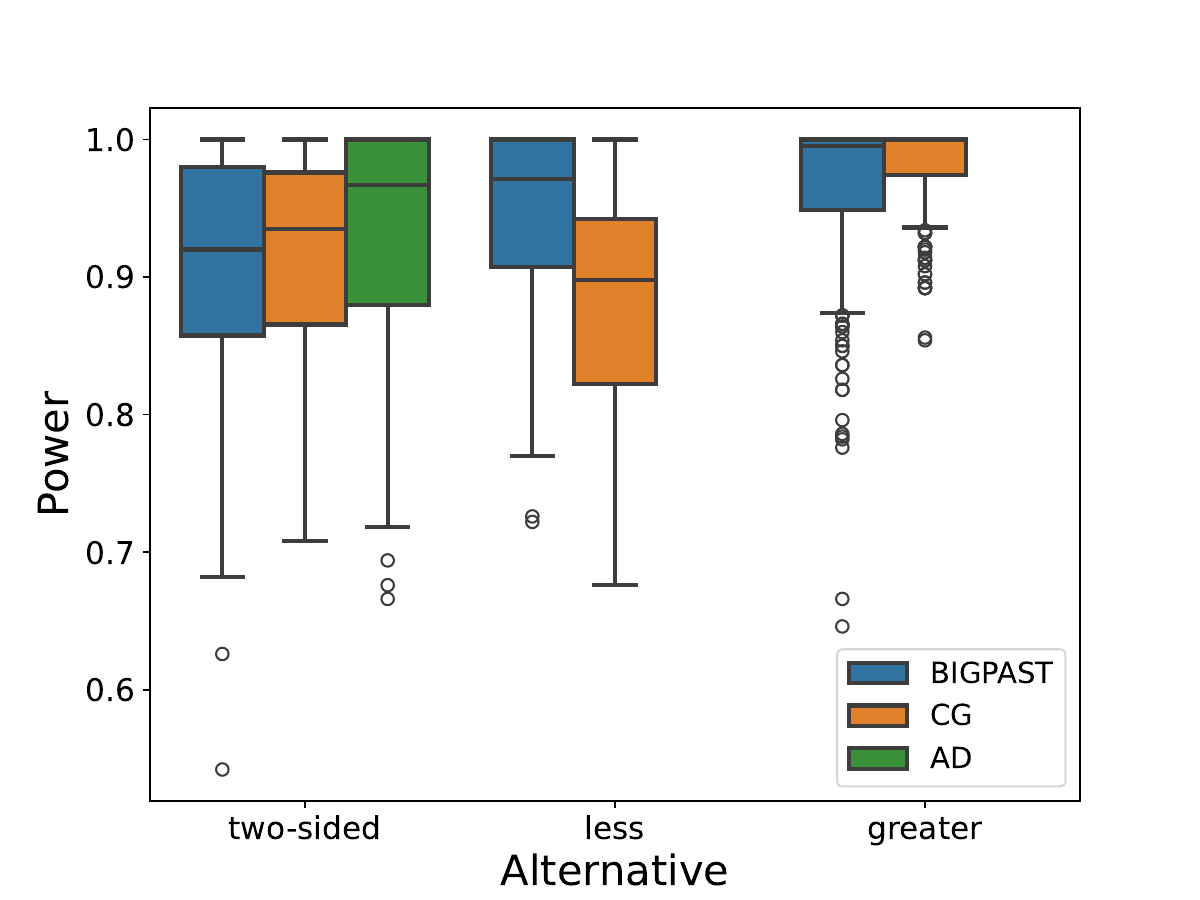}
		\caption{The power}\label{fig:figs/Poweralpha1df5.pdf}
	\end{subfigure}
	\begin{subfigure}[bt]{0.45\textwidth}
		\centering
		\includegraphics[width=\textwidth]{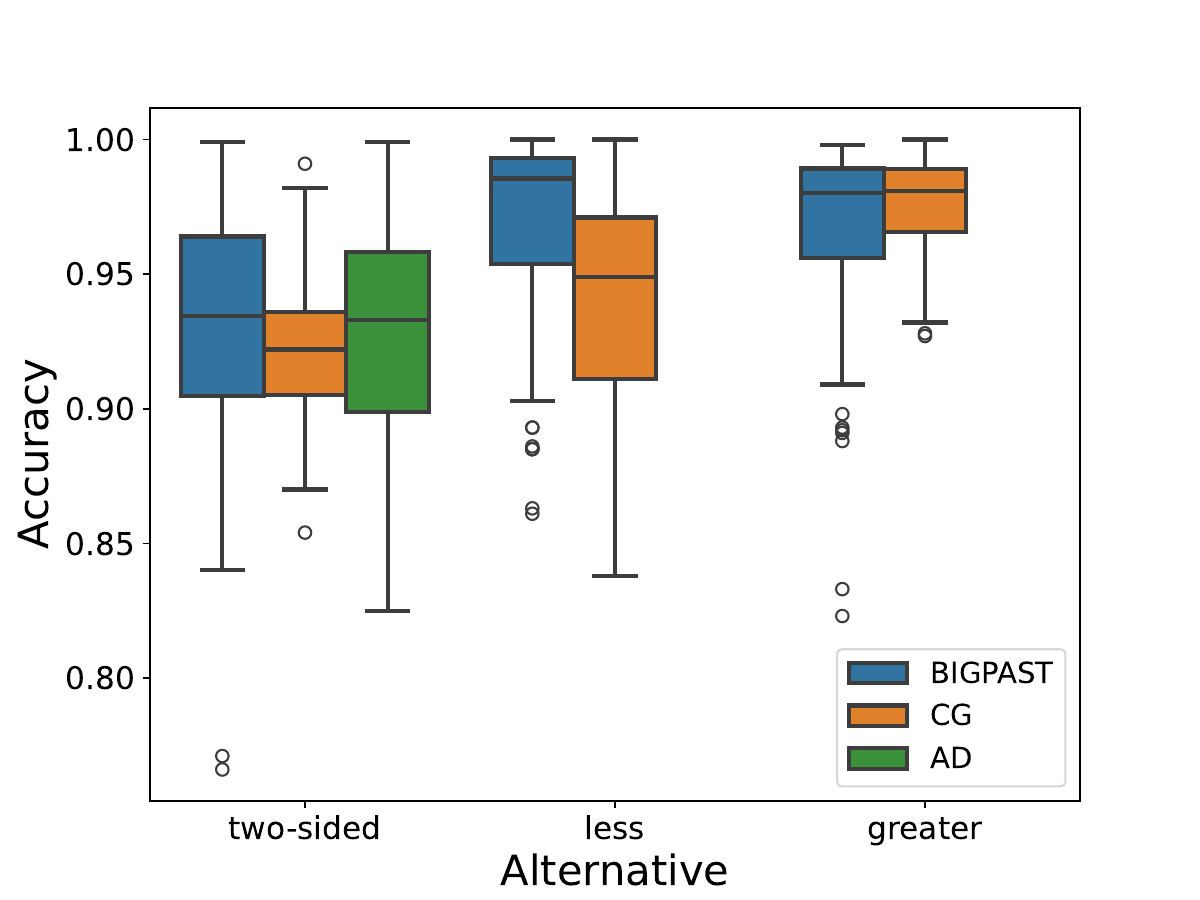}
		\caption{The accuracy}\label{fig:figs/Accuracyalpha1df5.pdf}
	\end{subfigure}
	\caption{The comparison results of BIGPAST, CG and AD approaches under three alternative hypotheses: `two-sided', `less' and `greater' respectively when \( \alpha=1 \) and \( \nu=5 \).~~\Cref{fig:figs/alpha1df5.pdf} shows the densities of skewed Student \( t \) and Normal distributions, the \( \mu \) and \( \sigma^{2} \) of Normal distribution are equal to the mean and variance of skewed Student \( t \) distribution. The Normal distribution is the theoretical assumption of the CG test when the sample comes from the skewed Student \( t \) distribution. TV distance is short for total variation distance. Each box plot summarises over 200 independent replications.}\label{fig:alpha1df5}
\end{figure}

\begin{figure}[ht]
	\centering
	\begin{subfigure}[bt]{0.45\textwidth}
		\centering
		\includegraphics[width=\textwidth]{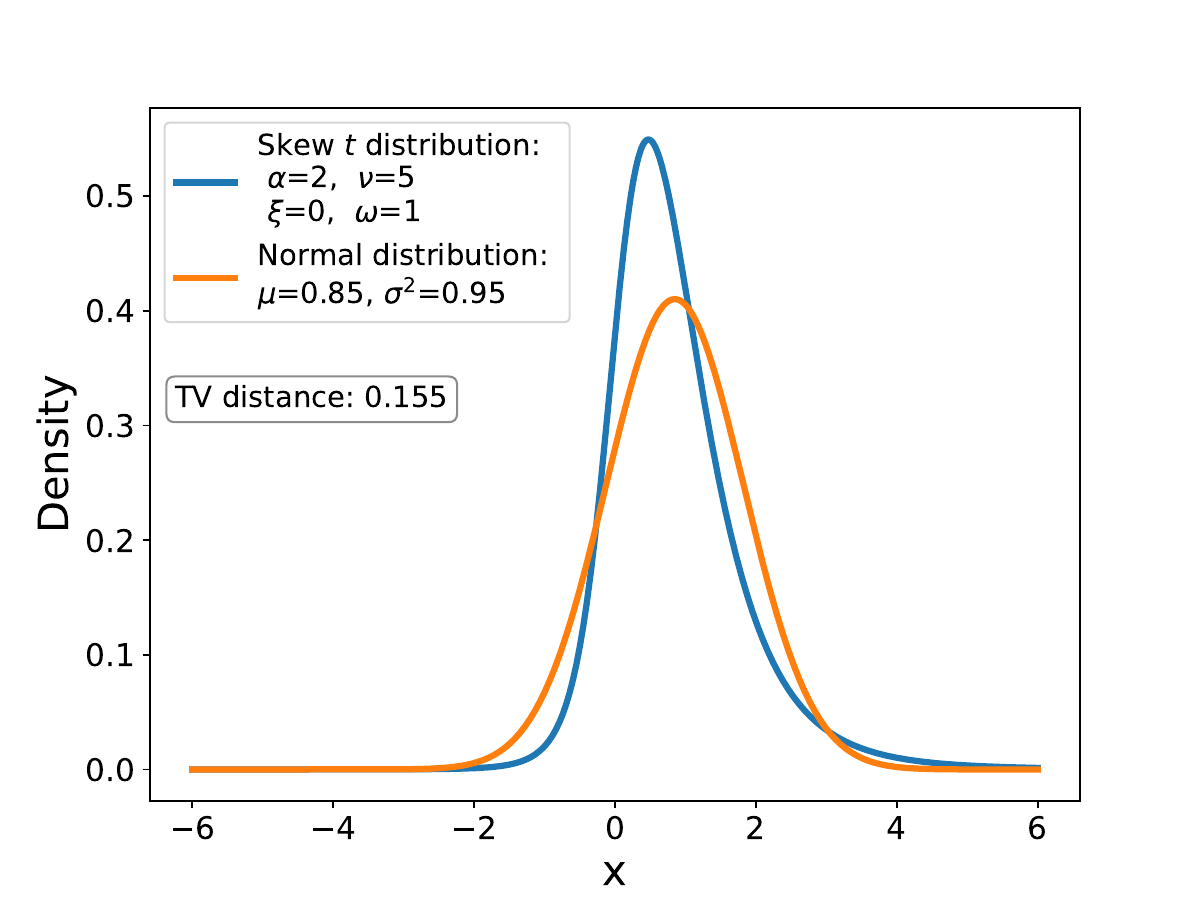}
		\caption{The density}\label{fig:figs/alpha2df5.pdf}
	\end{subfigure}
	\begin{subfigure}[bt]{0.45\textwidth}
		\centering
		\includegraphics[width=\textwidth]{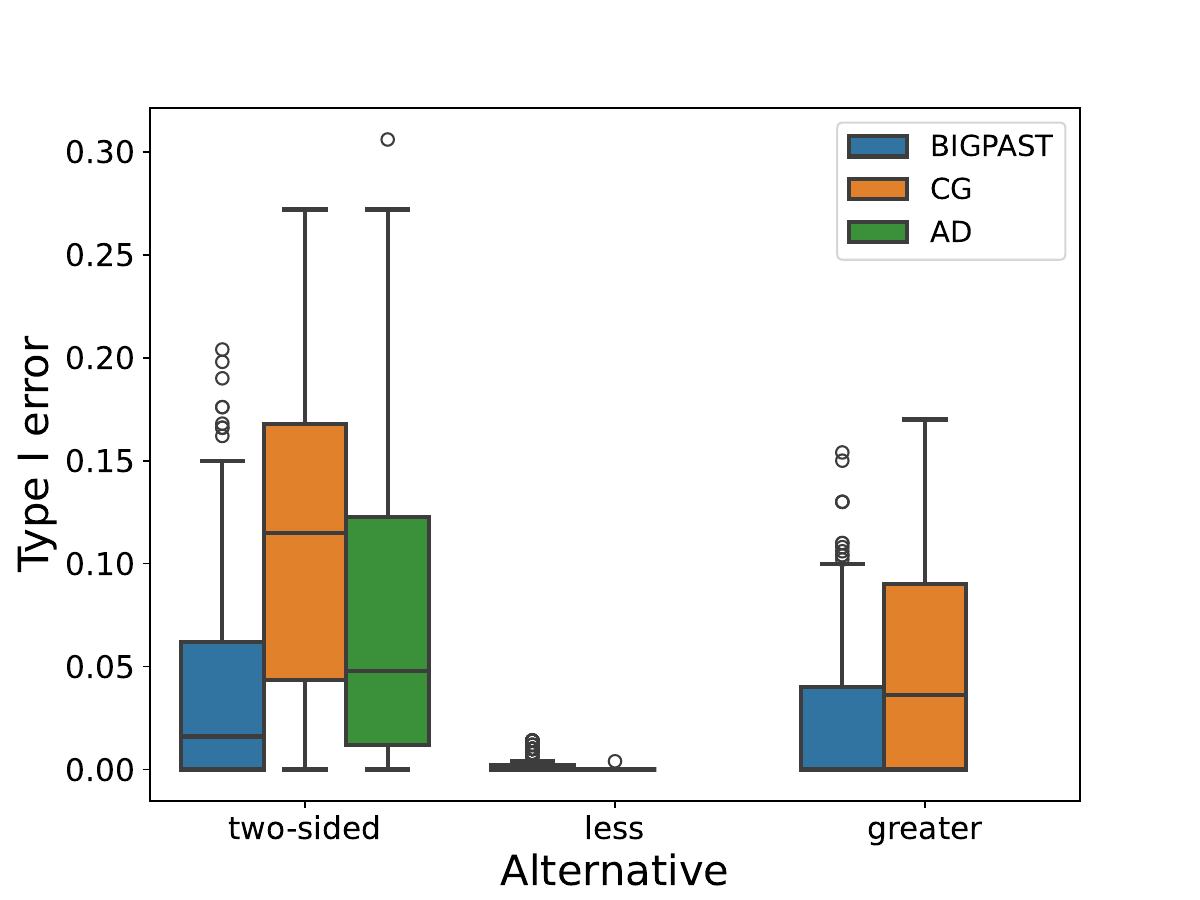}
		\caption{The type I error}\label{fig:figs/TypeIErroralpha2df5.pdf}
	\end{subfigure}
	\begin{subfigure}[bt]{0.45\textwidth}
		\centering
		\includegraphics[width=\textwidth]{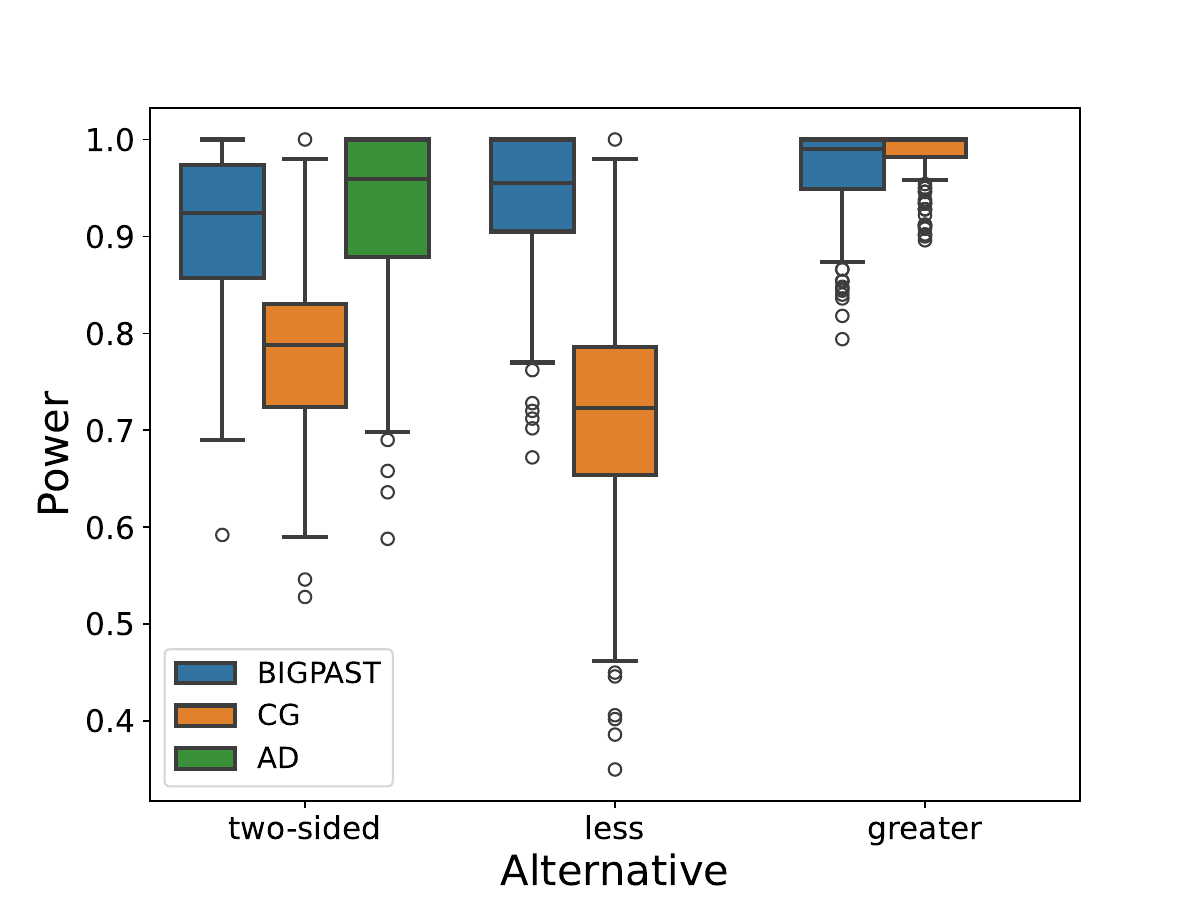}
		\caption{The power}\label{fig:figs/Poweralpha2df5.pdf}
	\end{subfigure}
	\begin{subfigure}[bt]{0.45\textwidth}
		\centering
		\includegraphics[width=\textwidth]{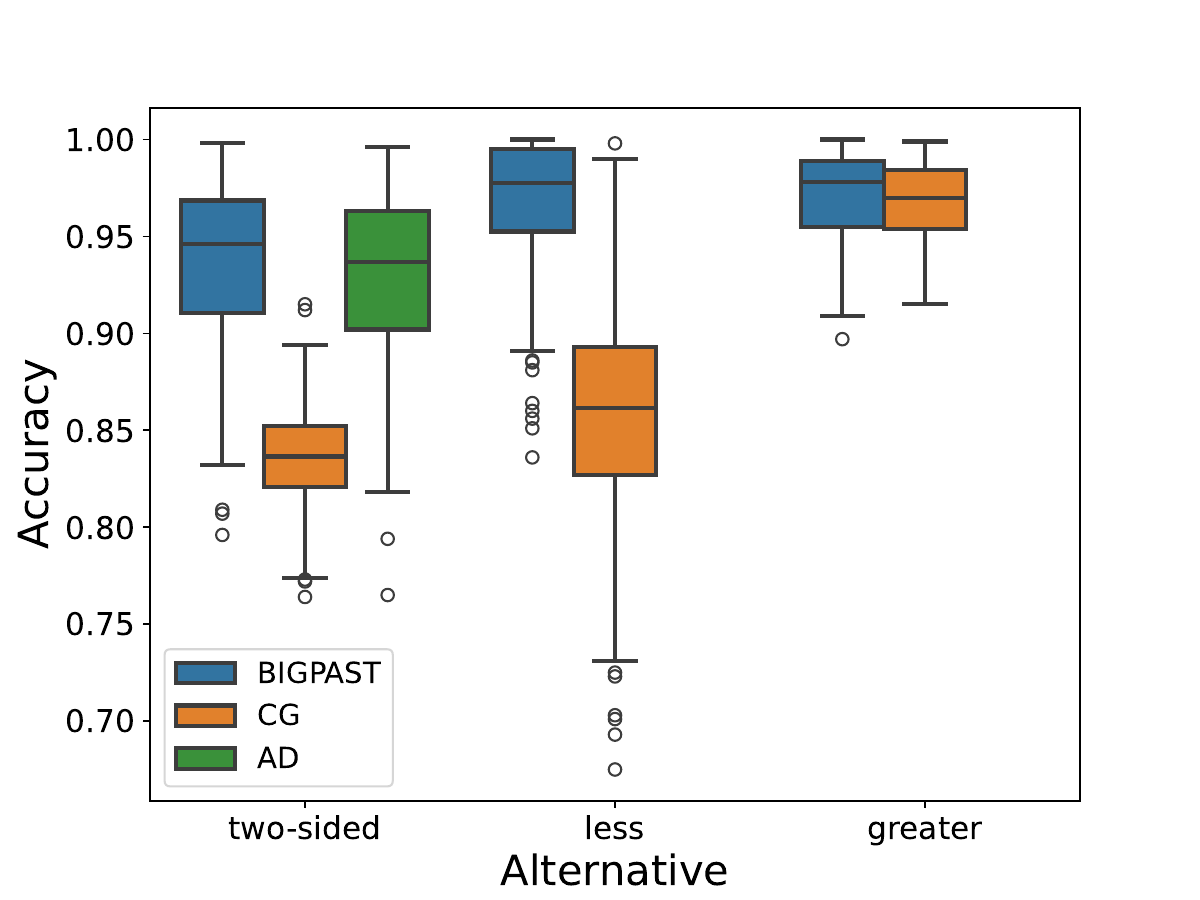}
		\caption{The accuracy}\label{fig:figs/Accuracyalpha2df5.pdf}
	\end{subfigure}
	\caption{The comparison results of BIGPAST, CG and AD approaches under three alternative hypotheses: `two-sided', `less' and `greater' respectively when \( \alpha=2 \) and \( \nu=5 \).~~\Cref{fig:figs/alpha2df5.pdf} shows the densities of skewed Student \( t \) and Normal distributions, the \( \mu \) and \( \sigma^{2} \) of Normal distribution are equal to the mean and variance of skewed Student \( t \) distribution. The Normal distribution is the theoretical assumption of the CG test when the sample comes from the skewed Student \( t \) distribution. TV distance is short for total variation distance. Each box plot summarises over 200 independent replications.}\label{fig:alpha2df5}
\end{figure}

In this section, we provide more simulation results in the main text. Specifically,~\Cref{tab:fpr_acc_less_comparison_mixed_less,tab:fpr_acc_less_comparison_mixed_greater,tab:fpr_acc_less_comparison_mixed_two_sided82} show the simulation results of Section 3.2 of the main text for the scenarios:
\begin{itemize}
	\item \textbf{\(\mathbf{c:d=50:50}\)} and one-sided test (less).
	\item \textbf{\(\mathbf{c:d=50:50}\)} and one-sided test (greater).
	\item \textbf{\(\mathbf{c:d=20:80}\)} and two-sided test.
\end{itemize}
~\Cref{tab:fpr_acc_less_comparison_mixed_two-sided_normal} show the simulation results of Section 3.2 of the main text for the normal assumption.~\Cref{fig:alpha1df10,fig:alpha1df5,fig:alpha2df5} show the results of Section 3.3 in main text for 3 combinations of  \( (\alpha, \nu) \): \( (1, 5) \), \( (2, 5) \), and \( (3, 5) \), respectively. Finally,~\Cref{tab:brain_region_index} lists the index of brain regions in our real data analysis.
\begin{table}[htbp]
	\centering
	\caption{The simulation results of false positive rate (FPR), true positive rate (TPR) and accuracy (ACC), derived from single-subject observations that consist of 50\% positive and 50\% negative, are presented under one-sided test (less). Each cell in the resulting table is calculated based on one million test outcomes. `CG', `CG-HA' and `AD' denote Crawford-Garthwaite Bayesian method, Crawford-Garthwaite Bayesian method based on hyperbolic arcsine transformation and Anderson-Darling methods, respectively.}\label{tab:fpr_acc_less_comparison_mixed_less}
	\begin{tabular}{cccccccc}
		\toprule
		& \( n \) & \( z \) & \( t \) & CG-HA & CG & AD & BIGPAST \\
		\midrule
		\multirow{4}{*}{FPR} & 50 & 0.0874 & 0.0788 & 0.0492 & 0.0788 & 0.2214 & 0.0473 \\
		& 100 & 0.0677 & 0.0630 & 0.0419 & 0.0630 & 0.1920 & 0.0389 \\
		& 200 & 0.0595 & 0.0571 & 0.0265 & 0.0571 & 0.1972 & 0.0223 \\
		& 400 & 0.0533 & 0.0520 & 0.0329 & 0.0520 & 0.1870 & 0.0144 \\
		\cmidrule{2-8}
		\multirow{4}{*}{TPR} & 50 & 0.9631 & 0.9574 & 0.9367 & 0.9574 & 0.8090 & 0.9190 \\
		& 100 & 0.9820 & 0.9802 & 0.9779 & 0.9802 & 0.8393 & 0.9571 \\
		& 200 & 0.9974 & 0.9968 & 0.9980 & 0.9968 & 0.8586 & 0.9790 \\
		& 400 & 0.9983 & 0.9982 & 0.9898 & 0.9982 & 0.8564 & 0.9819 \\
		\cmidrule{2-8}
		\multirow{4}{*}{ACC} & 50 & 0.9379 & 0.9393 & 0.9437 & 0.9393 & 0.7938 & 0.9359 \\
		& 100 & 0.9572 & 0.9586 & 0.9680 & 0.9586 & 0.8237 & 0.9591 \\
		& 200 & 0.9689 & 0.9698 & 0.9857 & 0.9698 & 0.8307 & 0.9784 \\
		& 400 & 0.9725 & 0.9731 & 0.9784 & 0.9731 & 0.8347 & 0.9837 \\
		\bottomrule
	\end{tabular}
\end{table}

\begin{table}[htbp]
	\centering
	\caption{The simulation results of false positive rate (FPR), true positive rate (TPR) and accuracy (ACC), derived from single-subject observations that consist of 50\% positive and 50\% negative, are presented under a one-sided test(greater). Each cell in the resulting table is calculated based on one million test outcomes. `CG', `CG-HA' and `AD' denote Crawford-Garthwaite Bayesian method, Crawford-Garthwaite Bayesian method based on hyperbolic arcsine transformation and Anderson-Darling methods, respectively.}\label{tab:fpr_acc_less_comparison_mixed_greater}
	\begin{tabular}{cccccccc}
		\toprule
		& \( n \) & \( z \) & \( t \) & CG-HA & CG & AD & BIGPAST \\
		\midrule
		\multirow{4}{*}{FPR} & 50 & 0.0003 & 0.0002 & 0.0023 & 0.0002 & 0.4511 & 0.0030 \\
		& 100 & 0.0000 & 0.0000 & 0.0020 & 0.0000 & 0.4654 & 0.0044 \\
		& 200 & 0.0000 & 0.0000 & 0.0003 & 0.0000 & 0.4699 & 0.0027 \\
		& 400 & 0.0000 & 0.0000 & 0.0001 & 0.0000 & 0.4701 & 0.0021 \\
		\cmidrule{2-8}
		\multirow{4}{*}{TPR} & 50 & 0.6402 & 0.5907 & 0.9103 & 0.5908 & 0.8229 & 0.8858 \\
		& 100 & 0.5445 & 0.5197 & 0.8975 & 0.5198 & 0.7792 & 0.9206 \\
		& 200 & 0.5824 & 0.5693 & 0.9138 & 0.5694 & 0.8491 & 0.9628 \\
		& 400 & 0.5389 & 0.5323 & 0.9281 & 0.5323 & 0.8197 & 0.9670 \\
		\cmidrule{2-8}
		\multirow{4}{*}{ACC} & 50 & 0.8199 & 0.7953 & 0.9540 & 0.7953 & 0.6859 & 0.9414 \\
		& 100 & 0.7722 & 0.7599 & 0.9478 & 0.7599 & 0.6569 & 0.9581 \\
		& 200 & 0.7912 & 0.7847 & 0.9568 & 0.7847 & 0.6896 & 0.9800 \\
		& 400 & 0.7694 & 0.7661 & 0.9640 & 0.7661 & 0.6748 & 0.9825 \\
		\bottomrule
	\end{tabular}
\end{table}

\begin{table}[htbp]
	\centering
	\caption{The simulation results of false positive rate (FPR), true positive rate (TPR) and accuracy (ACC), derived from single-subject observations that consist of 20\% positive and 80\% negative, are presented under a two-sided test. Each cell in the resulting table is calculated based on one million test outcomes. `CG', `CG-HA' and `AD' denote the Crawford-Garthwaite Bayesian method, the Crawford-Garthwaite Bayesian method based on hyperbolic arcsine transformation and the Anderson-Darling methods, respectively.}\label{tab:fpr_acc_less_comparison_mixed_two_sided82}
	\begin{tabular}{cccccccc}
		\toprule
		& \( n \) & \( z \) & \( t \) & CG-HA & CG & AD & BIGPAST \\
		\midrule
		\multirow{4}{*}{FPR} & 50 & 0.1383 & 0.1237 & 0.0498 & 0.1237 & 0.1170 & 0.0492 \\
		& 100 & 0.1341 & 0.1261 & 0.0485 & 0.1262 & 0.0699 & 0.0411 \\
		& 200 & 0.1218 & 0.1174 & 0.0367 & 0.1174 & 0.0641 & 0.0303 \\
		& 400 & 0.1294 & 0.1273 & 0.0396 & 0.1273 & 0.0468 & 0.0268 \\
		\cmidrule{2-8}
		\multirow{4}{*}{TPR} & 50 & 0.6622 & 0.6131 & 0.8243 & 0.6130 & 0.8519 & 0.8446 \\
		& 100 & 0.6439 & 0.6222 & 0.9149 & 0.6222 & 0.9589 & 0.9270 \\
		& 200 & 0.6275 & 0.6160 & 0.9165 & 0.6161 & 0.9650 & 0.9327 \\
		& 400 & 0.6036 & 0.5978 & 0.9196 & 0.5979 & 0.9805 & 0.9761 \\
		\cmidrule{2-8}
		\multirow{4}{*}{ACC} & 50 & 0.8218 & 0.8236 & 0.9250 & 0.8236 & 0.8768 & 0.9295 \\
		& 100 & 0.8215 & 0.8235 & 0.9442 & 0.8235 & 0.9358 & 0.9525 \\
		& 200 & 0.8281 & 0.8293 & 0.9539 & 0.8293 & 0.9417 & 0.9623 \\
		& 400 & 0.8172 & 0.8177 & 0.9523 & 0.8178 & 0.9586 & 0.9738 \\
		\bottomrule
	\end{tabular}
\end{table}

\begin{table}[htbp]
	\centering
	\caption{The simulation results of false positive rate (FPR), true positive rate (TPR) and accuracy (ACC), derived from single-subject observations that consist of 50\% positive and 50\% negative, are presented under a two-sided test. The underlying density is the Normal density function. Each cell in the resulting table is calculated based on one million test outcomes. `CG' and `AD' represent the Crawford-Garthwaite Bayesian and Anderson-Darling methods, respectively.}\label{tab:fpr_acc_less_comparison_mixed_two-sided_normal}
	\begin{tabular}{ccccccc}
		\toprule
		& \( n \) & \( z \) & \( t \) & CG & AD & BIGPAST \\
		\midrule
		\multirow{4}{*}{FPR} & 50 & 0.0467 & 0.0322 & 0.0322 & 0.0943 & 0.0313 \\
		& 100 & 0.0372 & 0.0286 & 0.0286 & 0.0523 & 0.0266 \\
		& 200 & 0.0324 & 0.0280 & 0.0280 & 0.0541 & 0.0231 \\
		& 400 & 0.0243 & 0.0220 & 0.0220 & 0.0392 & 0.0195 \\
		\cmidrule{2-7}
		\multirow{4}{*}{TPR} & 50 & 0.9353 & 0.9041 & 0.9041 & 0.9125 & 0.8854 \\
		& 100 & 0.9504 & 0.9358 & 0.9359 & 0.9302 & 0.9248 \\
		& 200 & 0.9617 & 0.9542 & 0.9541 & 0.9621 & 0.9450 \\
		& 400 & 0.9737 & 0.9700 & 0.9700 & 0.9568 & 0.9596 \\
		\cmidrule{2-7}
		\multirow{4}{*}{ACC} & 50 & 0.9443 & 0.9360 & 0.9359 & 0.9091 & 0.9270 \\
		& 100 & 0.9566 & 0.9536 & 0.9536 & 0.9389 & 0.9491 \\
		& 200 & 0.9647 & 0.9631 & 0.9631 & 0.9540 & 0.9610 \\
		& 400 & 0.9747 & 0.9740 & 0.9740 & 0.9588 & 0.9701 \\
		\bottomrule
	\end{tabular}
\end{table}
\clearpage
\newpage
\subsection*{Other results in~\Cref{subsec:Model_misspecification_error}}\label{subsec:other_results_in}
\clearpage
\newpage

\begin{table}[htbp]
	\centering
	\caption{The index of brain region of the Desikan-Killiany cortical atlas.}\label{tab:brain_region_index}
	\begin{tabular}{|c|l|c|l|}
		\hline
		\textbf{Index} & \textbf{label\_name} & \textbf{Index} & \textbf{label\_name} \\
		\hline
		0 & lh-bankssts & 34 & rh-bankssts \\
		1 & lh-caudalanteriorcingulate & 35 & rh-caudalanteriorcingulate \\
		2 & lh-caudalmiddlefrontal & 36 & rh-caudalmiddlefrontal \\
		3 & lh-cuneus & 37 & rh-cuneus \\
		4 & lh-entorhinal & 38 & rh-entorhinal \\
		5 & lh-frontalpole & 39 & rh-frontalpole \\
		6 & lh-fusiform & 40 & rh-fusiform \\
		7 & lh-inferiorparietal & 41 & rh-inferiorparietal \\
		8 & lh-inferiortemporal & 42 & rh-inferiortemporal \\
		9 & lh-insula & 43 & rh-insula \\
		10 & lh-isthmuscingulate & 44 & rh-isthmuscingulate \\
		11 & lh-lateraloccipital & 45 & rh-lateraloccipital \\
		12 & lh-lateralorbitofrontal & 46 & rh-lateralorbitofrontal \\
		13 & lh-lingual & 47 & rh-lingual \\
		14 & lh-medialorbitofrontal & 48 & rh-medialorbitofrontal \\
		15 & lh-middletemporal & 49 & rh-middletemporal \\
		16 & lh-paracentral & 50 & rh-paracentral \\
		17 & lh-parahippocampal & 51 & rh-parahippocampal \\
		18 & lh-parsopercularis & 52 & rh-parsopercularis \\
		19 & lh-parsorbitalis & 53 & rh-parsorbitalis \\
		20 & lh-parstriangularis & 54 & rh-parstriangularis \\
		21 & lh-pericalcarine & 55 & rh-pericalcarine \\
		22 & lh-postcentral & 56 & rh-postcentral \\
		23 & lh-posteriorcingulate & 57 & rh-posteriorcingulate \\
		24 & lh-precentral & 58 & rh-precentral \\
		25 & lh-precuneus & 59 & rh-precuneus \\
		26 & lh-rostralanteriorcingulate & 60 & rh-rostralanteriorcingulate \\
		27 & lh-rostralmiddlefrontal & 61 & rh-rostralmiddlefrontal \\
		28 & lh-superiorfrontal & 62 & rh-superiorfrontal \\
		29 & lh-superiorparietal & 63 & rh-superiorparietal \\
		30 & lh-superiortemporal & 64 & rh-superiortemporal \\
		31 & lh-supramarginal & 65 & rh-supramarginal \\
		32 & lh-temporalpole & 66 & rh-temporalpole \\
		33 & lh-transversetemporal & 67 & rh-transversetemporal \\
		\hline
	\end{tabular}
\end{table}
\end{document}